\documentclass[11pt]{article}
\usepackage[letterpaper, left=.75in, top=.75in, right=.75in, bottom=.75in,nohead,includefoot, verbose,ignoremp]{geometry}
\usepackage{charter,graphicx,natbib,amsmath,amssymb,amsfonts,hypernat,caption,placeins}

	\usepackage[usenames,dvipsnames,x11names]{xcolor}
	\usepackage[pdftex,pagebackref=true]{hyperref}
	\hypersetup{
	colorlinks,
	linkcolor=RoyalBlue1,
	urlcolor=RoyalBlue2,
	citecolor=RoyalBlue3
	}

\def\bnu{\mbox{\boldmath$\nu$}}
\def\bomega{\mbox{\boldmath$\omega$}}
\def\bxi{\mbox{\boldmath$\xi$}}
\def \xb {\mathbf{x}}
\def \zb {\mathbf{z}}
\def \yb {\mathbf{y}}
\def \mb {\mathbf{m}}

\def \Hb {\mathbf{H}}

\def \Y {\mathbf{Y}}
\def \Yo {\mathbf{Y}_o}
\def \X {\mathbf{X}}

\def\eqn#1{equation~(\ref{#1})}

\bibpunct{[}{]}{,}{n}{}{,}

\begin{document}
 
\title{Bayesian Analysis of Immune Response Dynamics\\ with Sparse Time Series Data}

\author{
	 Fernando V. Bonassi\footnote{Google Youtube. \href{mailto:bonassi@gmail.com}{bonassi@gmail.com}},\, 
       Cliburn Chan\footnote{Duke University Medical Center. \href{mailto:cliburn.chan@duke.edu}{cliburn.chan@duke.edu}}\, 
	and 
	Mike West\footnote{Duke University. \href{mailto:mw@stat.duke.edu}{mw@stat.duke.edu}}
	 		\footnote{Research partly supported by a grant from the National Science Foundation [DMS-1106516]
			and grants from the National  \newline\indent\quad Institutes of Heath [NIH P50-GM081883 and 5P30 AI064518]. 
			 Any opinions, findings and conclusions or  recommendations  \newline\indent\quad expressed in this work are those of 
			the authors and do not necessarily reflect the views  of the NSF or NIH. }
}

\maketitle

\begin{abstract}
In vaccine  development,
the temporal  profiles of relative  abundance of  subtypes of immune cells (T-cells)
is key to understanding vaccine efficacy.   Complex and expensive experimental studies
 generate very sparse time series data on this {\em immune response}.
Fitting multi-parameter dynamic models
of the immune response dynamics-- central to evaluating  mechanisms underlying vaccine efficacy--
is challenged by data sparsity.  The research reported here addresses this challenge.  For
 HIV/SIV vaccine studies in macaques, we:  (a) introduce novel dynamic models of progression of
cellular populations over time with relevant, time-delayed components reflecting the vaccine response;
(b)  define an effective Bayesian model fitting strategy that couples Markov chain Monte Carlo (MCMC) with
Approximate Bayesian Computation (ABC)-- building on the complementary strengths of the
two approaches, neither of which is effective alone; (c) explore questions of
information content in the sparse time series for each of the model parameters, linking into
experimental design and model simplification for future experiments;  and
(d) develop, apply and compare the analysis with samples from
a recent HIV/SIV  experiment, with novel insights and conclusions about the progressive response to the
vaccine, and how this varies across subjects.



\end{abstract}

\section{Introduction}

In vaccine design and development,
experiments in non-human primates (NHPs) are necessary preludes to  human clinical trials.
Our focus here is on a key case in point, that of new HIV/SIV immunization strategies.
While becoming more prevalent, NHP vaccination experiments are expensive and complex, so typically
generate small sample sizes and a limited number of longitudinal observations~\citep{pmid23615116,morgan2008use,johnston2000role}.
Mathematical modeling of the immune response dynamics to vaccination can provide insight into the mechanisms underlying vaccine efficacy and help predict the likely human response to the immunization strategy~\citep{elemans2011don,pmid18586297,pmid17202215,scherer2002mathematical}.
To capture the complexity of the immune response, such mathematical models are often represented using coupled systems of nonlinear ordinary differential equations (ODEs).
We are then faced with the challenges of model fitting with multiple model parameters on very sparse data.
There is also a key interest in variability in these characterizing parameters across subjects,
since NHPs come from outbred populations so that the immune response is  likely to vary across individuals.

We address these general questions in a specific vaccine design study,
involving evaluation of efficacy of a replicating Adenovirus Simian Immunodeficiency Virus (SIV) vector~\citep{pmid18694354} introduced via different mucosal routes in a collection of
macaques~\citep{pmid22441384}.  We summarize the   context and experiment, and   address questions of
dynamic modeling of the vaccine response   of each macaque.
Beginning with coupled systems of ODEs
for the vaccine response,  we discretize to define a class of non-linear, discrete-time state-space models for the
data defined by cell subtype frequencies.   This model framework builds on prior models in the field,
while going much further with innovations in time-delay modeling and discretization. We also describe
a mapping of model parameters to {\em dimensionless parameters} for fundamental characterization and
comparison across subjects.

Standard computational methods for model fitting--  analytic approximations
coupled with either Markov chain Monte Carlo~\citep{Niemi2010,West1997}, or sequential Monte Carlo
including Approximate Bayesian Computation (ABC)~\citep[e.g.,][]{Liu2001,toni2009,Prado2010,Bonassi2011}-- are fundamentally
challenged due to non-linearities and data sparsity.
After exploring  such  approaches, we have defined  with a creative coupling of MCMC
and ABC methods that builds on the complementary strengths of the two approaches.

In any modeling context where data is very sparse, questions should arise about which of the parameters are
informed upon, at all, and the extent of the information content of the data for each parameter.   We discuss
and investigate this here, evaluating the prior-to-posterior mapping via entropy and variance to
quantify {\em \lq\lq learnability''} of parameters from the sparse time series data. This is useful in
interpreting posterior summaries and comparing across experimental subjects,  and aids thinking
about potential model simplifications and design for future experiments.

Application to the macaque data is discussed and evaluated in the context of the goals of the experiment.
The precision of inferences on temporal profiles in the immune response is highlighted along with these
questions of model specification and potential parameter redundancy.  We  see real differences between
some of the response-characterizing parameters across individuals, and strong evidence from the
overall study that there is a significant time delay in memory T-cells responding to the vaccine interventions.
These broad summaries alone represent novel and meaningful findings in the HIV/SIV vaccine response
field. Moreover, the analyses suggest that the dynamics of the immune response depends on the vaccination route,(i.e., the
way in which the individual is vaccinated). This is a result that is not obvious from inspection of the time series data alone, and
is a novel and potential important finding in its own right, suggesting the need-- and likely design principles-- for
further experiments.

\section{Basics of Vaccine Biology and Experiment}

We performed a longitudinal study to compare the mucosal immune response when a prime-boost vaccine using replication-competent viral vector was given via different routes. Replication-competent Adenovirus (Ad5) with recombinant Simian Immunodeficiency Virus (SIV) Gag genes (Ad5hr-SIV{\em gag}) and a GFP marker for tracking was administered to Indian Rhesus macaques via intra-nasal/intra-tracheal, intra-vaginal and intra-rectal routes. The macaques were vaccinated twice with the replicating Ad recombinants (encoding SIV env/rev and gag, plus Ad-GFP) at weeks 0 and 12, and boosted with SIV envelope protein plus adjuvant at weeks 24 and 36 of the study.  Relative abundance of functionally and maturationaly
different T-cell subtypes were estimated  from flow cytometric assays~\citep[e.g.,][]{Chan2008,Chan2010,Lin2012} on samples from gut mucosa and peripheral blood. A small number of the macaques also had 2-4 measurments of GFP-labelled Adnovirus infected macrophages in the gut mucosa.
The summary estimates of cell subtype proportions at each time point assayed provide the raw time series for each subject
macaque.  The key cellular subtype of interest here is  the  antigen-specific memory T-cells, and models will address its
relation to the level of abundance of the adenovirus-infected macrophages over time.

Replication-competent vectors continue to propagate themselves after vaccination and can sustain a long-term immune response against the SIV genes encoded by the vector. Hence, we expect that the population size of the antigen-specific memory T-cell population is driven by the size of the replicating vector population. Laboratory experiments suggested that the Adenovirus vector resided in host macrophages in the gut mucosa. Our
development of nonlinear models of the dynamics of the Ad5-infected macrophage cells (using GFP-expression as a marker for Ad5 infection)  in the rectal mucosa, coupled with  the antigen-specific memory T-cell response in the blood, aim at evaluating the effects of different vaccination routes on the host response to a replication-competent Ad5 vector.

\section{Conceptual/Qualitative Dynamic Modeling Framework}

We begin with  stylized thinking in terms  of coupled ordinary differential equations (ODEs) to describe the qualitative
temporal changes in antigen and cellular concentrations in the immune system~\citep[e.g.,][]{DeBoer2003}. This is then extended
in practicable statistical model development that involves discrete-time representations of the conceptual differential equations systems, and time-delays in the immune response to vaccine intervention. The discretized system of model equations is overlaid with stochastic components to realistically reflect measurement error and unmodeled structure. The model includes noise and mechanistic rate parameters, initial conditions, and-- critically-- missing ({\em latent}) elements of the data time series on subsets of cell frequencies.

\subsection{Basic ODE Model}

The motivating ODE model describes evolution in continuous time $t$ of Adenovirus ($V_t$) and T-cell memory ($M_t$) populations, following immunization of a subject with the adenovirus Ad5hr-SIV{\em gag}. With subject-specific parameters, take
\begin{equation}
\label{eq:odemodel_abc_mcmc}
\begin{array}{lcl}
 {\displaystyle \frac{dV_t}{dt}} & = & \beta V_t \left( 1 -  {\displaystyle \frac{V_t}{K_V}} \right),  \vspace{0.35cm} \\
{\displaystyle \frac{dM_t}{dt}} & = & \alpha V_t + \rho V_t M_t \left( 1 - {\displaystyle  \frac{M_t}{K_M}} \right).
\end{array}
\end{equation}
The viral population is described by the first equation, a logistic growth form.
In the second equation, the first term has $\alpha=\phi N$ where $N$ represents na\"{i}ve T-cells; the rate of generation of new na\"{i}ve cells that recognize viral epitopes in the presence of a persistent Ad5 infection is fixed at $\alpha = \phi N$, and the virus drives their differentiation into into memory cells. The second term
reflects  density-dependent proliferation of pre-existing memory cells. Finally,
$V_t$ and $M_t$ have natural carrying capacities $K_V$ and $K_M$, respectively.

\subsection{Non-dimensionalization}

The essential structure of the model, and a reduced number of parameters, arises via a map to the
{\em non-dimensional}  version of the  model. As will be seen later, this has a critical practical role in quantifying learnability of model parameters. Let $\mu$, $\nu$ and $\lambda$ be scaling variables--
positive, but otherwise arbitrary and to be chosen-- that define
\begin{equation*}
v_t = V_t/\mu, \quad m_t = M_t/\nu \quad\textrm{and}\quad s = t/\lambda.
\end{equation*}
Then, via the chain rule,
\begin{equation*}
\frac{dv_s}{ds} = \frac{\lambda}{\mu}\frac{dV_t}{dt}
\quad\textrm{and}\quad
\frac{dm_s}{ds} = \frac{\lambda}{\nu}\frac{dM_t}{dt}.
\end{equation*}
Substituting in \eqn{eq:odemodel_abc_mcmc} gives, after some algebra,
\begin{align*}
\frac{dv_s}{ds} &
= \lambda \beta v_s \left( 1 - \frac{v_s}{K_V/\mu} \right), \\
\frac{dm_s}{ds} &
=\frac{ \lambda \alpha \mu}{ \nu} v_s + \lambda \rho \mu v_s m_s \left( 1 - \frac{m_s}{K_M/\nu} \right).
\end{align*}
Now choose to set $\mu=K_V$, $\nu=K_M$ and $\lambda=1/\beta$, so that
equation~(\ref{eq:odemodel_abc_mcmc})  maps to the  {\em dimensionless model}
\begin{equation}
\label{eq:odemodel_dimensionless}
\begin{array}{lcl}
 {\displaystyle\frac{dv_s}{ds}}&= &v_s(1 - v_s),\vspace{0.35cm} \\
 {\displaystyle\frac{dm_s}{ds}} &= &\eta v_s + \psi v_sm_s(1 - m_s)
\end{array}
\end{equation}
with $\eta = \lambda \alpha \mu/\nu$ and $\psi = \lambda \rho \mu.$
These   {\em dimensionless parameters} $(\eta,\psi)$ characterize the model; they aid in exploring  essential questions of information content of observed data, and in comparisons across subjects
in our later analyses.

\section{Initial Discrete-time Stochastic Model}

We have sparse observations on $V_t,M_t$ at a small number of time points, and of course these are
subject to  measurement errors, the nature of these errors being partly quantified in the flow cytometry assays. There is also a need for stochastic elements in the time evolution of $V_t,M_t$ to account for model
misspecification, i.e., structure not captured by the model. We now move to the practicable discretized form of
the ODE conceptual model and overlay these extensions.

Let $\xb_t = (V_t , M_t)'$ be the {\em latent} state vector of virus and memory frequencies representing the true underlying population densities for one individual at time $t$.
We use reparametrizations  $\delta=\beta/K_V$ and $\gamma=\rho/K_M$ to replace $K_V,K_M$ by new rate constants $\delta,\gamma.$ Note from \eqn{eq:odemodel_abc_mcmc} that this yields linear forms in $\beta,\delta,\alpha,\rho,\gamma$ and
this simplifies analysis; we can then recover $K_V=\beta/\delta,K_M=\gamma/\rho$ as desired.

Discretize to time scale $t=h, 2h, \ldots, T$
with increment $h>0,$  and suppose observations are made only at times $t_1 < t_2 < \cdots < t_n.$
Directly via Euler discretization of \eqn{eq:odemodel_abc_mcmc}, and allowing for measurement errors
and state evolution noise, we have
\begin{equation}\begin{array}{lcl}
V_t =  V_{t-h} + h (\beta -  \delta V_{t-h} ) V_{t-h} + \omega_{Vht} \vspace{0.3cm} \\
M_t = M_{t-h} + h \alpha V_{t-h}+ h (\rho V_{t-h}  -  \gamma V_{t-h} M_{t-h}  )  M_{t-h} + \omega_{Mht}
\end{array} \label{eq:statemodel1_abc_mcmc}\end{equation}
over all $ t=h,  \ldots, T.$
The evolution noise $\bomega_{ht} = (\omega_{Vht},\omega_{Mht})'$ represents stochastic noise in the state evolution as well as the model misfit. We take this as zero-mean Gaussian with diagonal variance matrix
having variances  $(\kappa^2_V,\kappa^2_M)$ that  are now additional
parameters.

Measurements are made at a set of times $t_1,\ldots, t_n.$  So we observe
\begin{equation}
\yb_j =  \xb_{t_j} + \bnu_j, \qquad j=1, \ldots, n,
\label{eq:obsn}
\end{equation}
where measurement errors $\bnu_j$ are  zero-mean Gaussian with diagonal variance matrix having entries $(\sigma^2_V, \sigma^2_M),$ additional parameters
subject to prior information from flow cytometry assays.
Sometimes only one of $V_t,M_t$ is measured at a particular $t_j$. Let
$t \in O_V$ and $t \in O_M$  be times measurements are made on $V_t,M_t$, respectively.
The time sets may or may not intersect.
Theoretically, we can reflect this  using measurement error variances $(\sigma^2_V/c_{Vt}, \sigma^2_M/c_{Mt}),$
where $c_{Vt}=1$ but $c_{Mt}\to 0$ when $t\in O_V$ but $t\notin O_M,$ and so forth.
This results in minor  technical changes to  the analysis  below.

\subsection{Extended Model: Incorporating Time Delays\label{sec:delays}}

The initial model form turns out to be unable to reflect a key feature of the data that relates to what
in retrospect is anticipated additional complexity in the immune response-- that of time delays in the
driving of cellular differentiation following vaccine introduction.  This is reflected in delay
 before appearance of Ad5-infected macrophages in the gut mucosa.
The basic model above is  therefore extended to allow for subject-specific time delays in the
effects of the viral population level on the memory cells. Specifically, \eqn{eq:statemodel1_abc_mcmc}
is extended to
\begin{equation} \begin{array}{lcl}
V_t & = & \begin{cases} V_{t-h} + \omega_{Vht}, & \text{if $t \le \tau_V$}, \\
				      V_{t-h} + h\partial\!V_{t-h}  + \omega_{Vht}, & \text{otherwise;} \end{cases} \vspace{0.1cm} \\
M_t & = & \begin{cases}  M_{t-h} + \omega_{Mht}, & \text{if $t \le \tau_V + \tau_M$,} \\
 				       M_{t-h} + h \partial\!M_{t-h}  + \omega_{Mht}, \phantom{........} & \text{otherwise.} \end{cases}
\end{array}\label{eq:statemodel2_abc_mcmc}\end{equation}
where
\begin{equation*} \begin{array}{lcl}
\partial\!V_{t-h} &=& (\beta -  \delta V_{t-h} ) V_{t-h},\\
\partial\!M_{t-h} &=& \alpha V_{t-h}+  (\rho V_{t-h}  -  \gamma V_{t-h} M_{t-h}  )  M_{t-h}.
\end{array}\end{equation*}
Here $\tau_V$ is the time delay on the direct impact of vaccination on $V_t,$ and
$\tau_M$ is the additional delay before $V_t$ impacts on memory cells. The
parameters are now
\begin{equation}\label{eq:params}
\Theta = \{\beta, \delta, \alpha, \rho, \gamma, V_0, \tau_V, \tau_M, \sigma^2_V, \sigma^2_M, \kappa^2_V, \kappa^2_M\}
\end{equation}  including the unknown, initial viral level $V_0$, all  noise, rate and delay parameters. We  also note the implied parameters of the dimensionless model
are just  $\lambda_V= \beta\tau_V$ and $\lambda_M= \beta\tau_M.$

\section{Bayesian Computation for Model Fitting}

With observed data $\Y = \{\yb_{t_j},\ j=1,\ldots n\},$ we aim to explore the posterior $p(\Theta, \X|\Y)$ where
$\X=\{ \xb_t,\ t=h,2h,\ldots,T\}$ is the  trajectory of latent states $V_t,M_t$ over the time period.  The model is a complicated, non-linear dynamic model and standard computational methods of MCMC and sequential Monte Carlo, including ABC, are notoriously challenged by the needs to define relevant proposal mechanisms, and high rejection rates associated with long series of  latent states.  Recent approaches representing current research frontiers focus heavily on biological systems applications~\citep[e.g.,][]{Niemi2010,Bonassi2011,Drovandi2011,Golightly2011,Wilk:stoc:2006}. The sparsity of data in our studies
exacerbates the difficulties.   MCMC is challenged by the nonlinearities and truncation to positive states in the model.
ABC methods are challenged by the high-dimensionality of the latent states and the interest in relatively diffuse priors for
model parameters, so that it is difficult to generate prior:model simulations in the region of the sparse observed data.
After detailed experimentation with these and other approaches, including other sequential methods, we have identified
a strategy of coupling MCMC with adaptive ABC that draws on their complementary strengths. The two stages are:
\begin{itemize}
\item[{\em (i)}]  Use MCMC coupled with traditional analytic approximation (local linearization) of the model
to define an approximate posterior sample. Knowing that this will be biased in ways that are difficult to assess,
\item[{\em (ii)}] Use this MCMC sample to define a proposal distribution for an efficient, adaptive ABC.
\end{itemize}
Step {\em (i)} is easy to implement and run to define a \lq\lq ballpark'' approximation to the posterior;  step
{\em (ii)} is then ideal for ABC accept/reject methods, as approximate posterior samples are far more likely than
prior:model simulations to generate synthetic data close to the observed data. Reweighting corrects ABC accepted
samples to define the refined posterior approximation.

\subsection{First stage analysis: MCMC} 
At each MCMC step, model parameters are resampled from complete conditional posteriors given
a current set of latent states $\X.$ The conditional linearity of the model in rate parameters $\beta,\delta,\alpha,\rho,\gamma$
and initial $V_0$; this
means that uniform or normal priors yield truncated normal conditional posteriors. Then, inverse gamma priors are
conditionally conjugate for the variances $\sigma_V^2,\sigma_M^2,\kappa_V^2,\kappa_M^2.$   The resampling of
delay parameters $\tau_V,\tau_M$ uses a Metropolis-Hastings random walk step.

Conditional on latest parameter samples,  the full trajectory $\X$ is resampled from an
approximation to its posterior conditional on parameters. We use local linearization (extended Kalman filtering) for forward
filtering of the states, followed by backward sampling to generate $\X.$
The extended Kalman filter approach \citep[Section 13.2]{West1997} sequentially updates conditional Gaussian
distributions for the states over time, and at each time uses the current mean of that Gaussian as a point of
local linearization of the non-linear model {\em at that time}.  This maps
to an approximating sequence of linear models, and standard forward filtering, backward sampling (FFBS)
\citep{Prado2010,West1997}  to generate the full state trajectory applies.

Start with the exact non-linear model of evolution of the state $\xb_t = (V_t,M_t)'$ in \eqn{eq:statemodel2_abc_mcmc}.
This can be cast as
\begin{equation}\label{eq:DLM}
\zb_t = \Hb_t(\zb_{t-h}) \ \zb_{t-h} + \bxi_t
\end{equation}
with {\em extended state} vector  $\zb_t$,  noise $\bxi_t$ and  matrix $ \Hb_t(\zb_{t-h})$ given by
$$
\begin{array}{rcl}
\zb_t & = &  (V_t, M_t, V_{t-1}, V_{t-2}, \ldots, V_{t- \tau_M + 1})',   \vspace{0.1cm}  \\
\bxi_t & = & (\omega_{Vt}, \omega_{Mt}, 0, \ldots, 0)',  \vspace{0.1cm} \\
\Hb_{t}(\zb_{t-h}) &=&   \begin{pmatrix}
a_t               & 0            & 0          & 0          & \cdots    & 0           & 0          \\
0                    & b_t    & 0          & 0          & \cdots    & 0           & c_t   \\
1                    & 0            & 0          & 0          & \cdots    & 0           & 0          \\
0                    & 0            & 1          & 0          & \cdots    & 0           & 0          \\
0                    & 0            & 0          & 1          & \cdots    & 0           & 0          \\
\vdots            & \vdots   & \vdots  & \vdots  & \ddots   & \vdots   & \vdots  \\
0                    & 0            & 0          & 0          & \cdots    & 1           & 0          \\
\end{pmatrix}
\end{array}
$$
where
$$
\begin{array}{rcl}
a_t&=& \begin{cases} 1,    &  \text{if $t \le \tau_V$,} \\
					    1 + \beta - \delta \zb_{t-h,1},    & \text{if $t>\tau_V$;} \end{cases}  \vspace{0.1cm} \\
\{ c_t,  b_t \} &=& \begin{cases}   \{ 0, 1 \},    &  \text{if $t \le \tau_V + \tau_M$,} \\
			                 \{ \alpha,  1 +  \rho \zb_{t-h,\tau_M+1} - \gamma \zb_{t-h,2} \ \zb_{t-h,\tau_M+1} \},   & \text{if $t > \tau_V + \tau_M$.} \end{cases}
\end{array}
$$
In linearized forward filtering,  write  $\mb_{t-h}$ for
the current estimate of $\zb_{t-h}$; this is the approximate posterior mean
conditional on data up to time $t-h$.  Linearization  adopts the
model at time $t$ as the {\em dynamic linear model} obtained by replacing $ \Hb_t(\zb_{t-h})$ in \eqn{eq:DLM}
with  $\Hb_t(\mb_{t-h}) .$ This is performed sequentially over $t=h,2h,\ldots,T,$  saving relevant summaries at
each step in order to then apply FFBS-based backward sampling generate/resample the
full trajectory $\X = \{ \xb_h,\xb_{2h},\ldots,\xb_T\}.$

\subsection{Second stage analysis: ABC}

MCMC  is easily implemented but  it is difficult to understand
errors and biases induced by the approximations. The
linearized model analysis does not impose   positivity constraints on $V_t,M_t$,  and other biases
are due to the inherent non-linearities.  This motivates  adjustments based on a second stage ABC approach.
The set of MCMC-based posterior samples for $\Theta, \X$ are used to generate a proposal distribution for  candidate
draws in an  ABC method. The ABC step refines the initial MCMC-based
approximation, aiming to correct biases; in complement, this drives  ABC analysis with an already useful
posterior approximation.

Denote the observed data by $\Yo$, and choose a {\em discrepancy} function $\delta(\Y,\Yo)$ measuring
distance from any candidate/synthetic observation set $\Y$ from $\Yo.$
Weighted ABC
using a proposal distribution with p.d.f. $g(\Theta)$ on the model parameters proceeds as follows:
\begin{itemize}
\item Sample a candidate parameter set $\Theta\sim g(\Theta);$
\item forward simulate the latent state $\X|\Theta$ from the exact dynamic model and, given this $\X,$
\item generate synthetic data $\Y$ at the observation times $t_1,\ldots,t_n;$
\item if $d(\Y,\Yo) < \epsilon$ for some small threshold $\epsilon>0$, accept the candidate $\Theta$; otherwise, try again.
\item Once a large sample of accepted draws is achieved, resample  with weights proportional to $\pi(\Theta)/g(\Theta)$
where $\pi(\Theta)$ is the prior p.d.f.
\end{itemize}
The result is a resampled set of parameters approximating $p(\Theta|\Yo).$

We define $g(\theta)$ as follows. For the rate parameters $\Lambda = (\beta, \delta, \alpha, \rho, \gamma)$-- for
which the posterior:prior contrast is expected to be greatest-- we use a truncated $5-$dimensional Gaussian mixture
distribution of the form
$g(\Lambda) \propto \sum_jN(\Lambda_j, h_{\Lambda}^2 S_{\Lambda}) I_{(\Lambda \in A)},$ where
the $\Lambda_j$ are the MCMC-based samples, $h_{\Lambda}$ is a bandwidth multiplier, $S_{\Lambda}$ is the sample
covariance matrix of MCMC draws, and $A$ is the hypercube region formed by the marginal ranges of each parameter
in the MCMC sample. This follows~\cite{Bonassi2011} and~\cite{West1992b,West1993b,West1993a}, with details there including bandwidth specification. The proposal distribution is completed as $g(\Theta)=g(\Lambda)g(\Sigma)$
where $g(\Sigma)$ is the product of the univariate priors  on each element of $\Sigma= \{ V_0, \tau_V, \tau_M, \sigma^2_V, \sigma^2_M, \kappa^2_V, \kappa^2_M\}.$

\begin{figure}[ht!]
  \centering
 \includegraphics[width=2.0in]{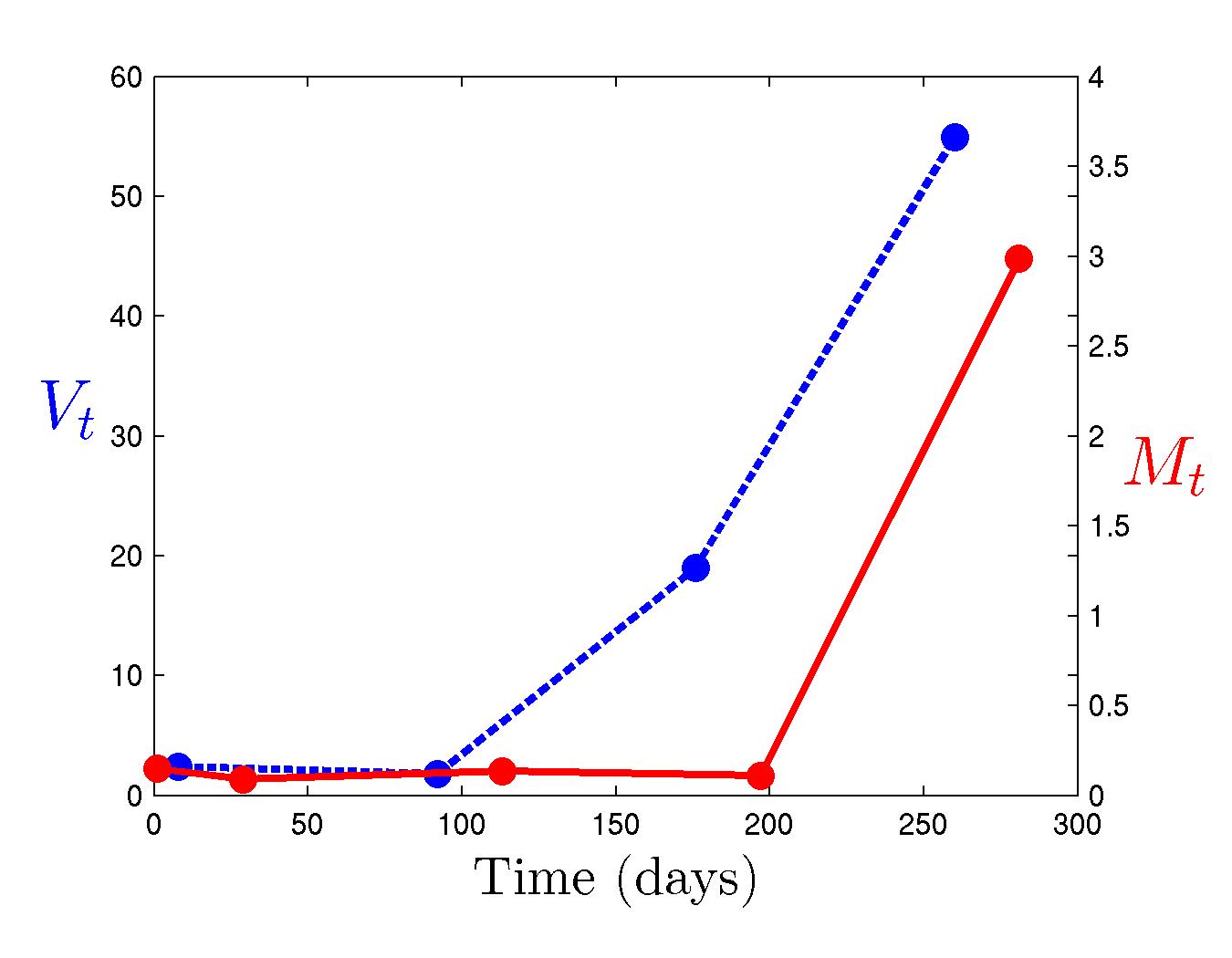}
 \includegraphics[width=2.0in]{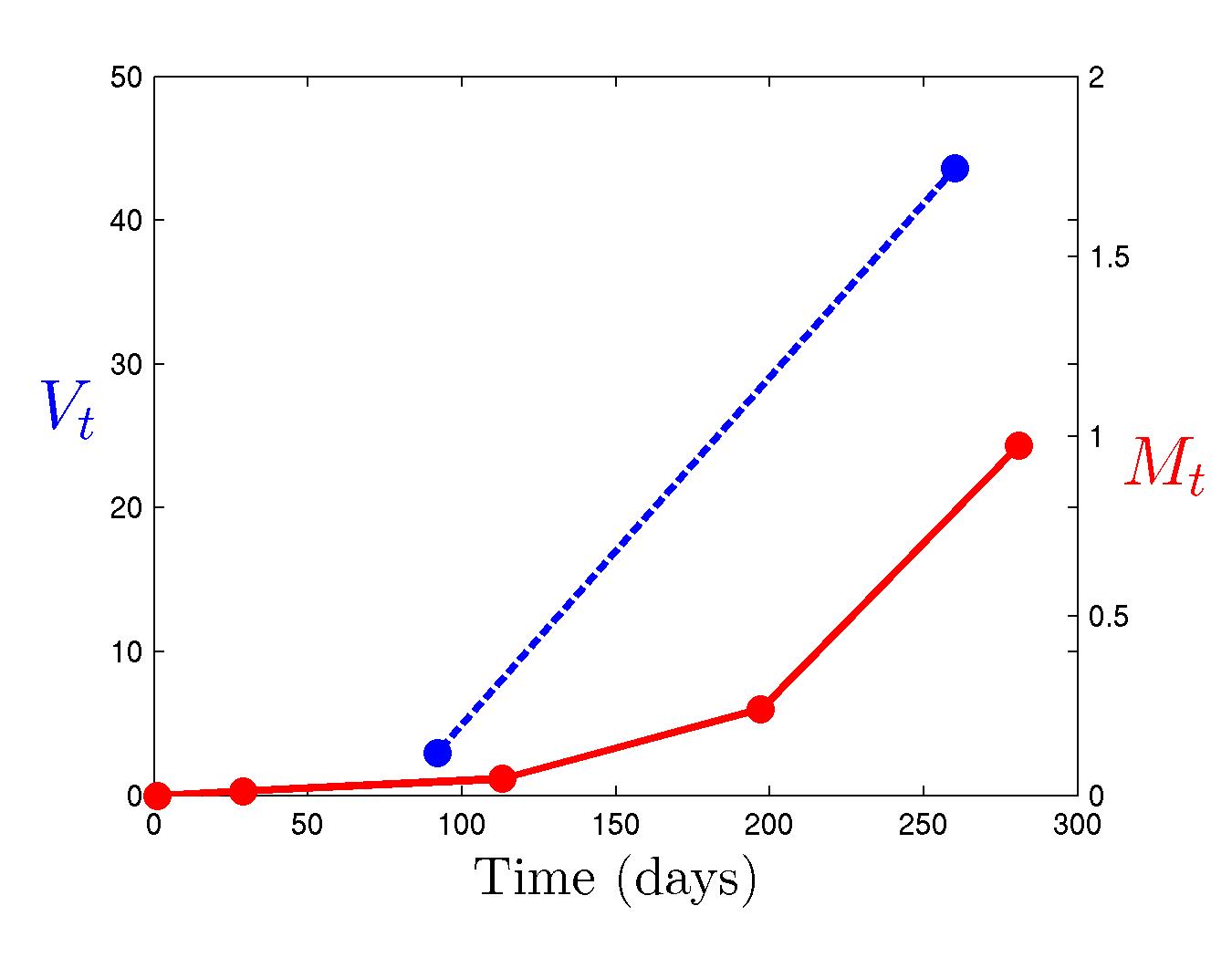} \\
 \includegraphics[width=2.0in]{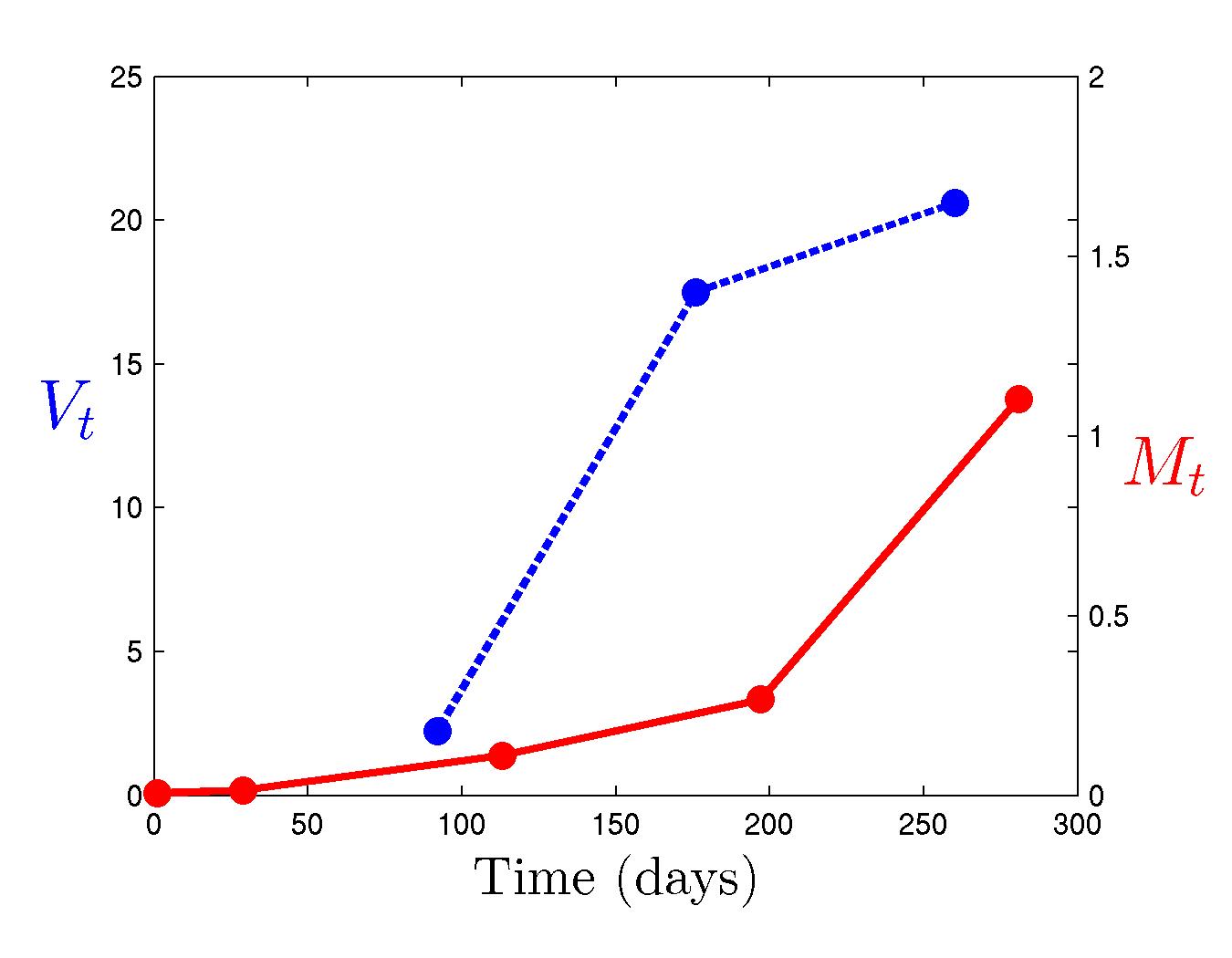}
 \includegraphics[width=2.0in]{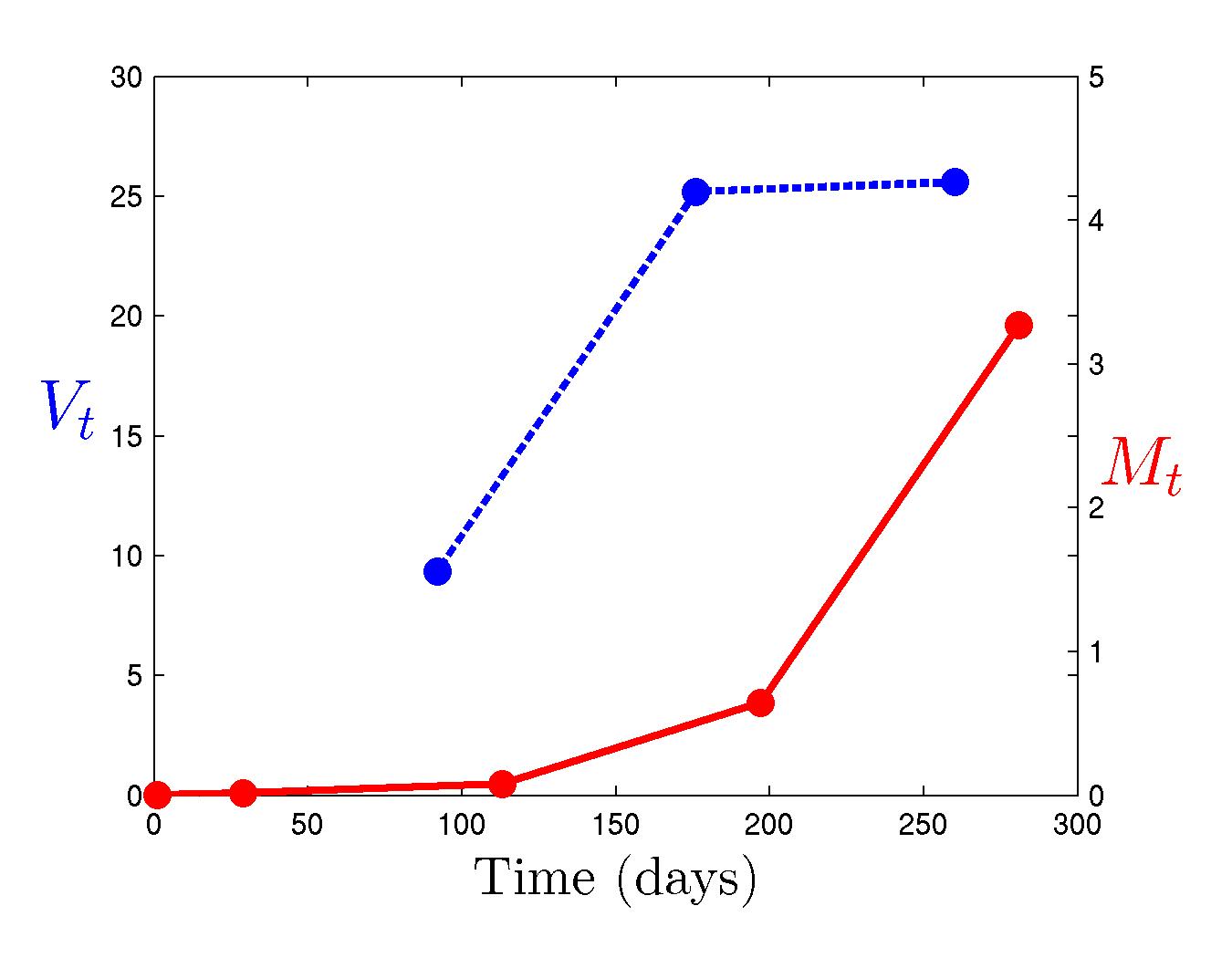}
\caption{\label{fig:plots_data_abc_mcmc} Observed data  $V_t$ (blue) and $M_t$ (red) for 4 of the 8
macaques (subject IDs 400, 401, 404, 405, respectively).}
  \centering
 \includegraphics[width=2.0in]{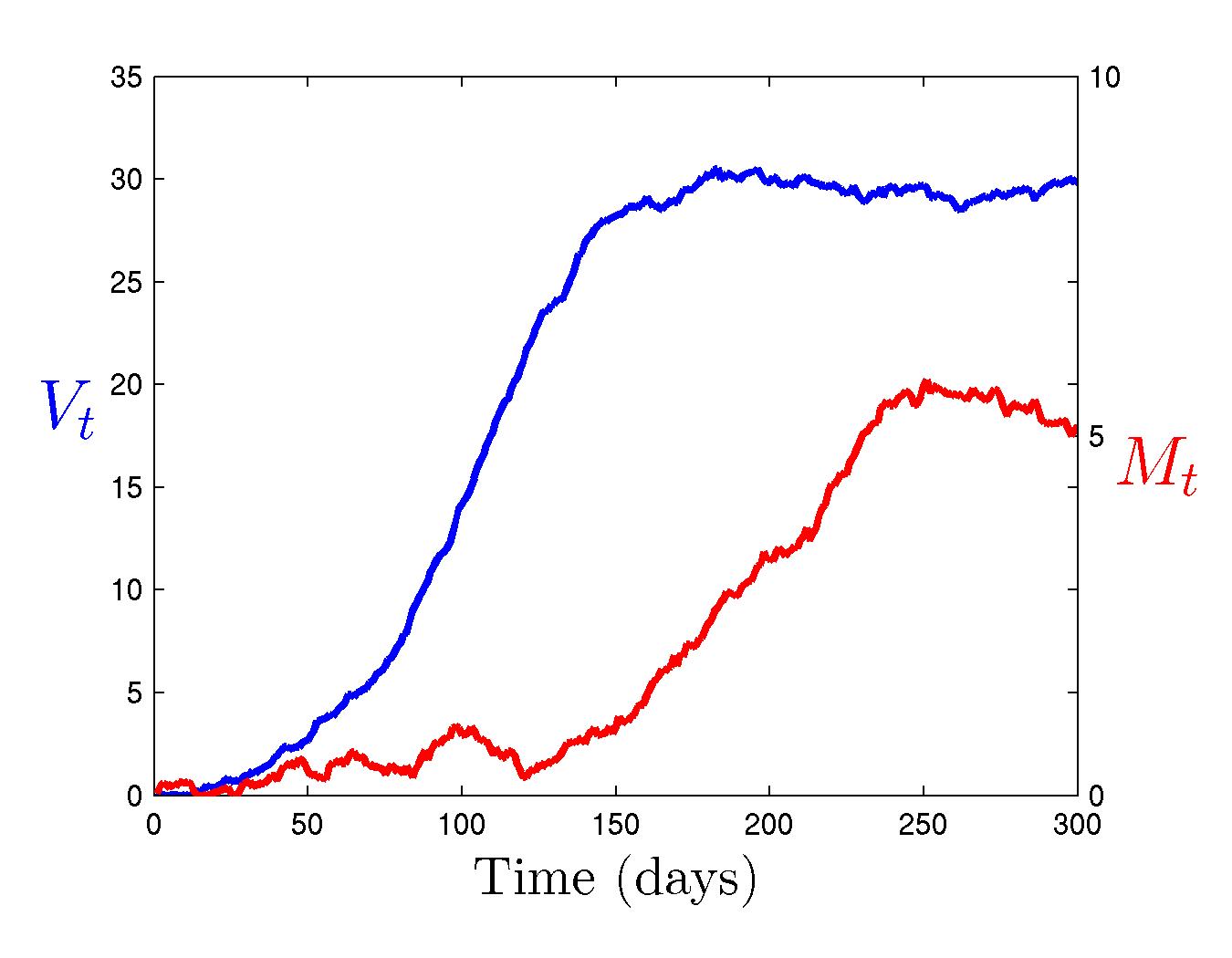}
 \includegraphics[width=2.0in]{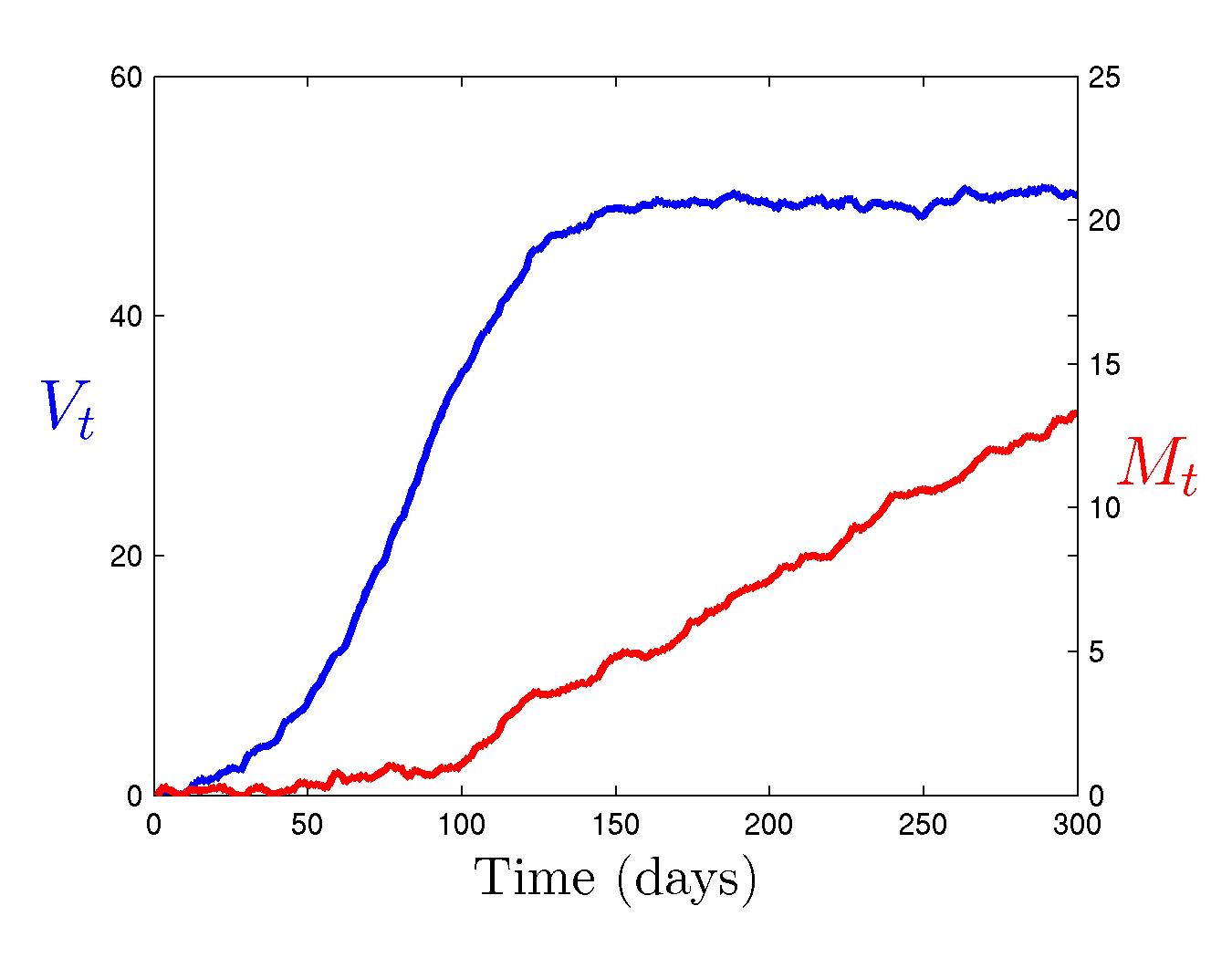} \\
 \includegraphics[width=2.0in]{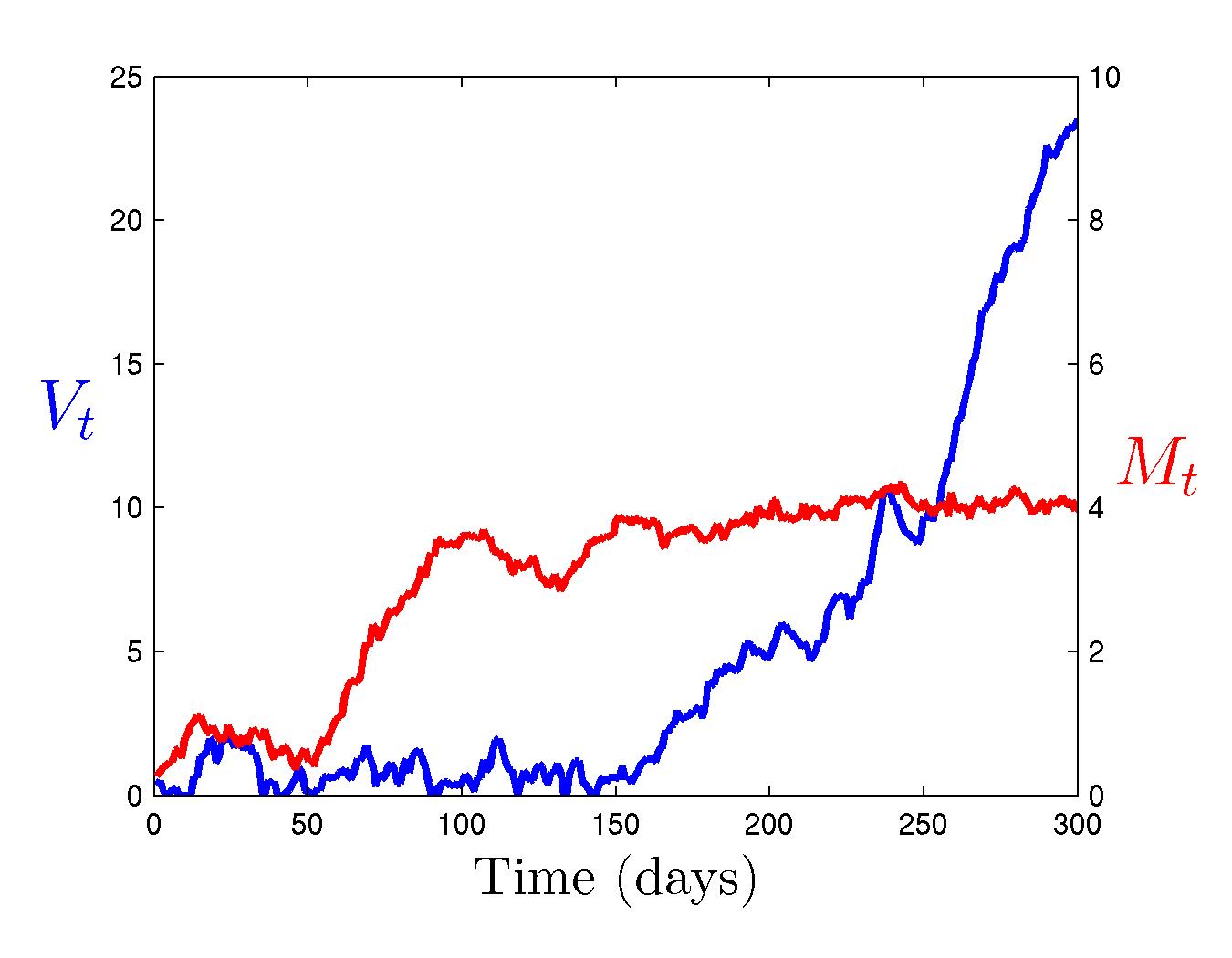}
 \includegraphics[width=2.0in]{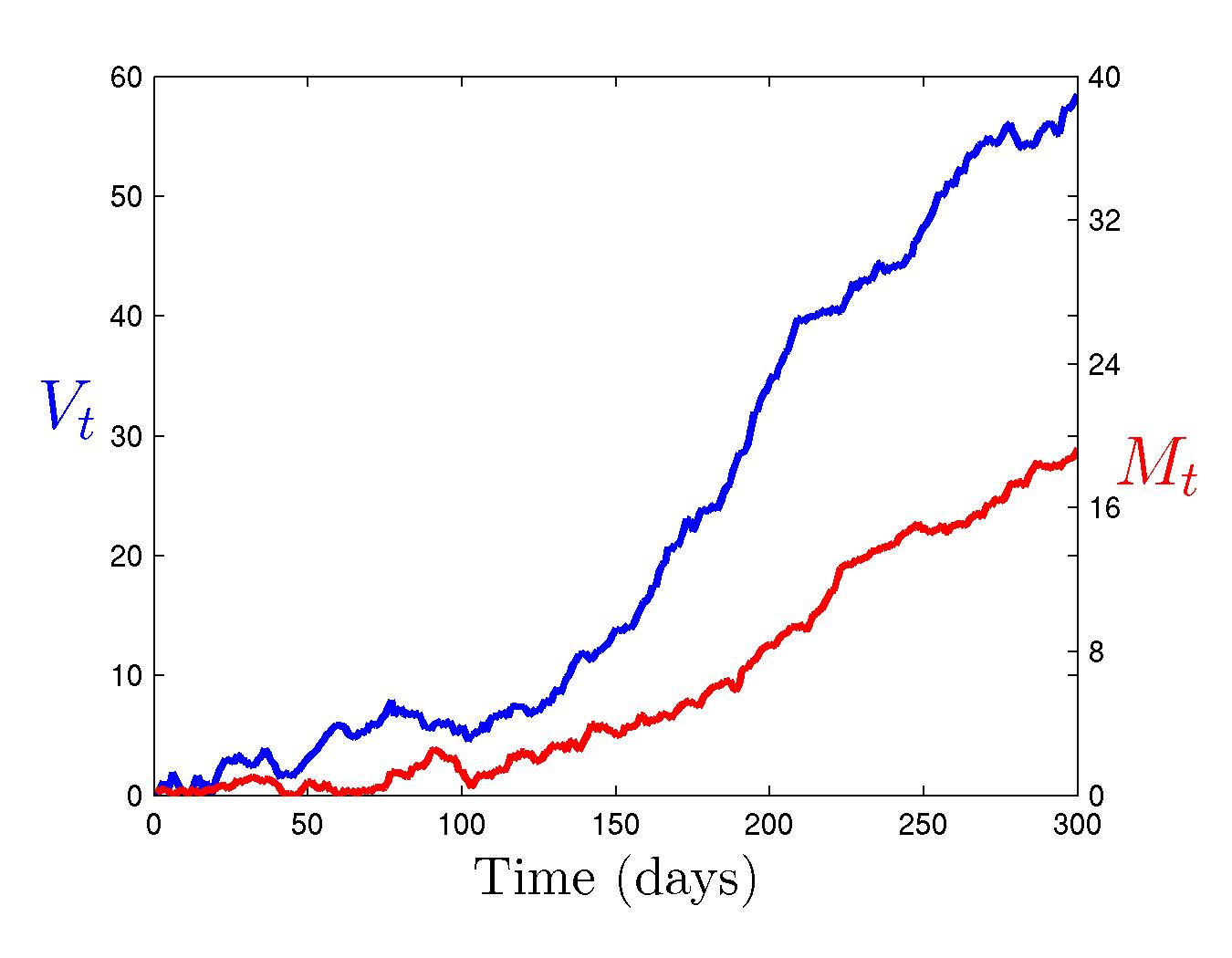}
\caption{\label{fig:plots_sde_abc_mcmc} Synthetic trajectories of $V_t$ (blue) and $M_t$ (red) from  $(V_0, M_0)=(0.5, 0.5)$, using
model parameters chosen to be consistent with posteriors from the macaque study.}
\end{figure}

\section{Analysis: HIV/SIV Study on Indian Rhesus Macaques}
Data on 8 subject macaques was generated from flow cytometric analysis of GFP-labeled macrophages (i.e., Ad5 infected) from rectal mucosa sampled at weeks 1, 13, 25 and 37,  together with the concentrations of gag-specific memory T-cells sampled at weeks 0, 4, 16, 28 and 40. Figure~\ref{fig:plots_data_abc_mcmc} shows the sparse time series of concentrations of viral population ($V_t$) and memory cells ($M_t$) on 4 macaques. The remaining 4 data sets are shown in the  Supplementary Material. 
We perform separate analyses for each subject (macaque) and explore comparisons of inferences across subjects.

\subsection{Model Predictions of Trajectories}
To appreciate the relevance of the discrete-time, stochastic model with delay effects, we show representative trajectories
in Figure~\ref{fig:plots_sde_abc_mcmc}. These are simply model simulations of $V_t,M_t$ using plug-in
parameter values that reflect the ranges we see in posteriors for some of the macaques, and is displayed
here just to communicate that the model generates trajectories whose forms evidently match those of the data sets.

\subsection{Priors}

For the rate parameters $(\beta, \delta, \alpha, \rho, \gamma)$ we start with independent $U(0,1)$ priors, then
impose constraints so that the carrying capacities $K_v=\beta/\delta$ and $K_M=\rho/\gamma$ are each less than 100\%.
Note that these are very diffuse priors.
Priors for the variance parameters are independent, scaled inverse chi-squared on 5 degrees of freedom; their
scale parameter are determined from analysis of data from other studies: $s_{\sigma^2_V} =0.4$, $s_{\sigma^2_M} =0.08$,
$s_{\kappa^2_V} =0.05$, $s_{\kappa^2_M} =0.01$. For the time discretization, we take $h=1$day  and the priors for
time delays $\tau_V,\tau_M$ are independent discrete uniforms on $\{1, \ldots, 50\}$.
The initial viral level $V_0$ has a $U(0,0.5)$ prior, and $M_0$ is fixed at a value based on initial  measurements for
each macaque.

\subsection{Bayesian Computations}

The two stage analysis ran MCMC chains 10,000 iterations with a burn-in of 3,000. The weighted ABC analysis generated
10,000 accepted particles. Following standard practice~\citep[e.g.,][]{Pritchard1999, Beaumont2002,Bonassi2011}, the threshold level $\epsilon$ was set based on the discrepancy value that separated the closest 5\%
of synthetic data sets to $\Yo,$ using $d(\Y,\Yo) = \sum_{t \in O_V} (y_{Vt}-y_{oVt})^2 / s_V^2 + \sum_{t \in O_M} (y_{Mt}-y_{oMt})^2 / s_M^2 $, where $s_V$ and $s_M$ are the standard deviations of the observed values $y_{oV},y_{oM}$, respectively.

\subsection{Some Posterior Summaries for Parameters}

\begin{figure}[ht!]
\centering
\begin{tabular}{cccc}
\hskip-0.1in\includegraphics[width=1.25in]{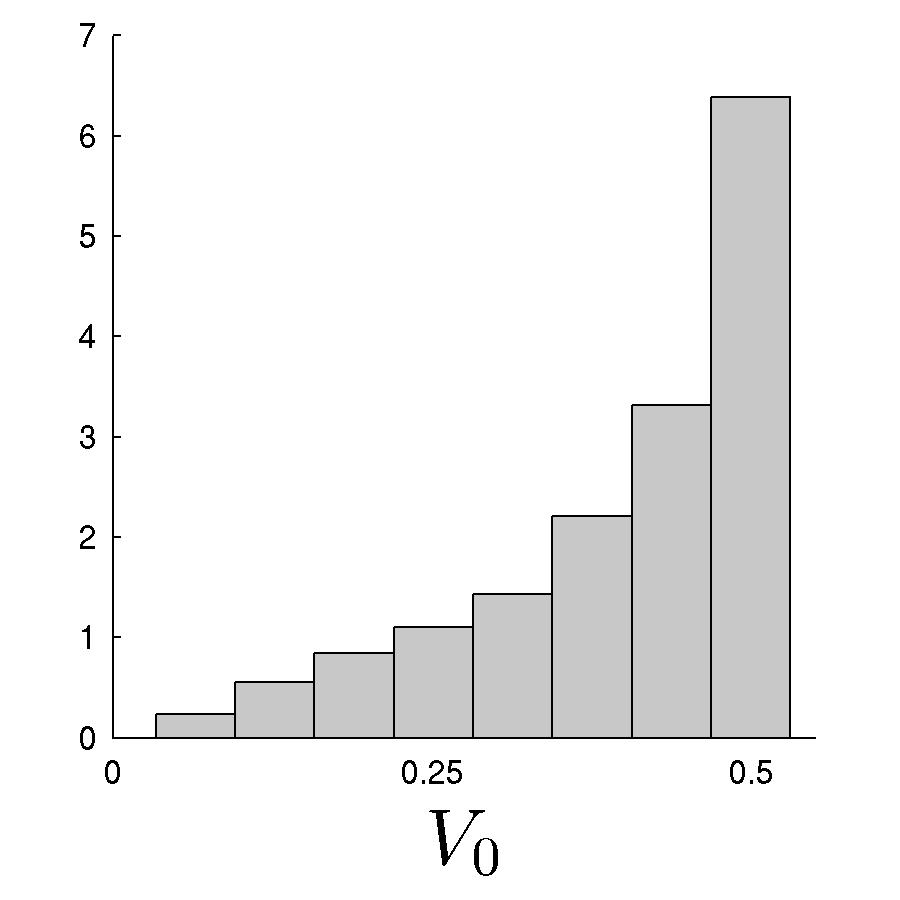}
\includegraphics[width=1.25in]{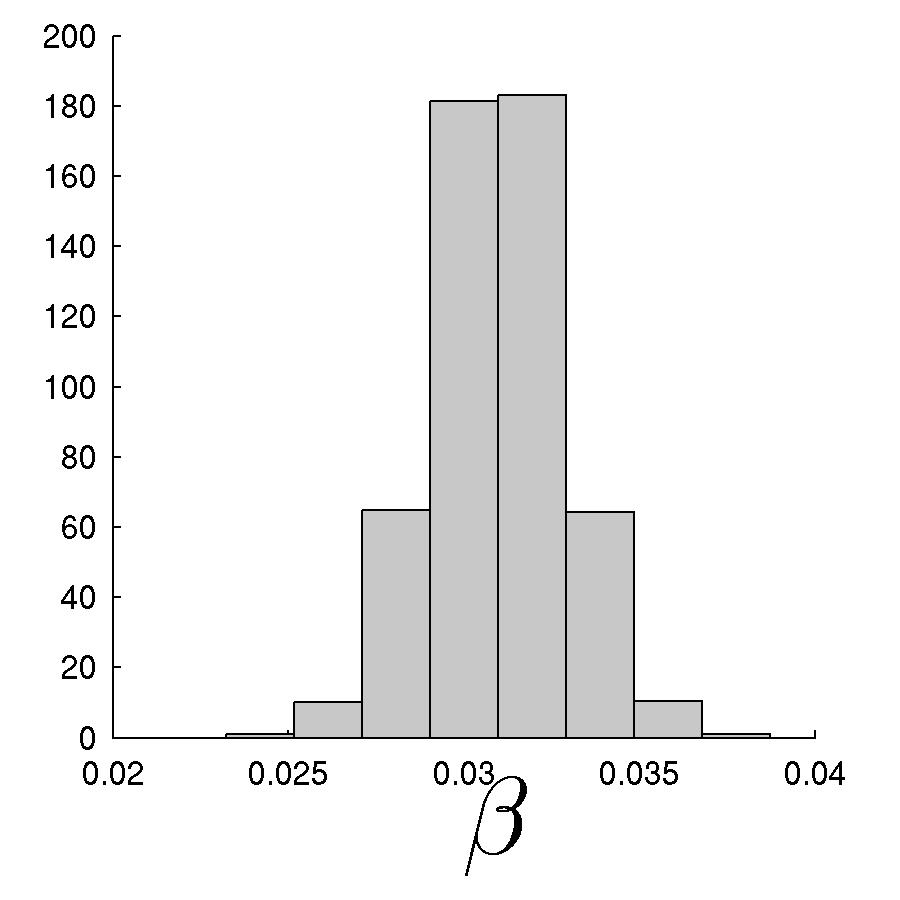}
\includegraphics[width=1.25in]{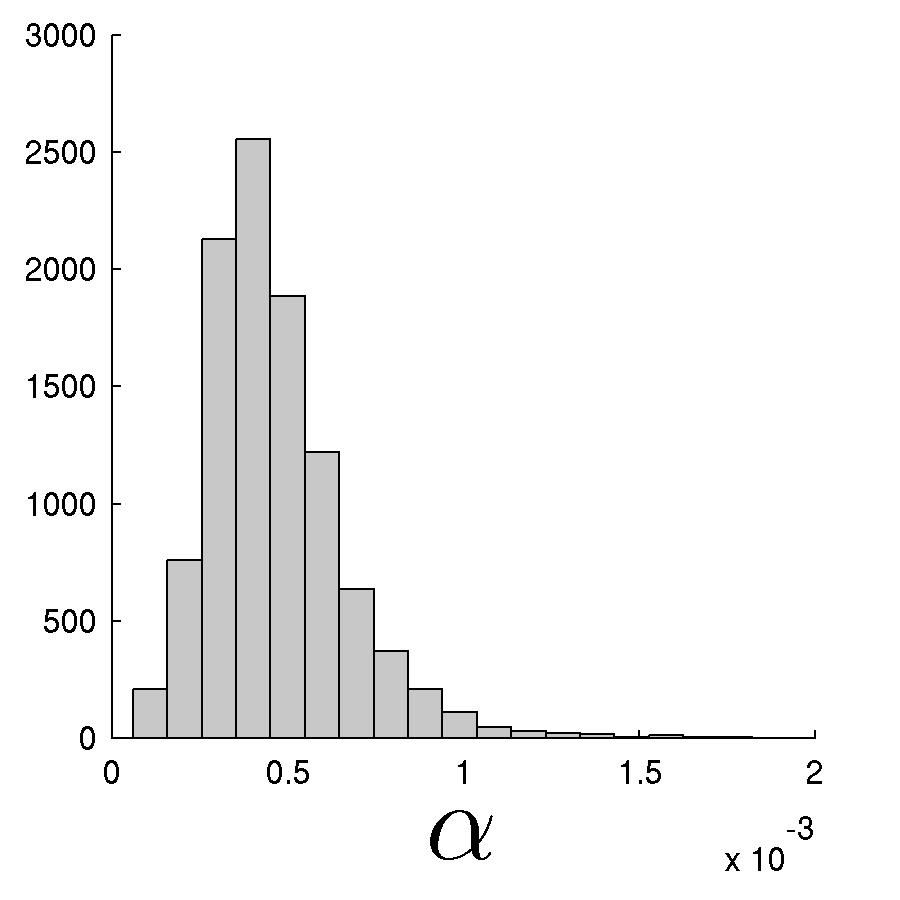}
\includegraphics[width=1.25in]{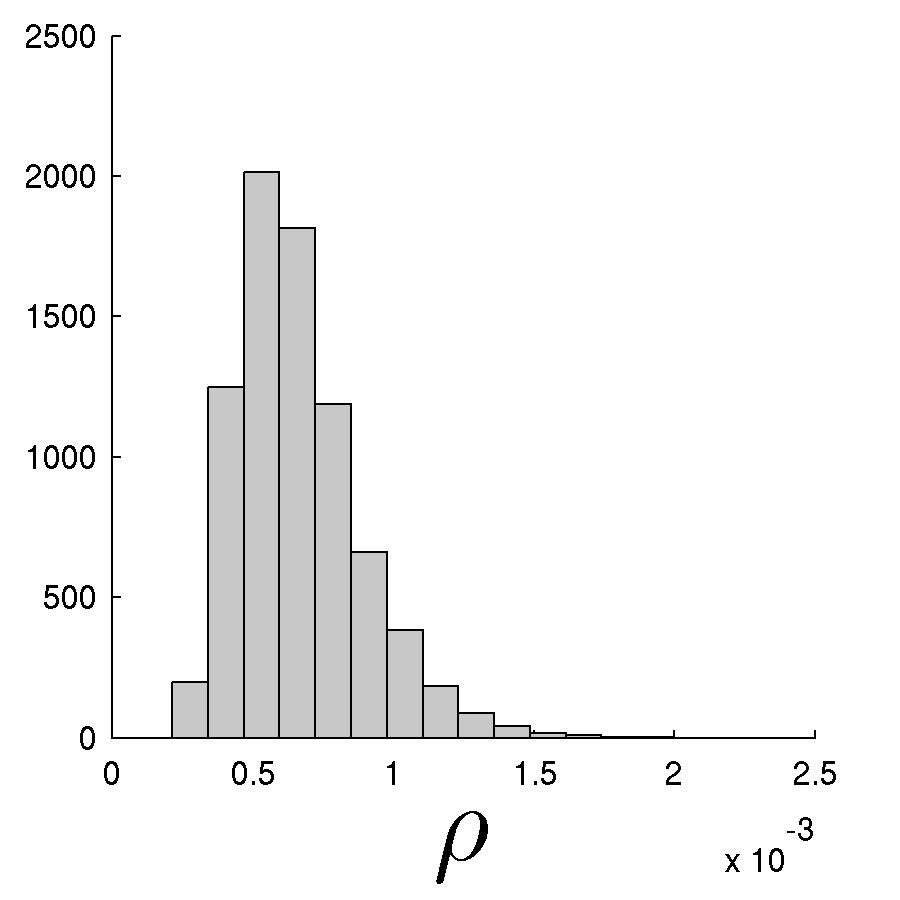}\\
\hskip-0.1in\includegraphics[width=1.25in]{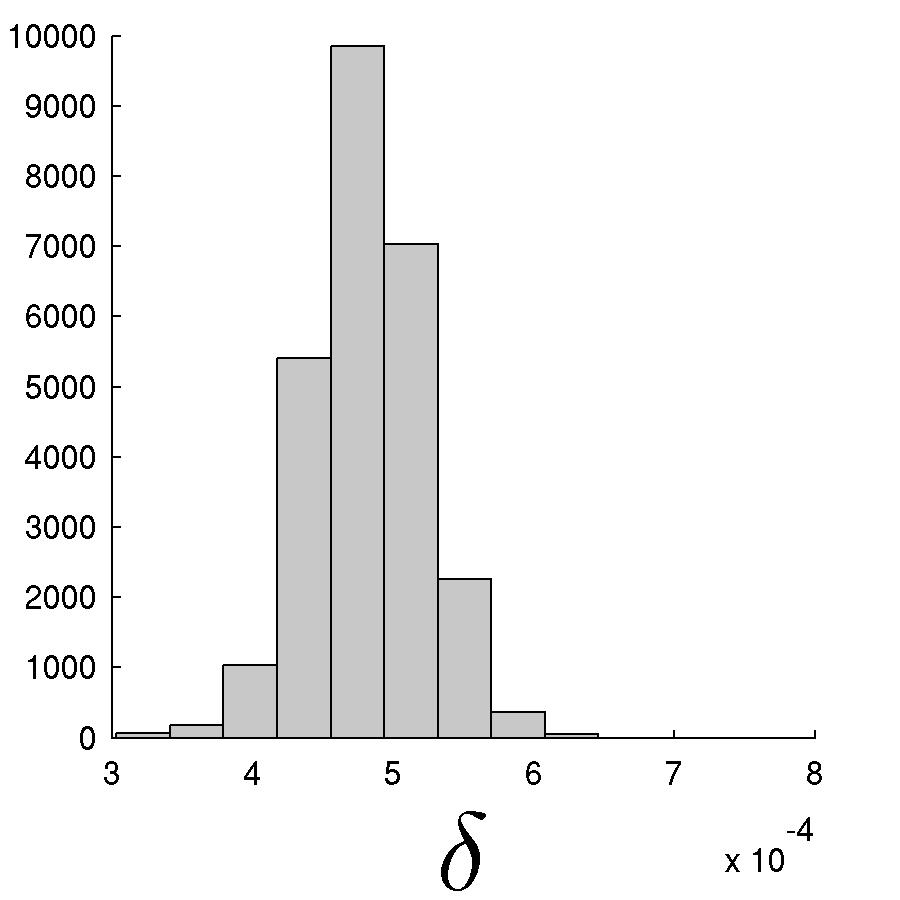}
\includegraphics[width=1.25in]{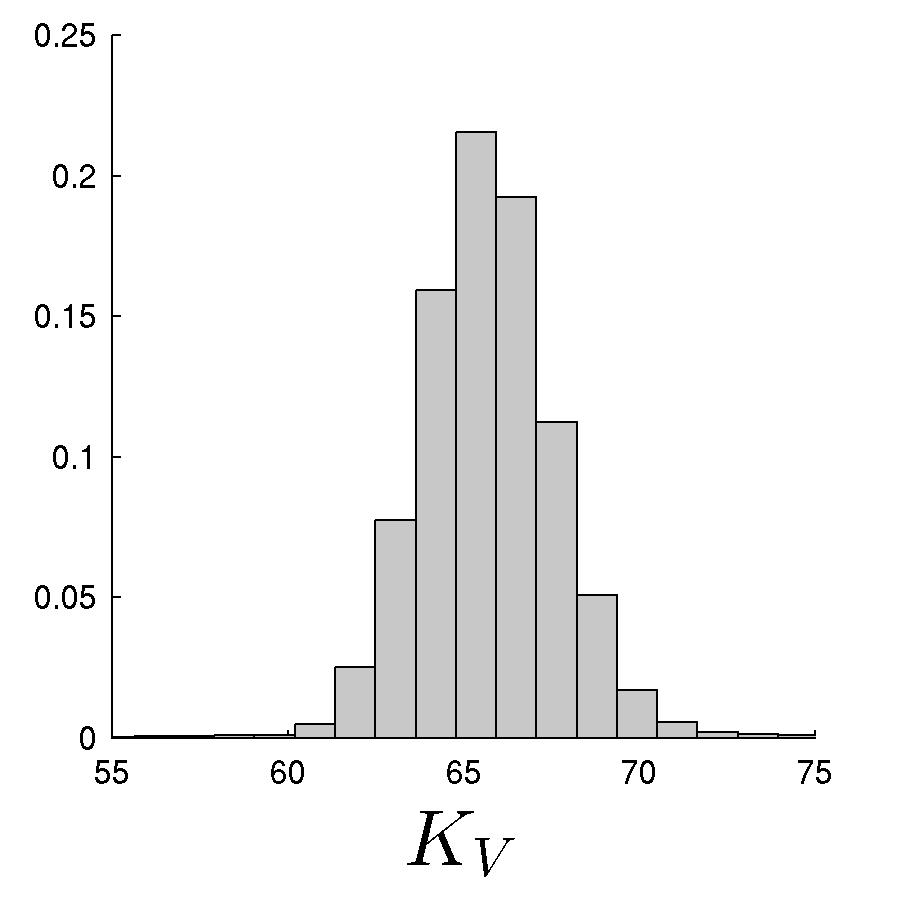}
\includegraphics[width=1.25in]{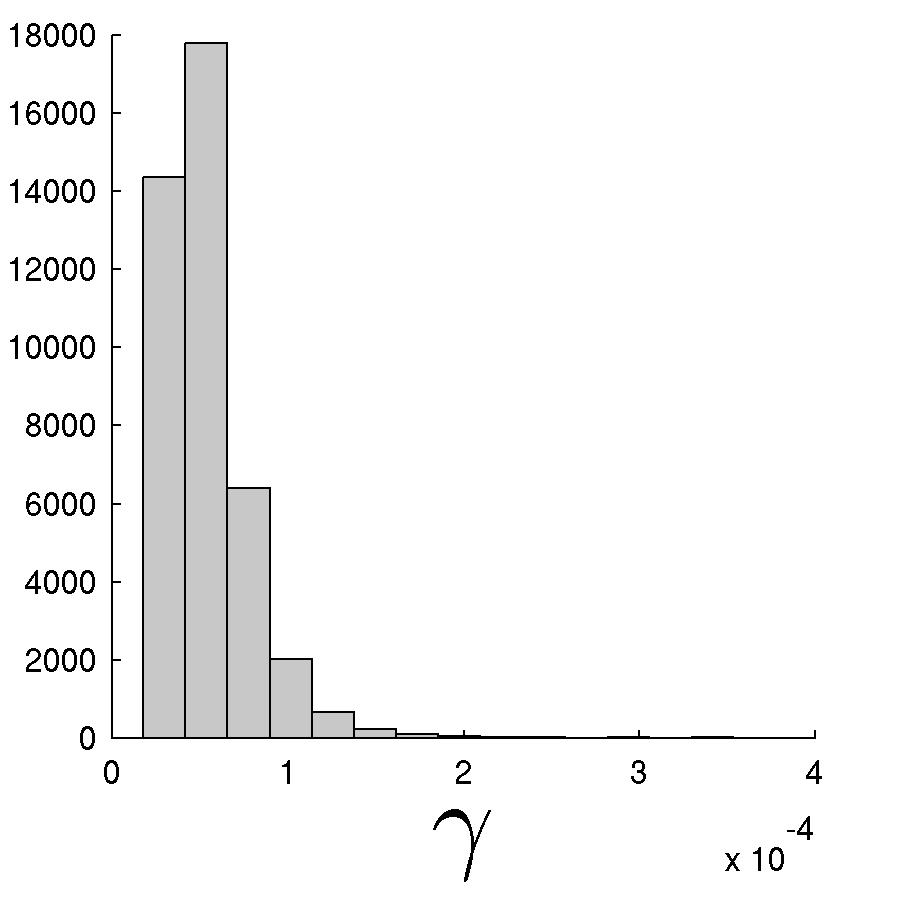}
\includegraphics[width=1.25in]{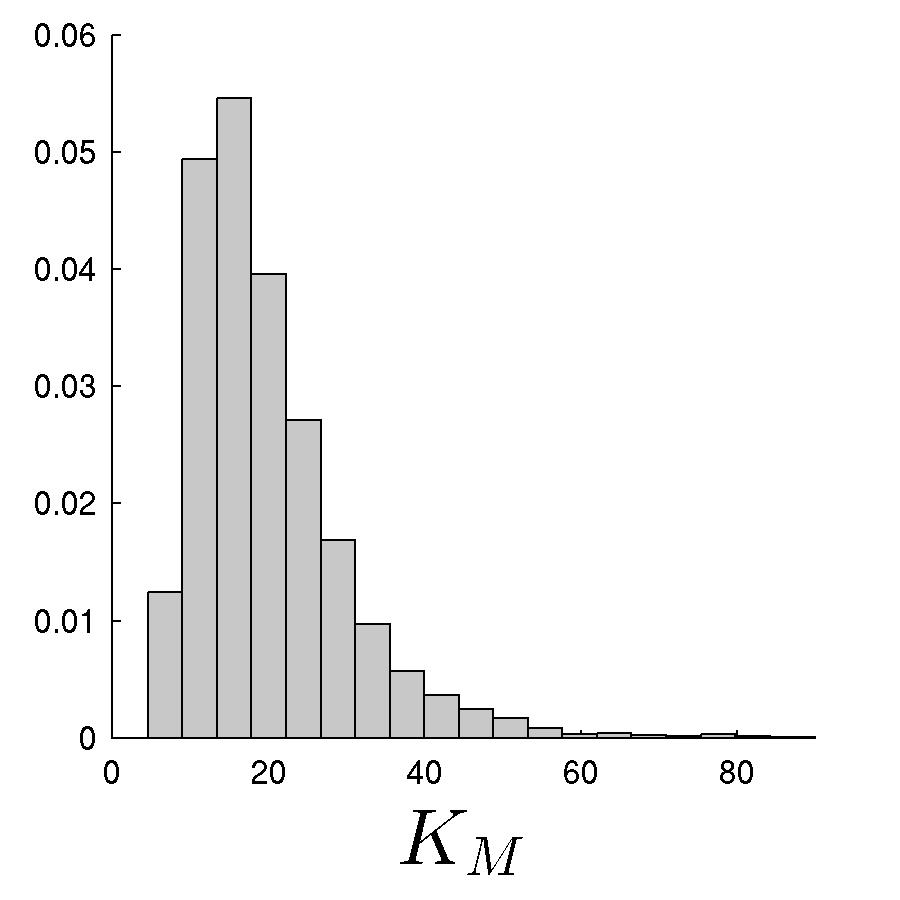}\\
\hskip-0.1in\includegraphics[width=1.25in]{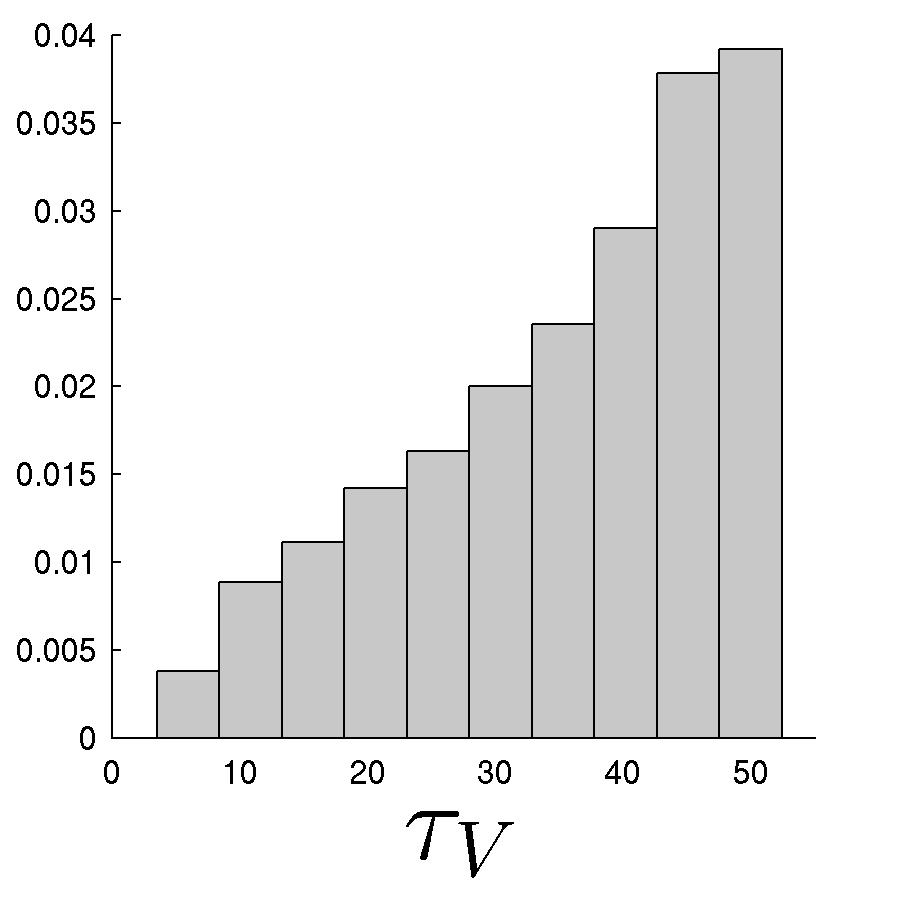}
\includegraphics[width=1.25in]{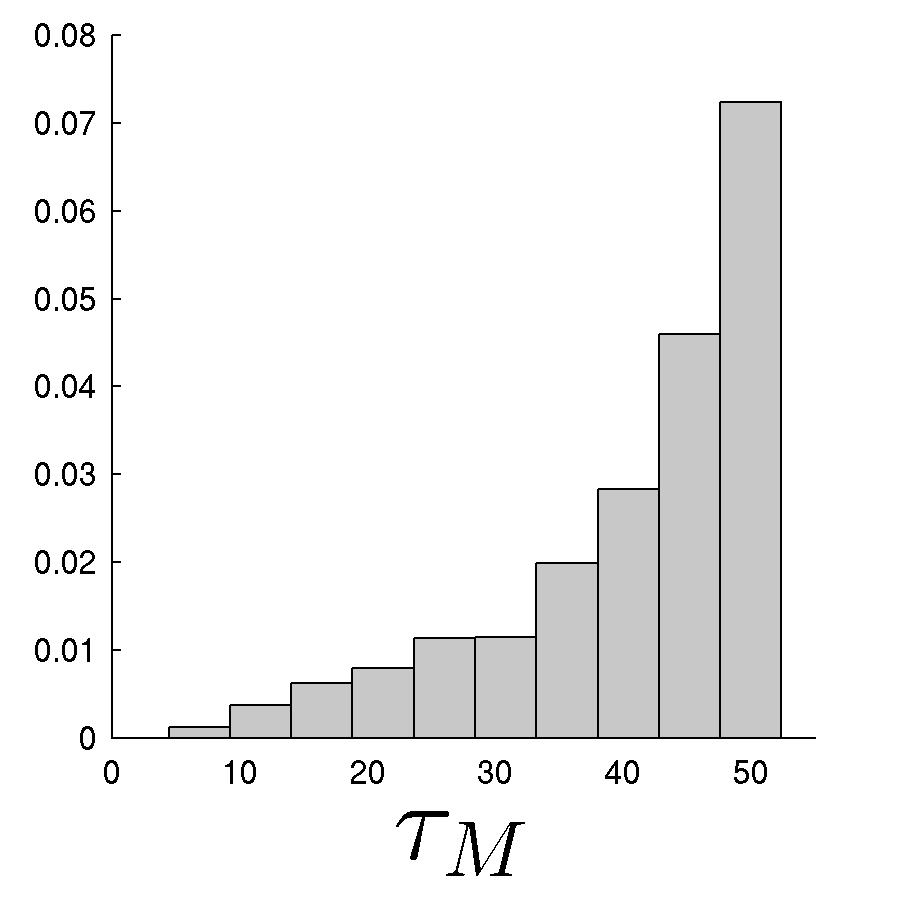} \\
\end{tabular}
\caption{\label{fig:post_id400} Macaque 400:  posterior margins for the 10 parameters as annotated.}
\end{figure}

\begin{figure}[ht!]
  \centering
\begin{tabular}{cccc}
\hskip-0.15in\includegraphics[width=1.25in]{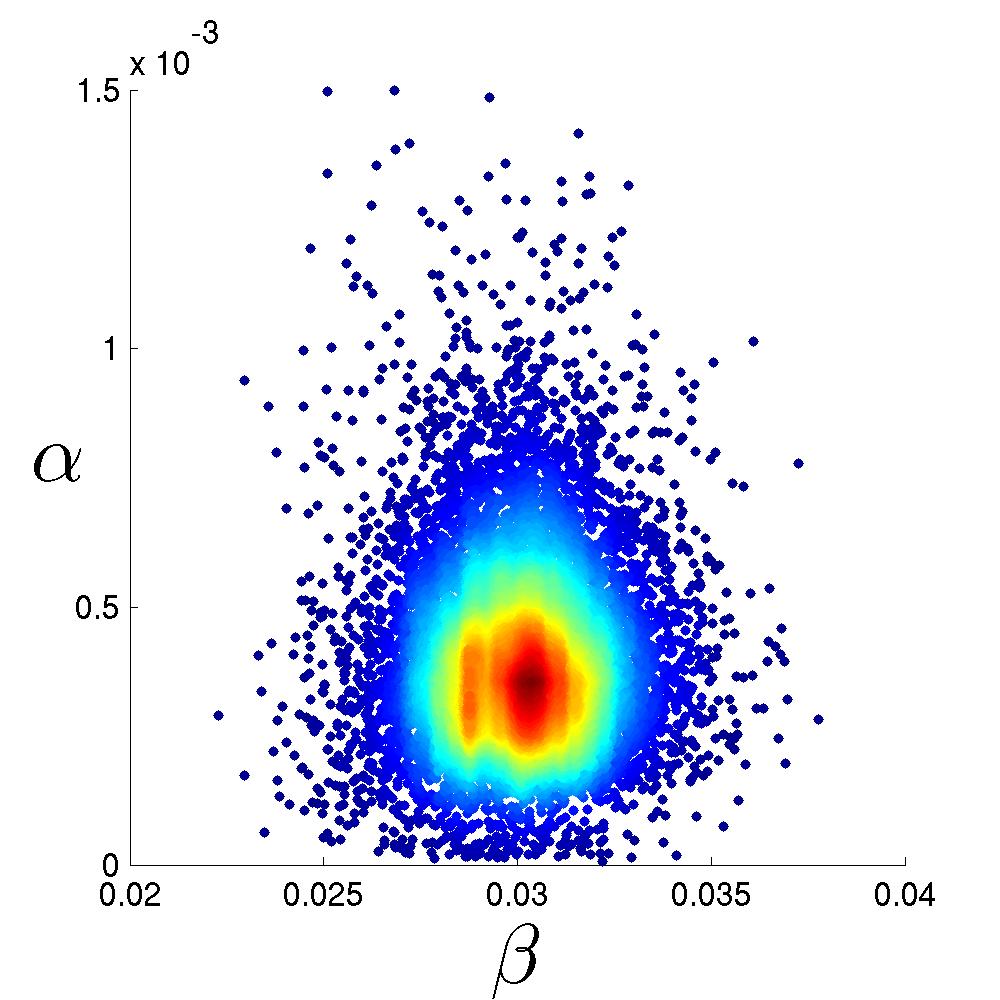}
\includegraphics[width=1.25in]{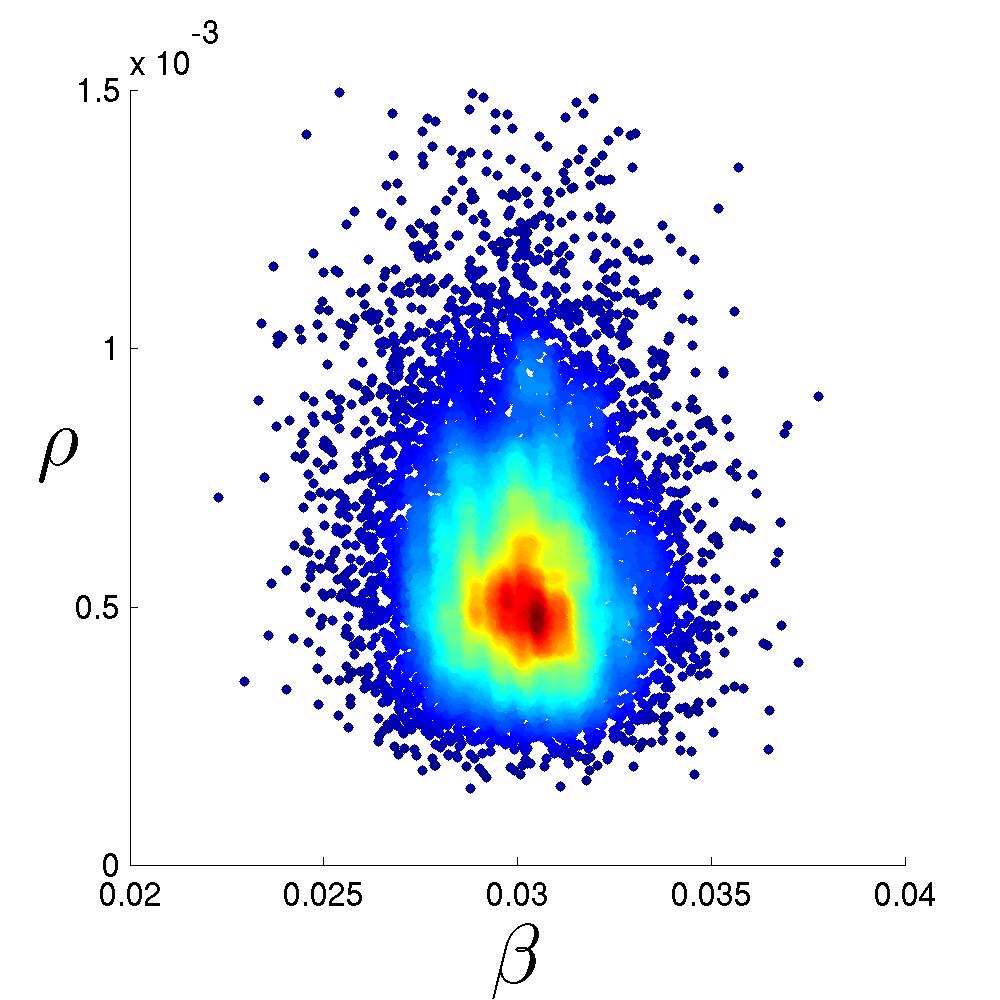}
\includegraphics[width=1.25in]{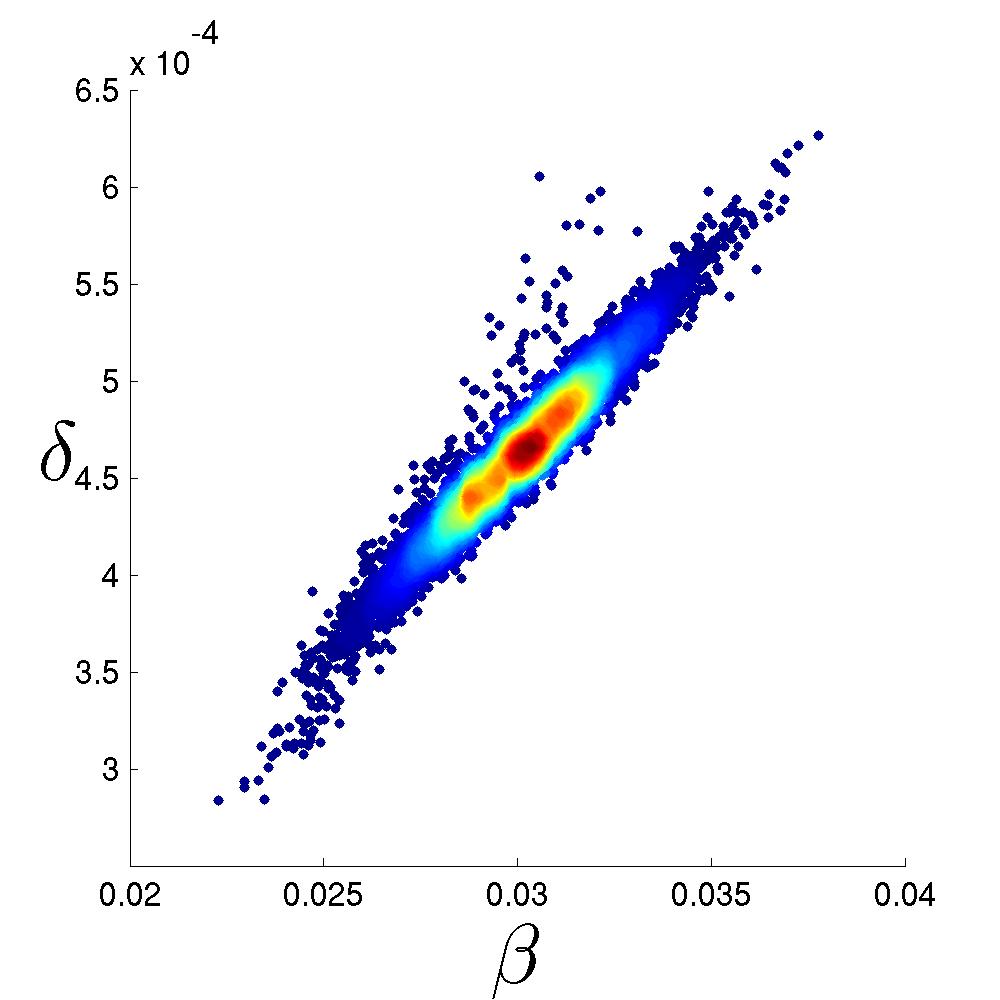}
\includegraphics[width=1.25in]{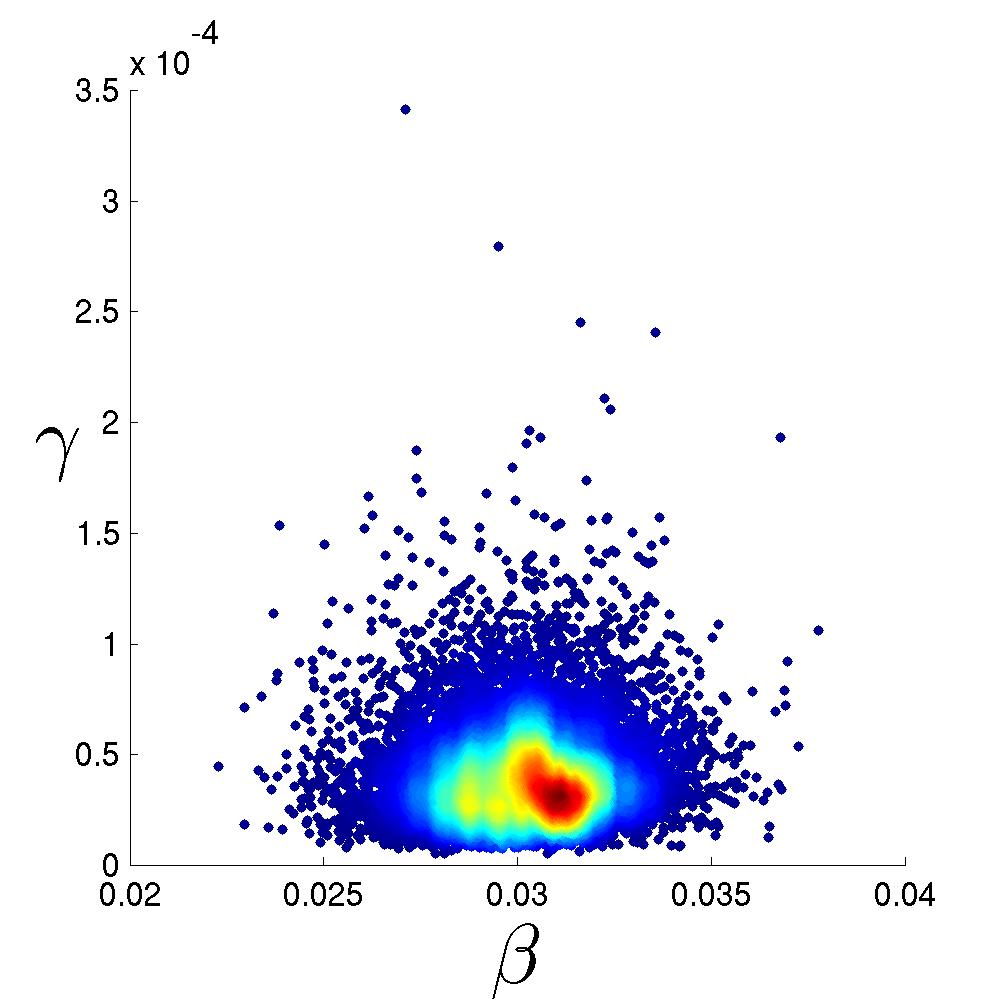}\\
\hskip-0.15in\includegraphics[width=1.25in]{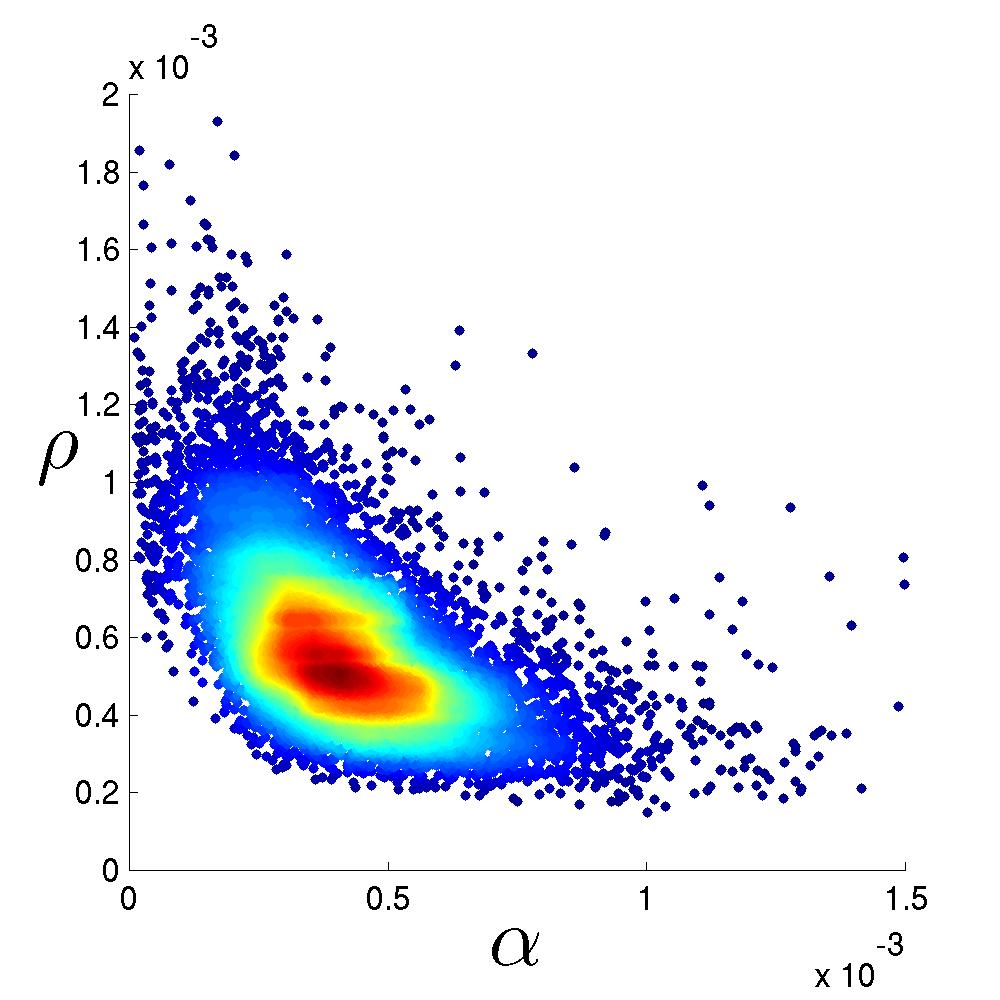}
\includegraphics[width=1.25in]{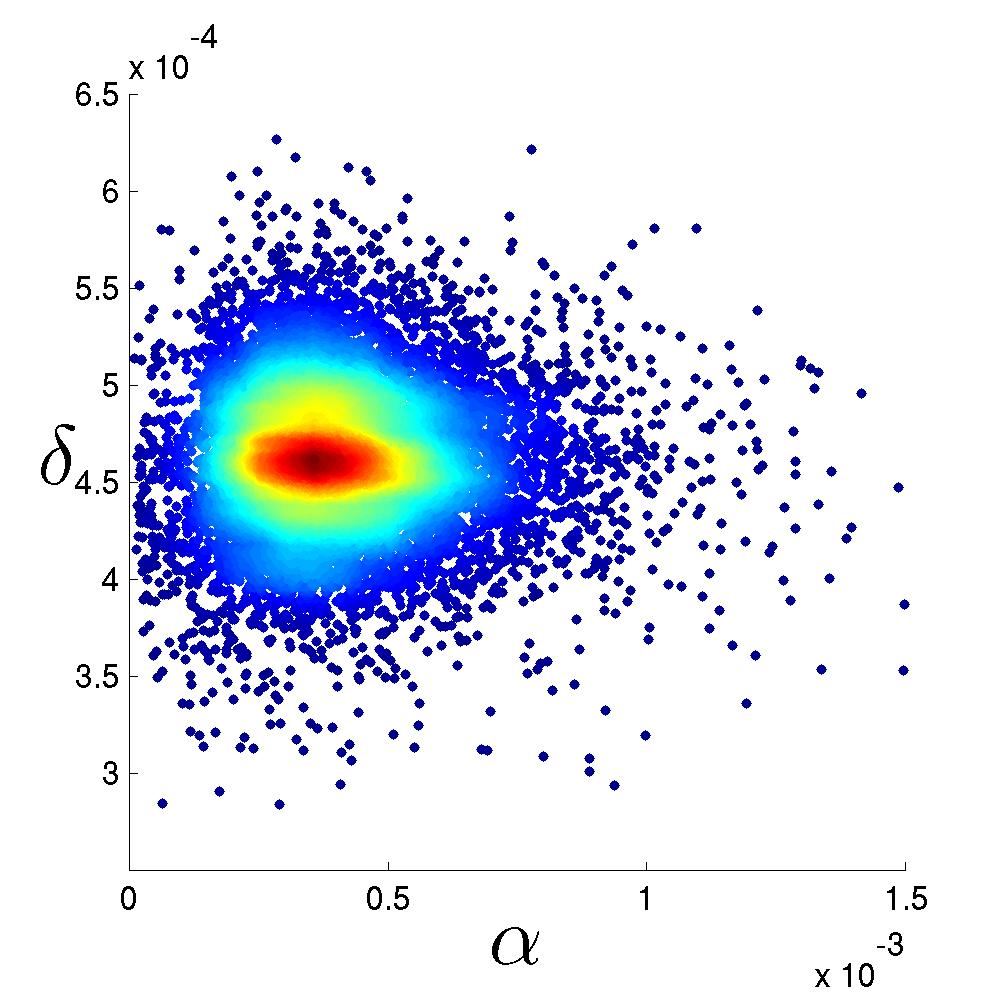}
\includegraphics[width=1.25in]{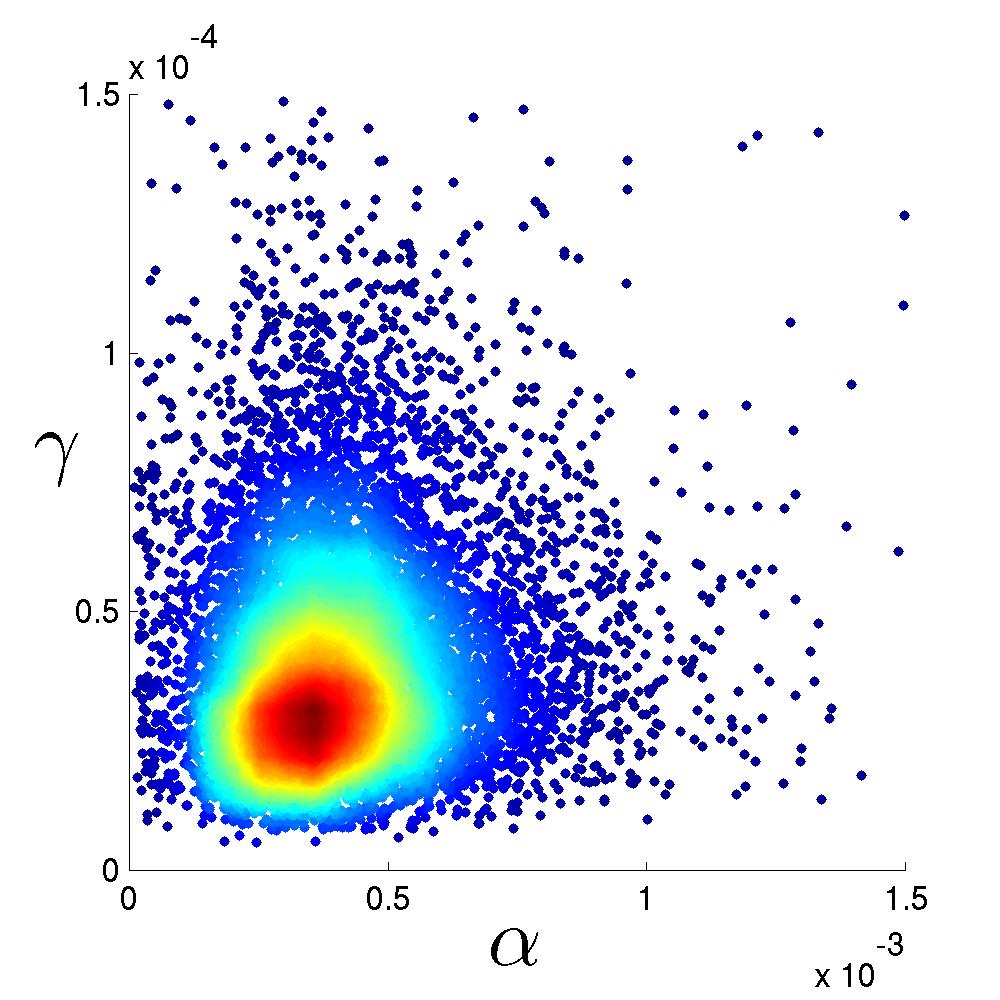}
\includegraphics[width=1.25in]{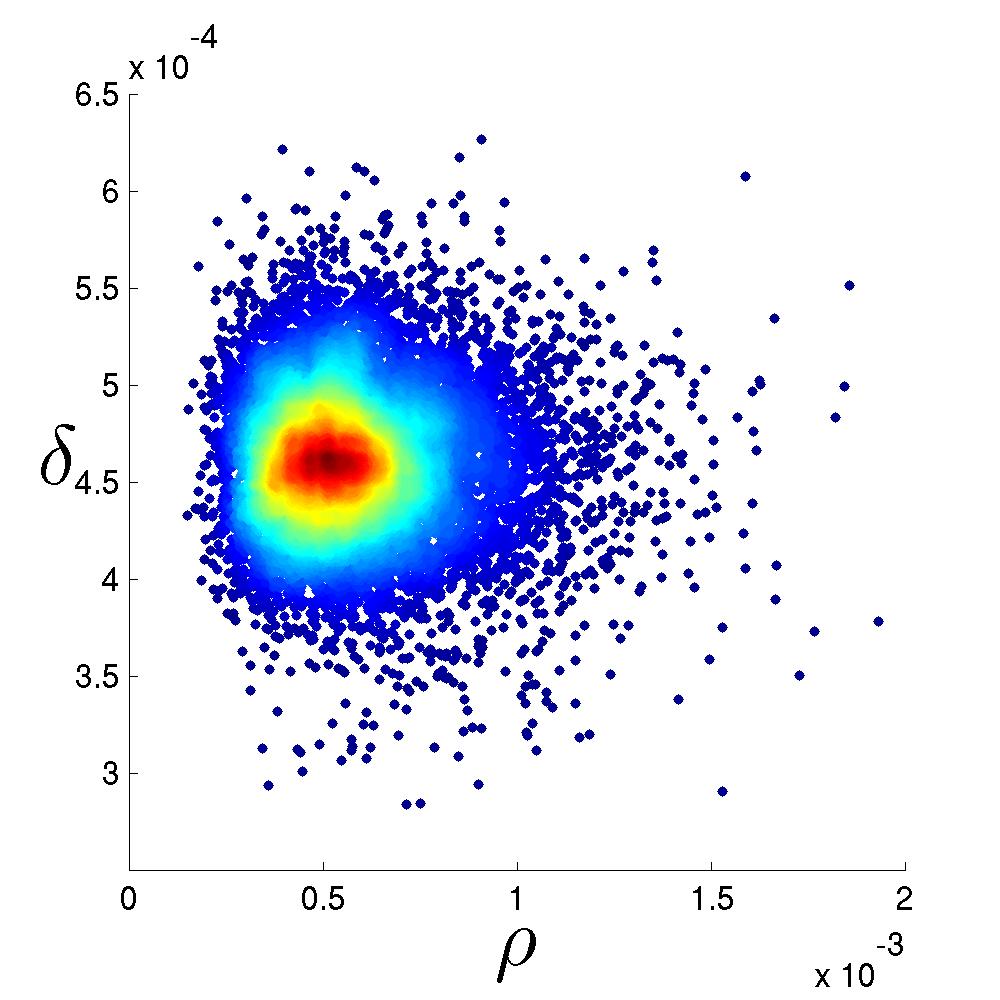}\\
\hskip-0.15in\includegraphics[width=1.25in]{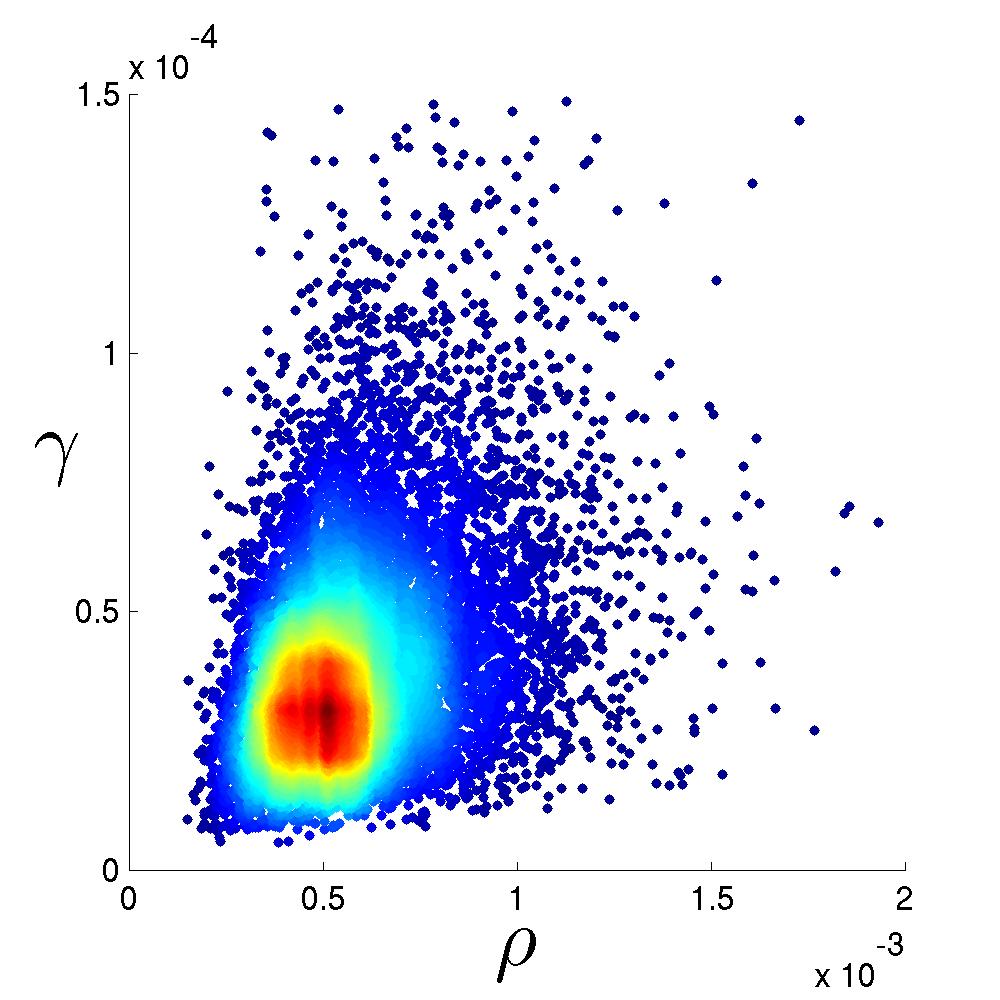}
\includegraphics[width=1.25in]{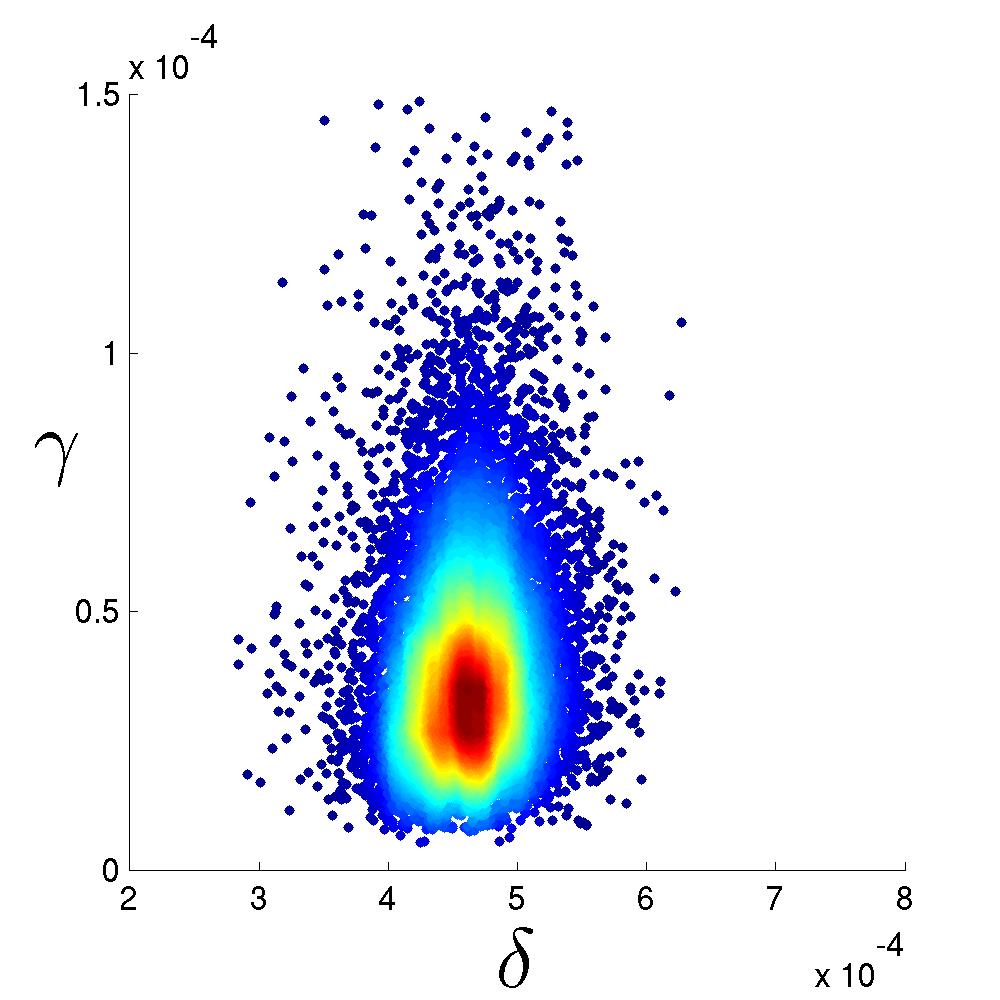}
\end{tabular}
\caption{\label{fig:post_scatter_id400} Macaque 400:  bivariate posterior margins for the 10 parameters as annotated.}
\end{figure}

Figure~\ref{fig:post_id400} summarizes marginal posteriors from analysis of data on one macaque;
these are typical of all 8 analyses. Evidently, some concentrated heavily in small regions relative to the diffuse initial $U(0,1)$ priors; e.g., see rate parameters $\beta, \alpha, \rho, \delta$ and $\gamma$. Margins for the carrying capacity parameters, $K_V$ and $K_M$ are also quite informative, while
those for $V_0$ and the delays $\tau_V,\tau_M$ are more diffuse while still contrasting somewhat with their
uniform priors.
Figure~\ref{fig:post_scatter_id400} shows summaries for selected bivariate margins, evidencing
some posterior dependencies between rate parameters in the posterior distribution.

\begin{figure}[ht!]
  \centering
\begin{tabular}{cccc}
\hskip-0.3in\includegraphics[width=1.3in]{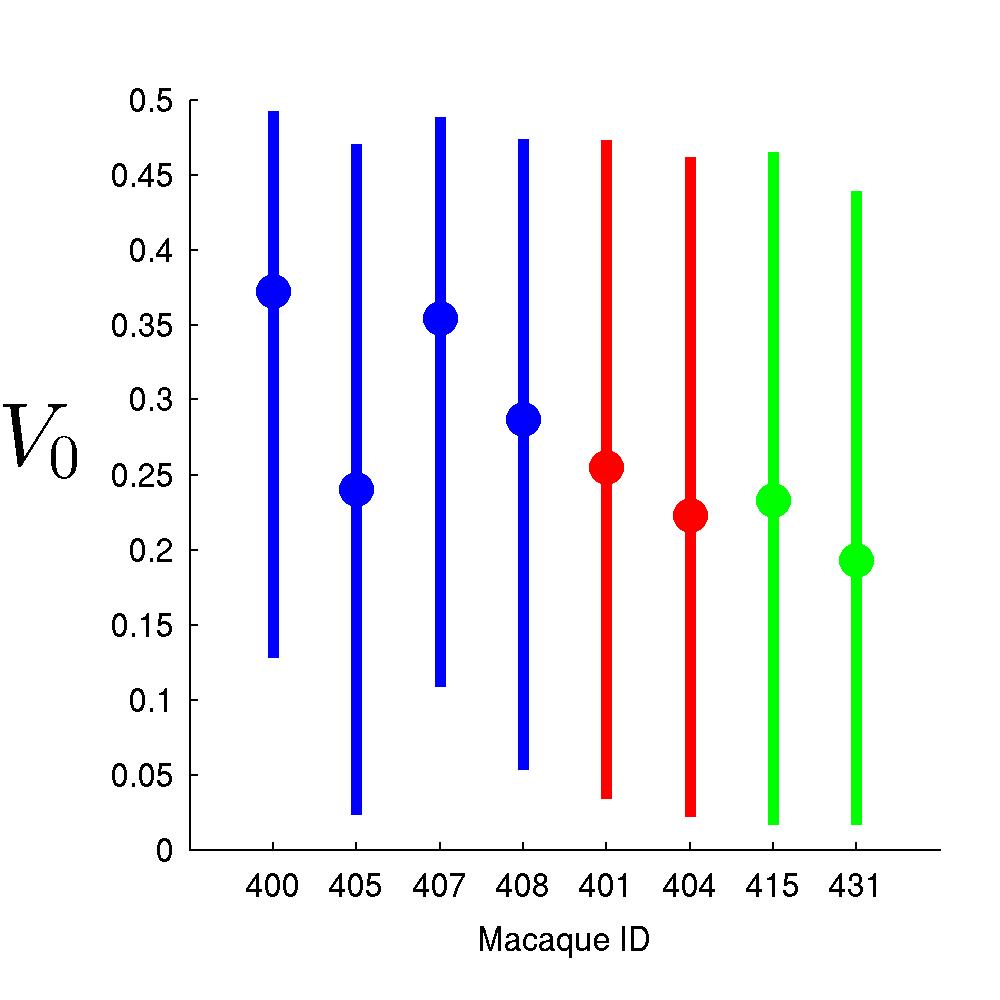}
\includegraphics[width=1.3in]{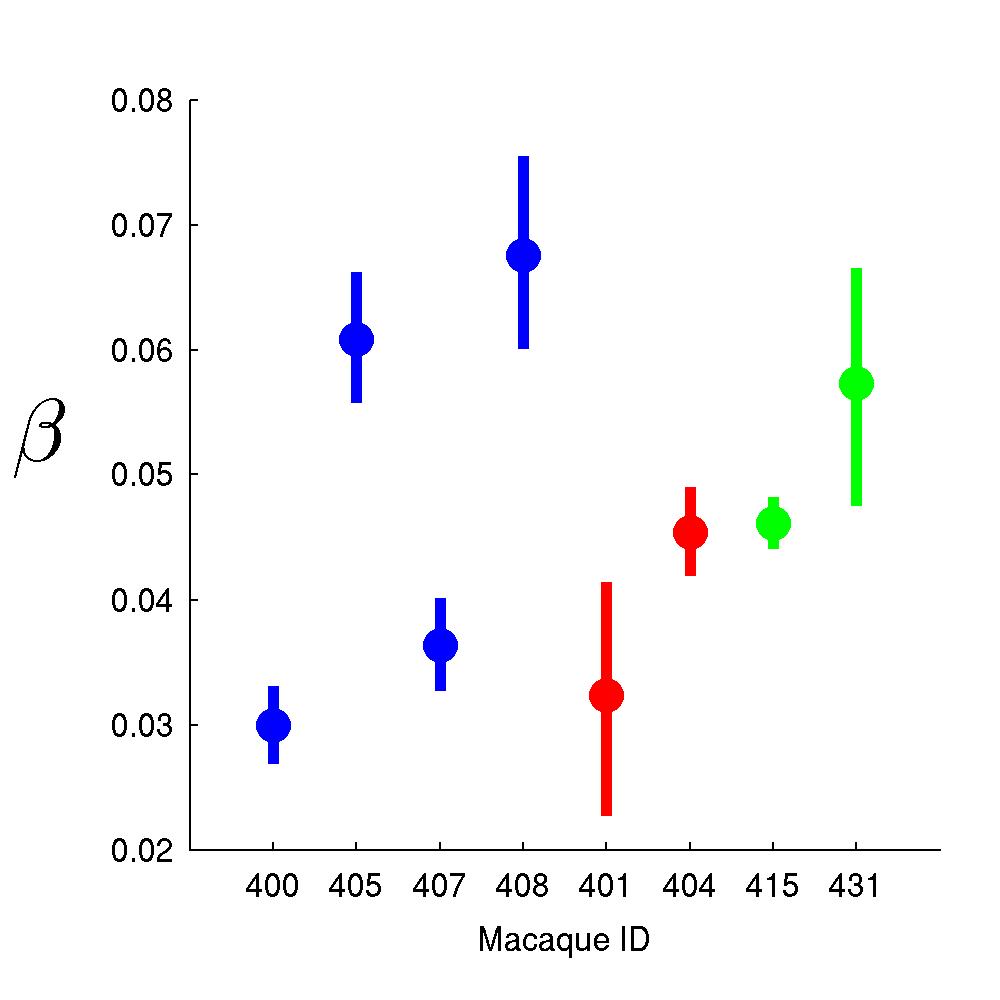}
\includegraphics[width=1.3in]{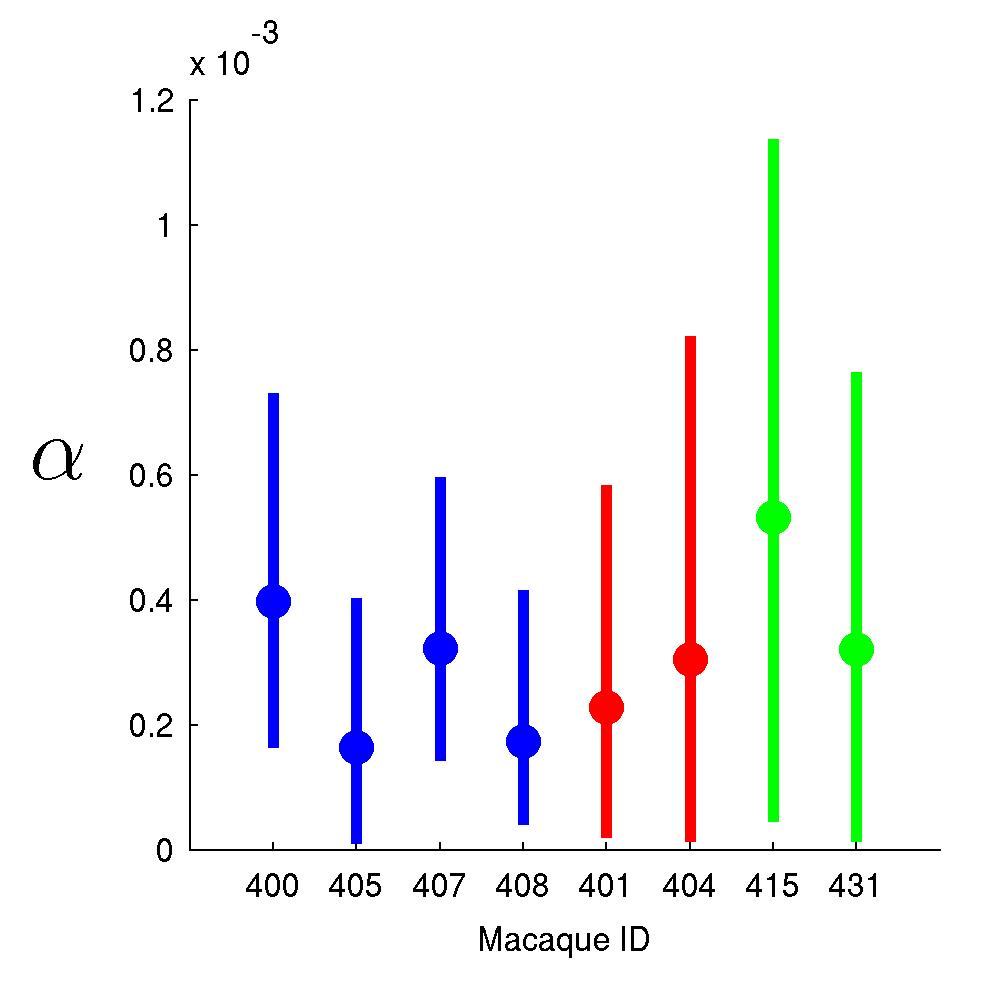}
\includegraphics[width=1.3in]{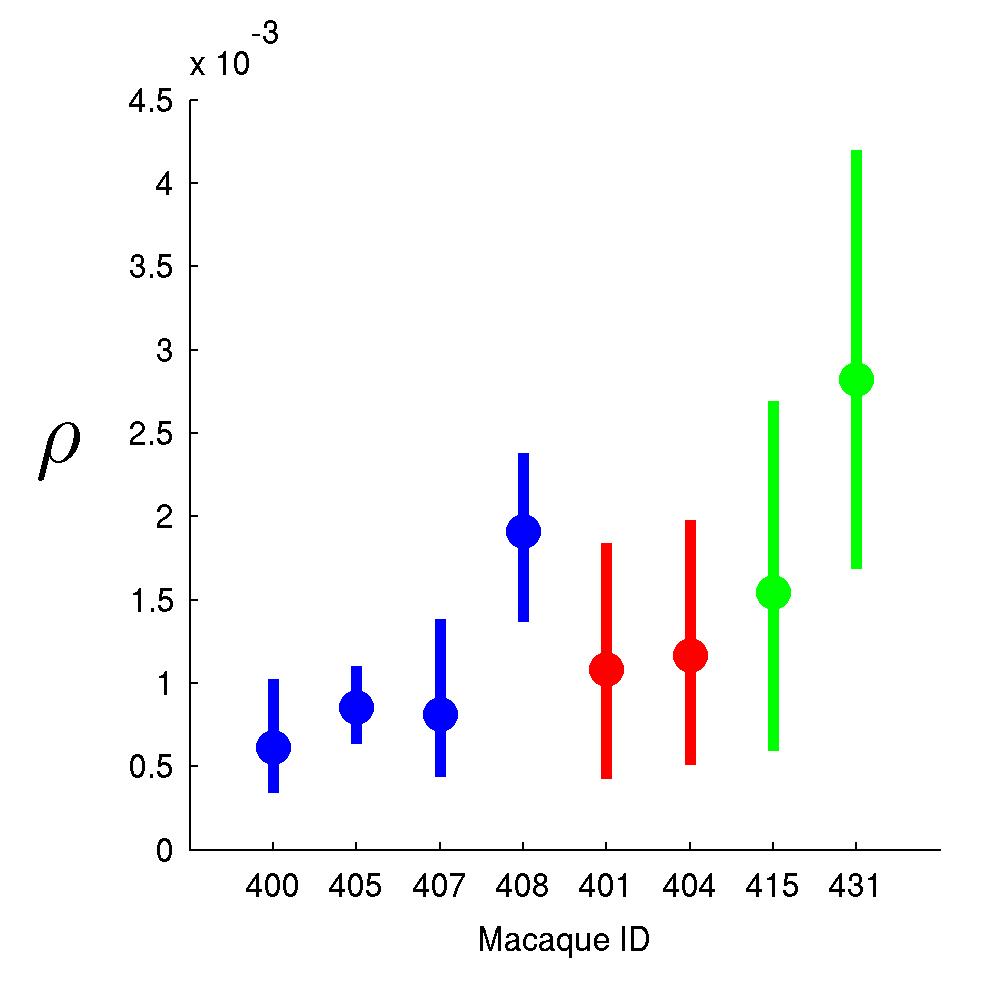}\\
\hskip-0.3in\includegraphics[width=1.3in]{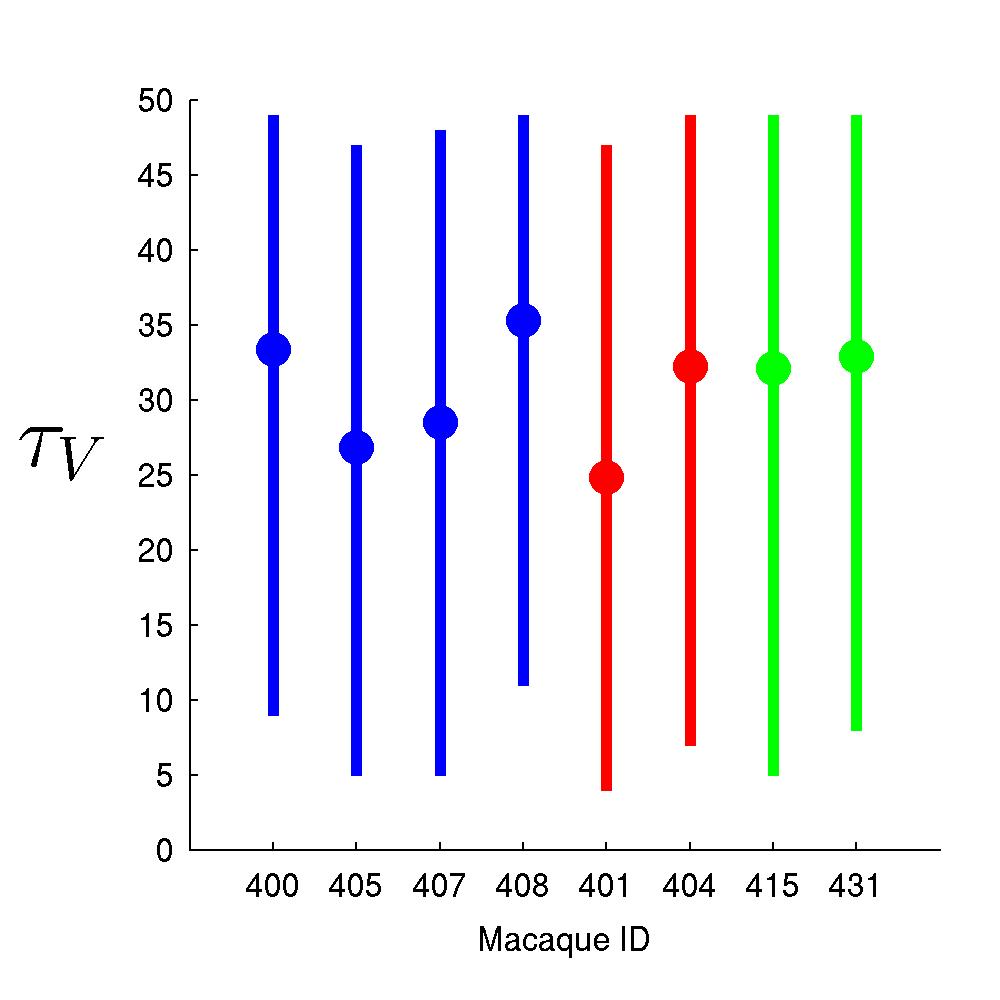}
\includegraphics[width=1.3in]{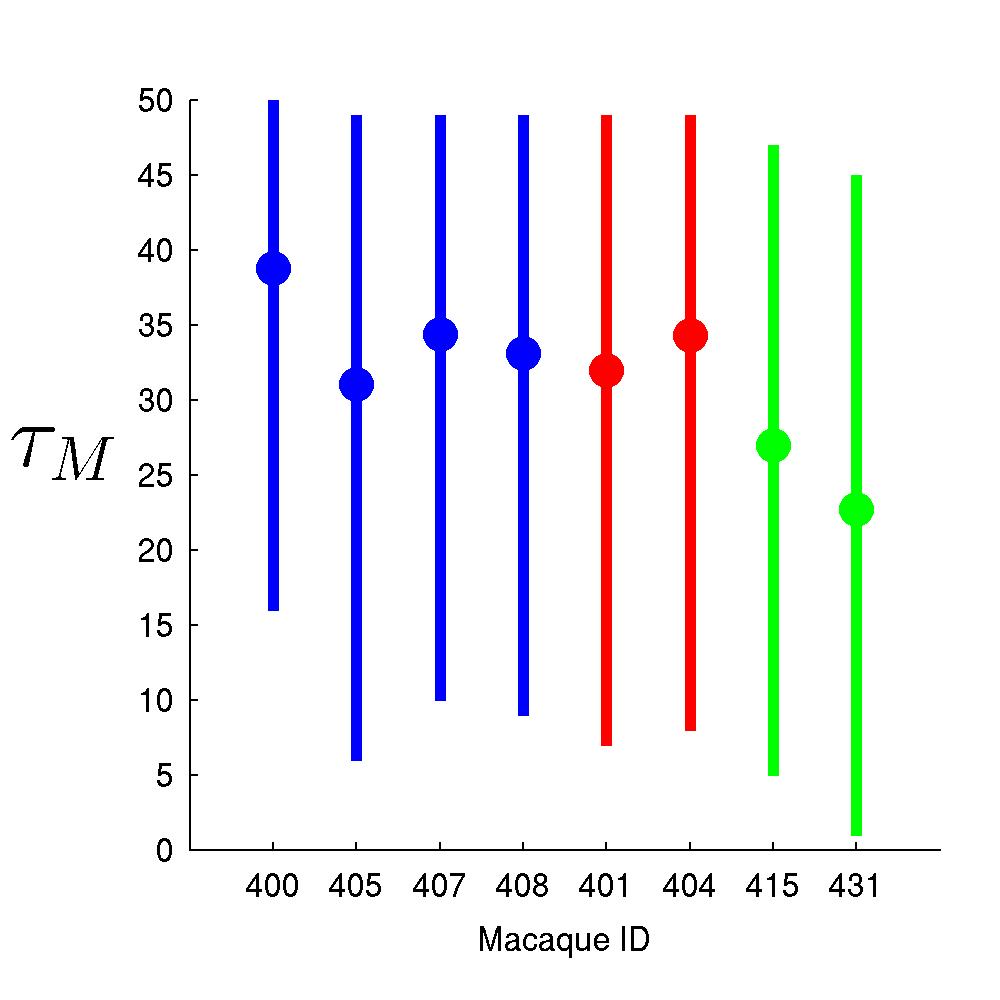}
\includegraphics[width=1.3in]{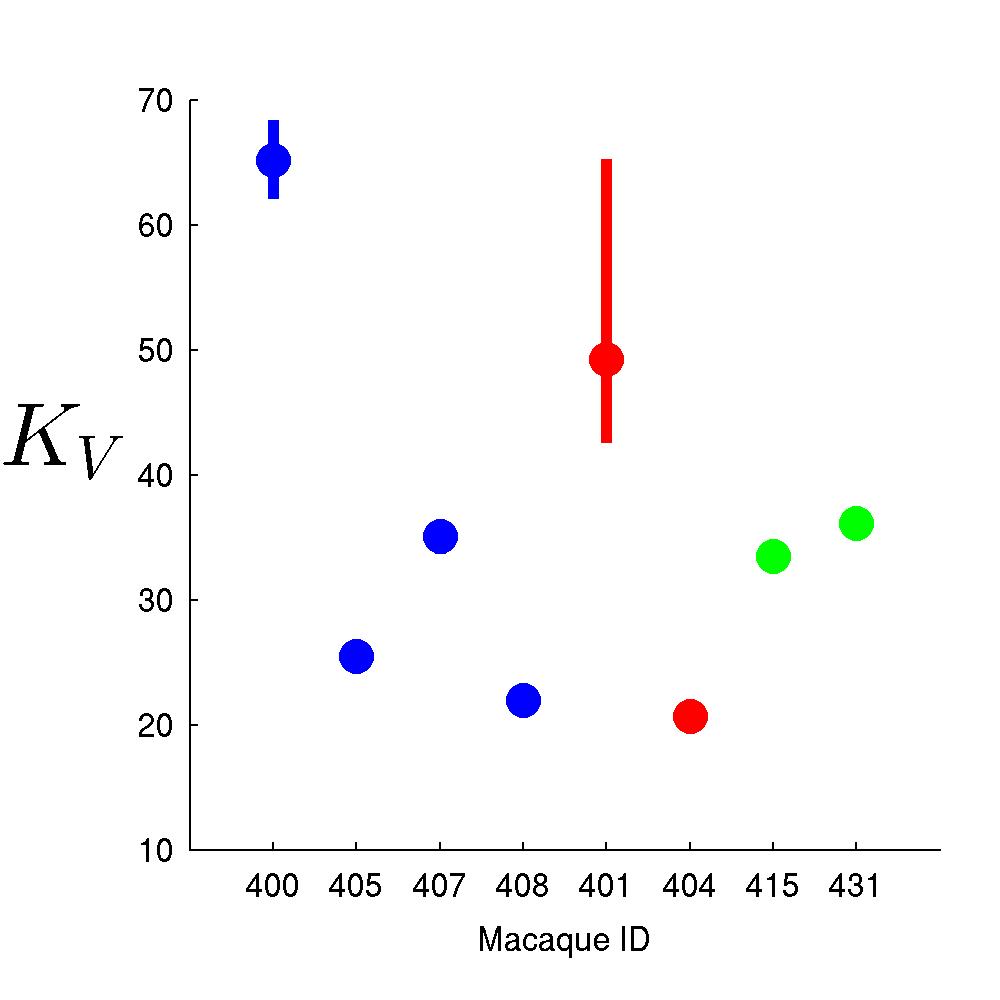}
\includegraphics[width=1.3in]{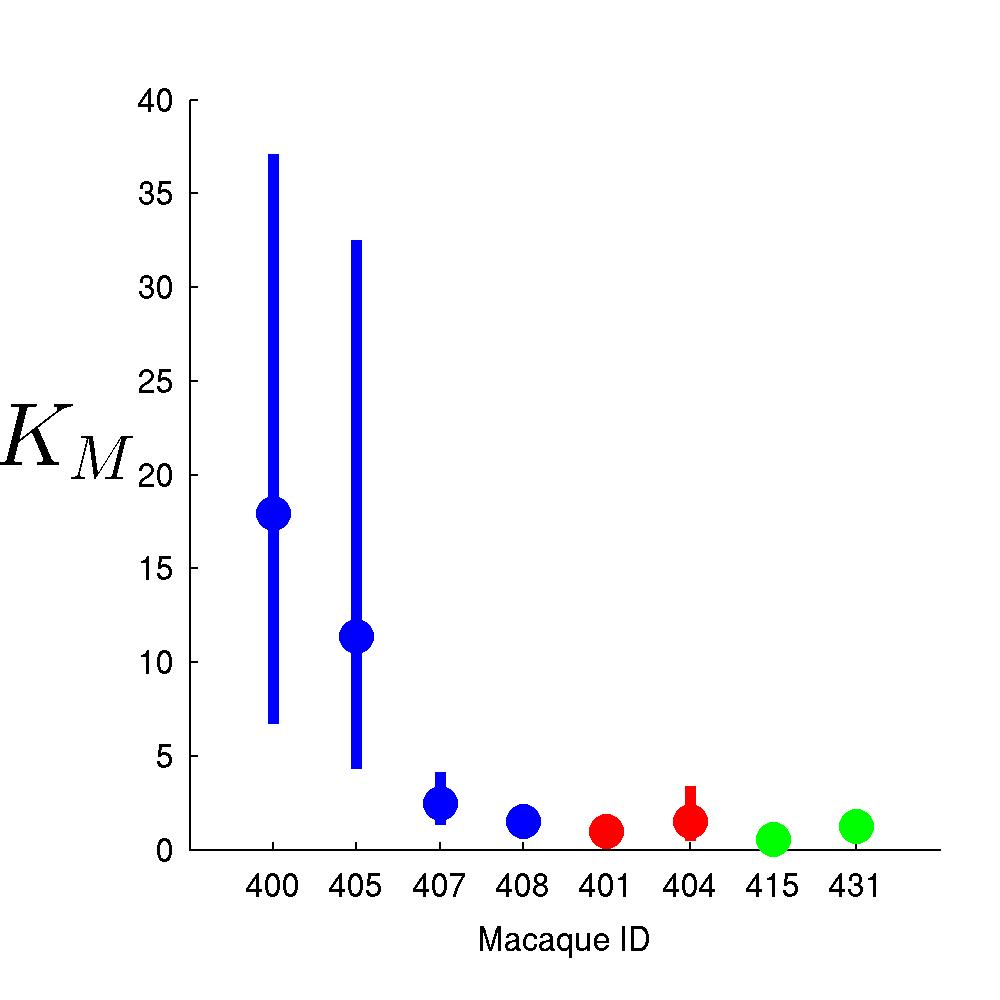}
\end{tabular}
\caption{\label{fig:summaries_abc_mcmc} Marginal posterior means and 90\% credible intervals for key parameters
on all 8 macaques. Color reflects vaccine immunization route:  intra-nasal/intra-tracheal (blue),
 intra-rectal (red),  intra-vagina (green).}
\end{figure}

\subsection{Comparisons of Subjects and Vaccination Routes}

Figure~\ref{fig:summaries_abc_mcmc} provides marginal posterior summaries to compare across the 8 macaques.
In general, there is substantial learning for the rate parameters $\beta, \alpha$, and $\rho$, and for the carrying capacity parameters, $K_V$ and $K_M$. The extent of learning is dependent on the conditions of the experiment, evidenced by the varying widths of credible intervals. For three parameter -- the initial viral level $V_0$ and  delays $\tau_V,\tau_M$--
such sparse time series data  is generally relatively uninformative: posteriors resemble priors on these quantities.

Color coding relates to immunization route of the vaccine administered--  intra-nasal/intra-tracheal, intra-rectal, or intra-vaginal. For $\rho,K_M,$  there is some indication of differences by immunization route, which
was unanticipated. Additional experiments directed towards comparing finer-scale immune responses across
immunization route are suggested as a result.

\subsection{MCMC+ABC Improves Predictive Model Fit}

\begin{figure}[ht!]
  \centering
\includegraphics[width=2.2in]{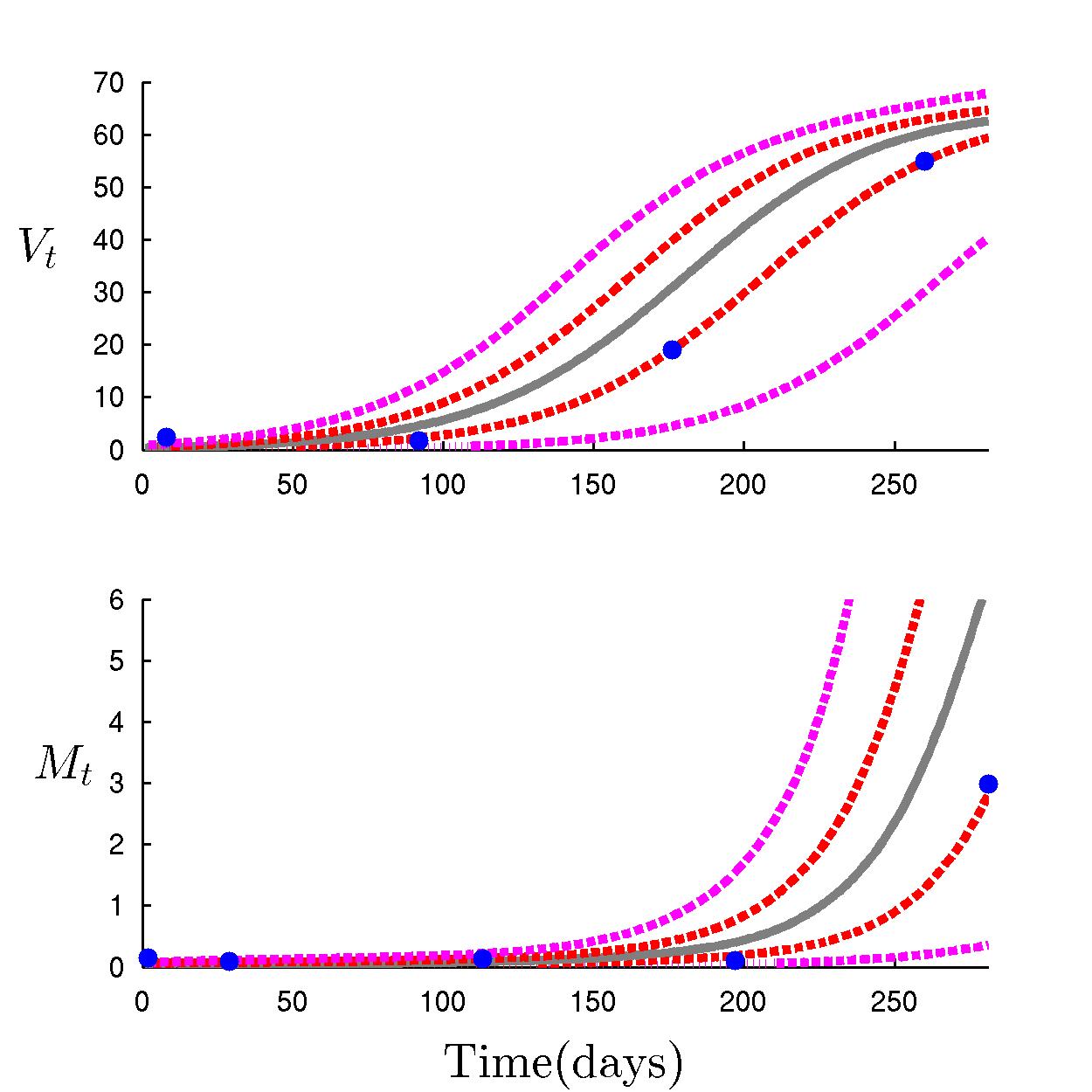}
\includegraphics[width=2.2in]{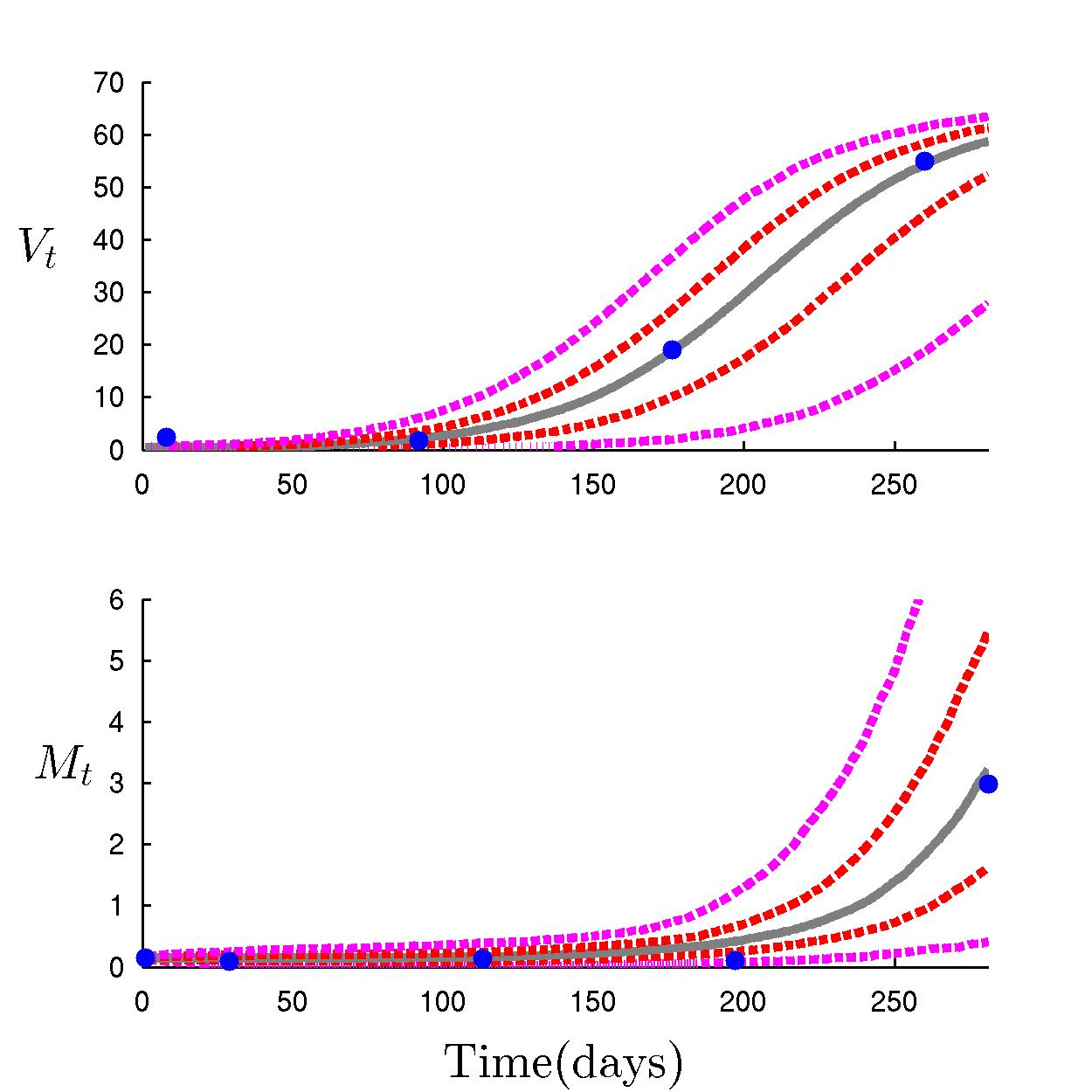} \\
\includegraphics[width=2.2in]{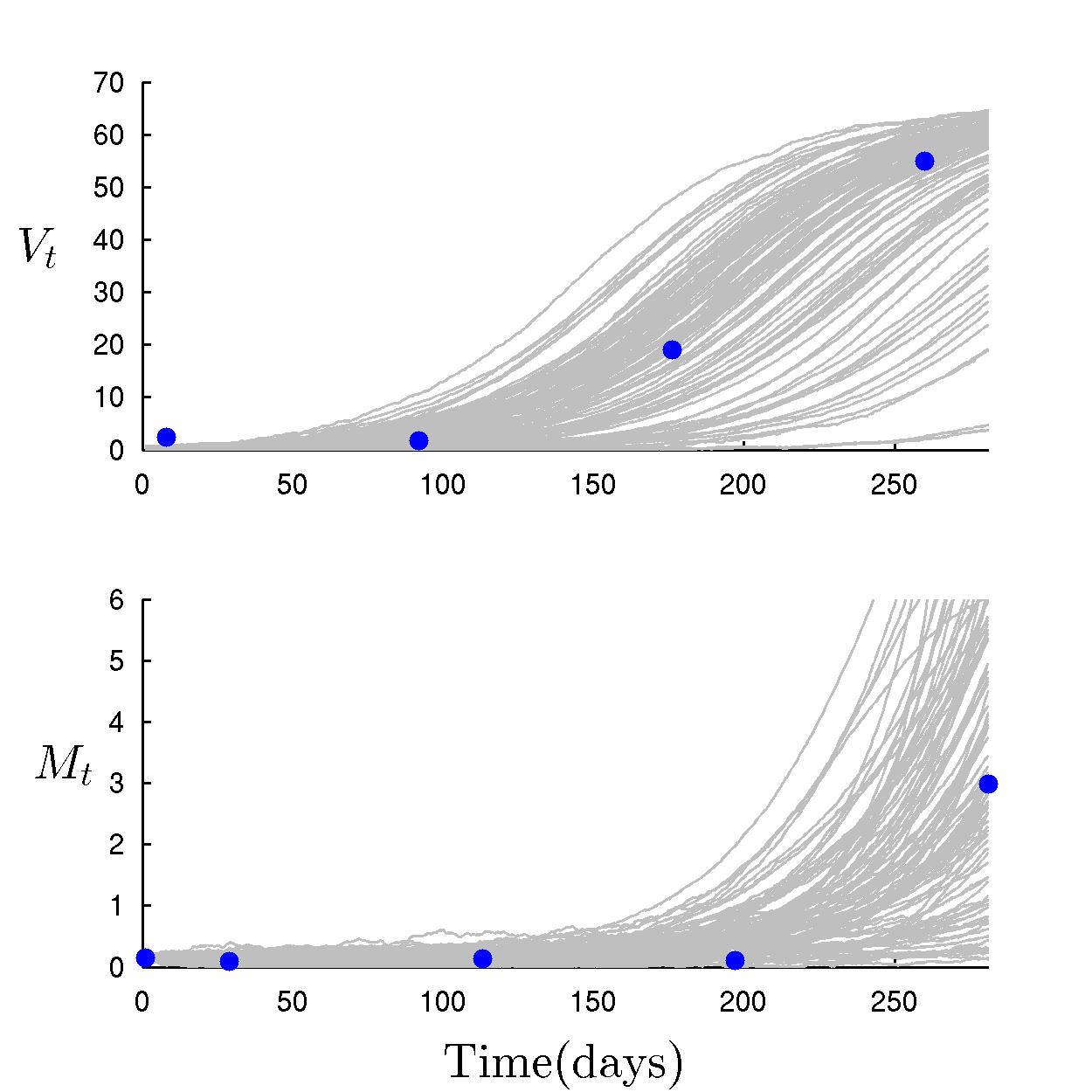}\\
\caption{\label{fig:predplots_id400} Posterior predictions for macaque \#400:
{\em Top left:}  Observed data (blue);
50\% (red), 95\% (magenta) credible bands and median (gray) of posterior predictions for the latent states
using only MCMC analysis.
{\em Top right:} Same format, but based on MCMC+ABC analysis.
{\em Lower center:}
100 sample trajectories from the posterior predictive distribution of the latent states from MCMC+ABC analysis.}
\end{figure}

Posterior predictions  assess and compare model fits; these demonstrate the positive benefit of
 the ABC procedure in refining the first stage MCMC analysis. For each data set,  final posterior parameter samples
were used to generate synthetic latent state  trajectories.  Figures~\ref{fig:predplots_id400}  shows aspects of these predictions for macaque \#400. These posterior predictive trajectories were simulated
{\em (i)}  based on the first stage MCMC alone,  and {\em (ii)} from the full MCMC+ABC analysis. This shows that the MCMC analysis results in a relatively good concordance with the data; however, systematic bias in overestimating the latent states is clear-- likely a result of linearization and lack of truncation in the MCMC analysis. The figure shows
how the ABC step is able to \lq\lq correct'' this,  resulting in clear improvements in predictive fit.

Repeating this comparative analysis for the other 6 data sets leads to similar conclusions; see graphs in the  Supplementary Material.
For some cases, very clear improvements are evident.  For a few, such as macaque \#401,
there is less evidence of systematic bias in the MCMC analysis, and then the ABC step results in minimal adjustment.

\subsection{Comparisons via Dimensionless Model Parameters}

\begin{figure}[ht!]
\centering
\begin{tabular}{cccc}
\hskip-0.1in\includegraphics[width=1.25in]{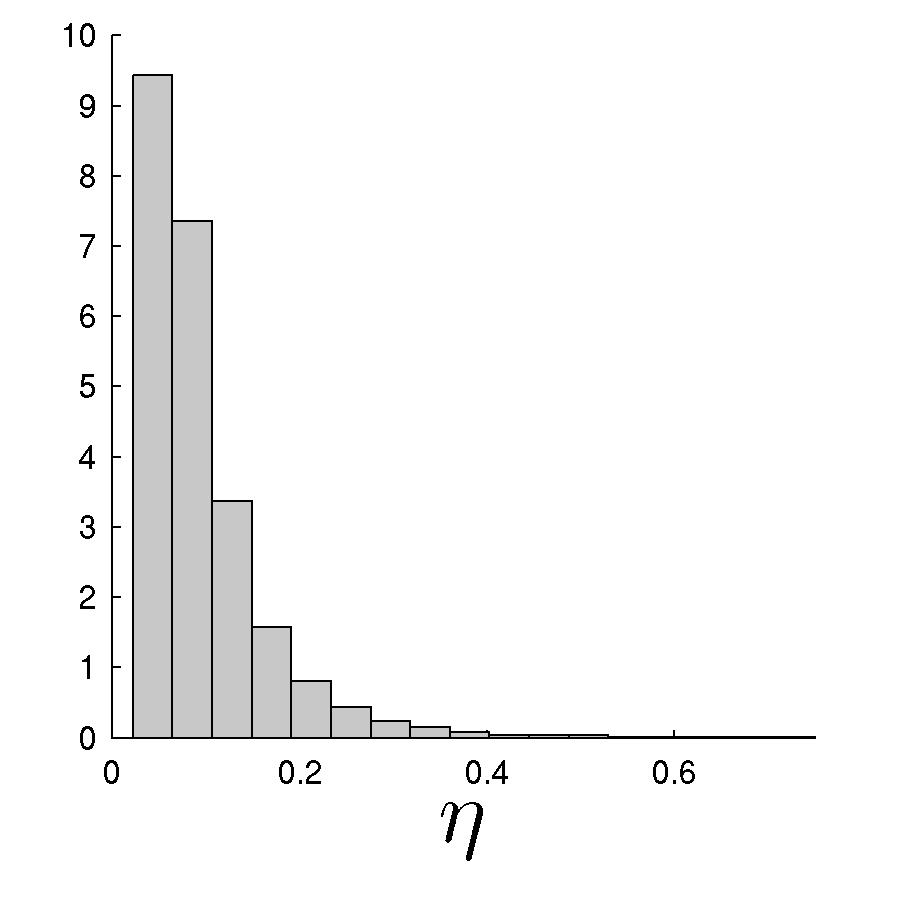}
\includegraphics[width=1.25in]{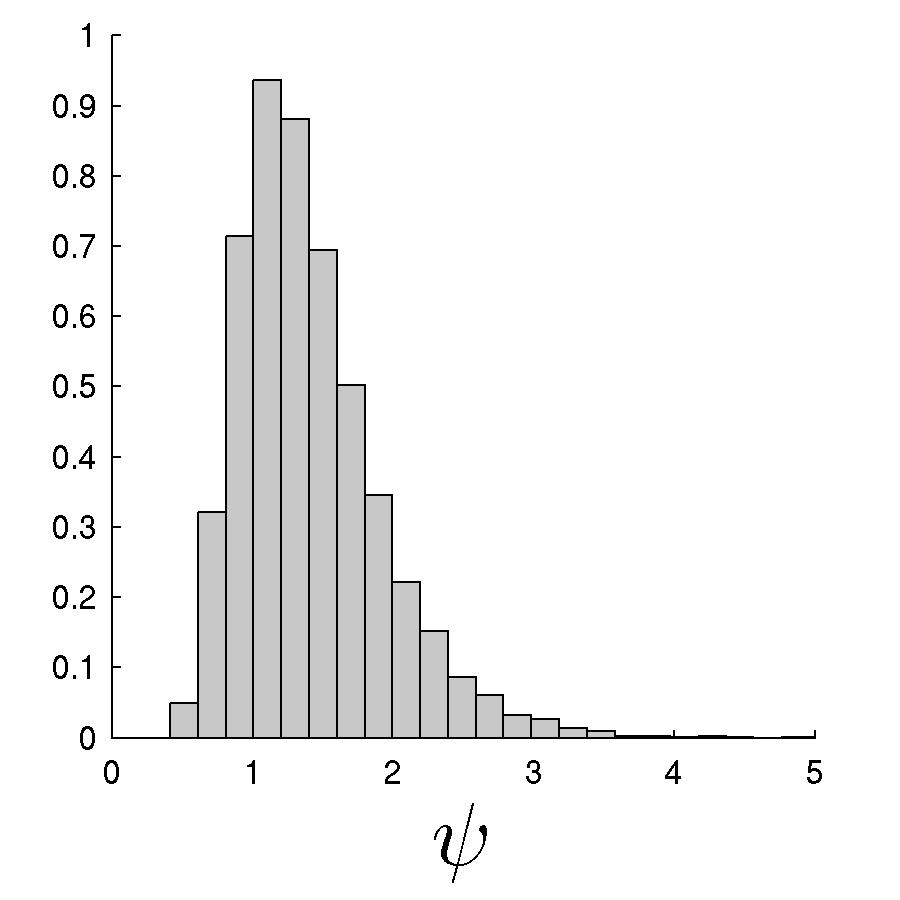}
\includegraphics[width=1.25in]{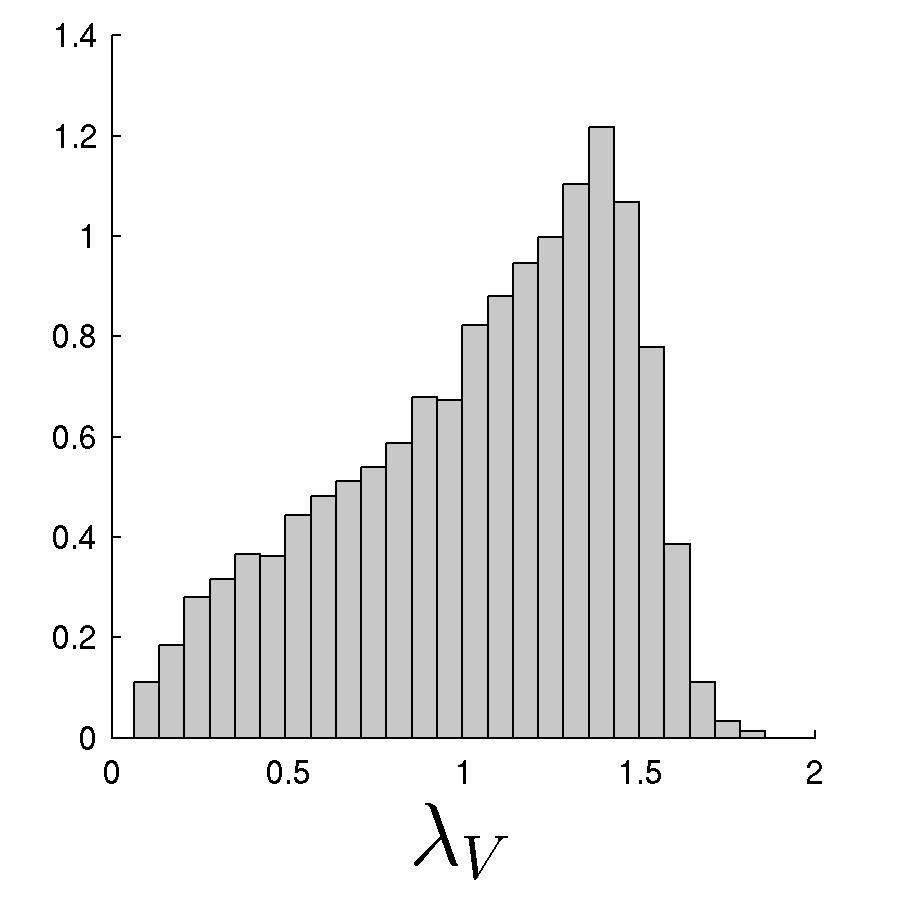}
\includegraphics[width=1.25in]{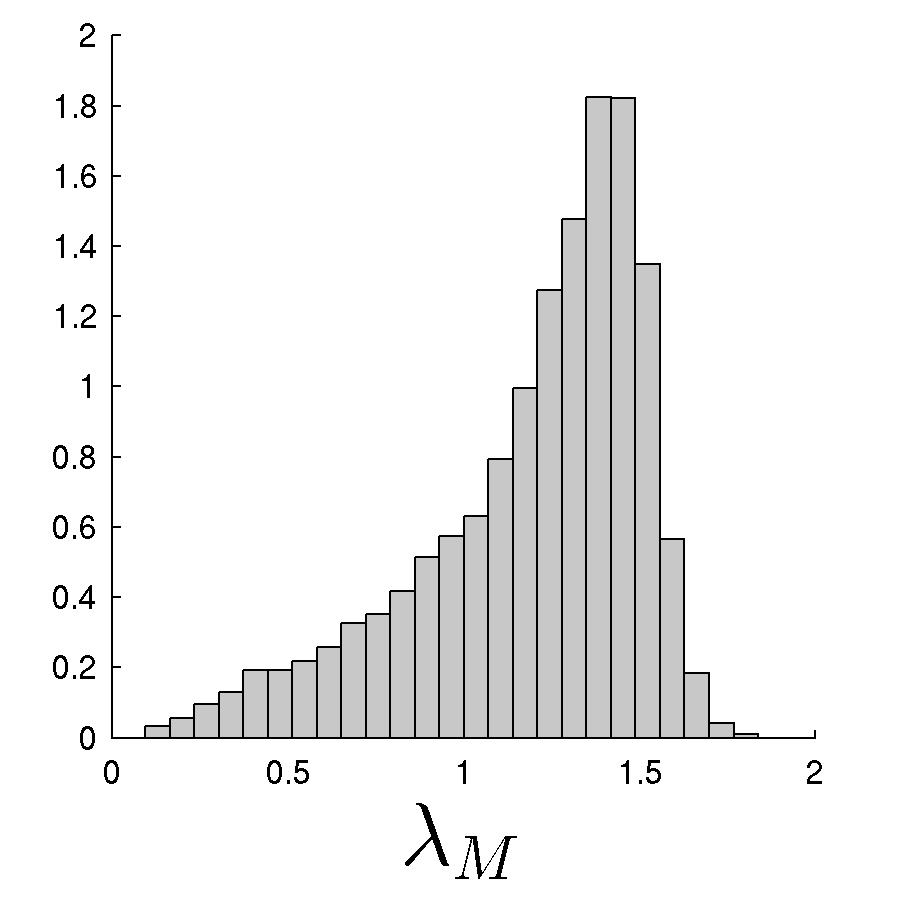}\\
\end{tabular}
\caption{\label{fig:dimensionless_marginal} Macaque 400:  posterior margins for the for the dimensionless model parameters as annotated.}
\end{figure}

\begin{figure}[ht!]
  \centering
\begin{tabular}{cccc}
\hskip-0.15in\includegraphics[width=1.25in]{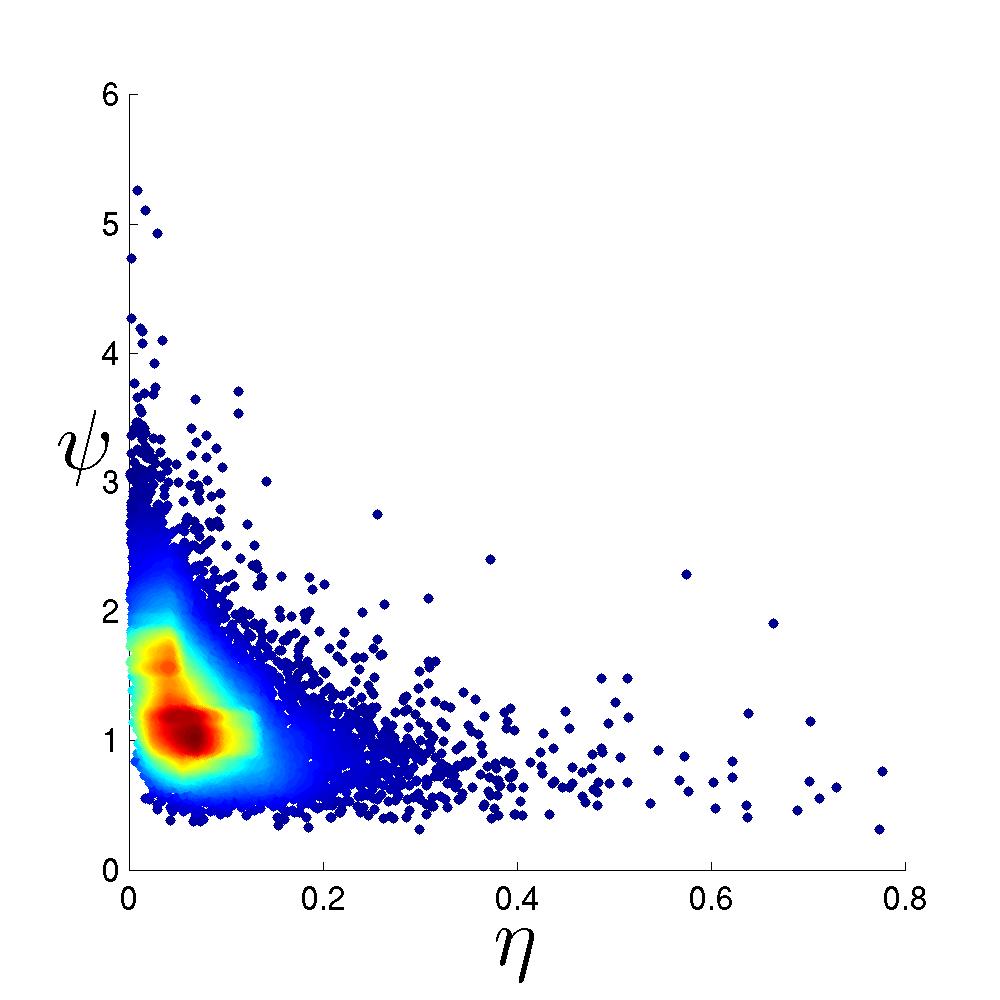}
\includegraphics[width=1.25in]{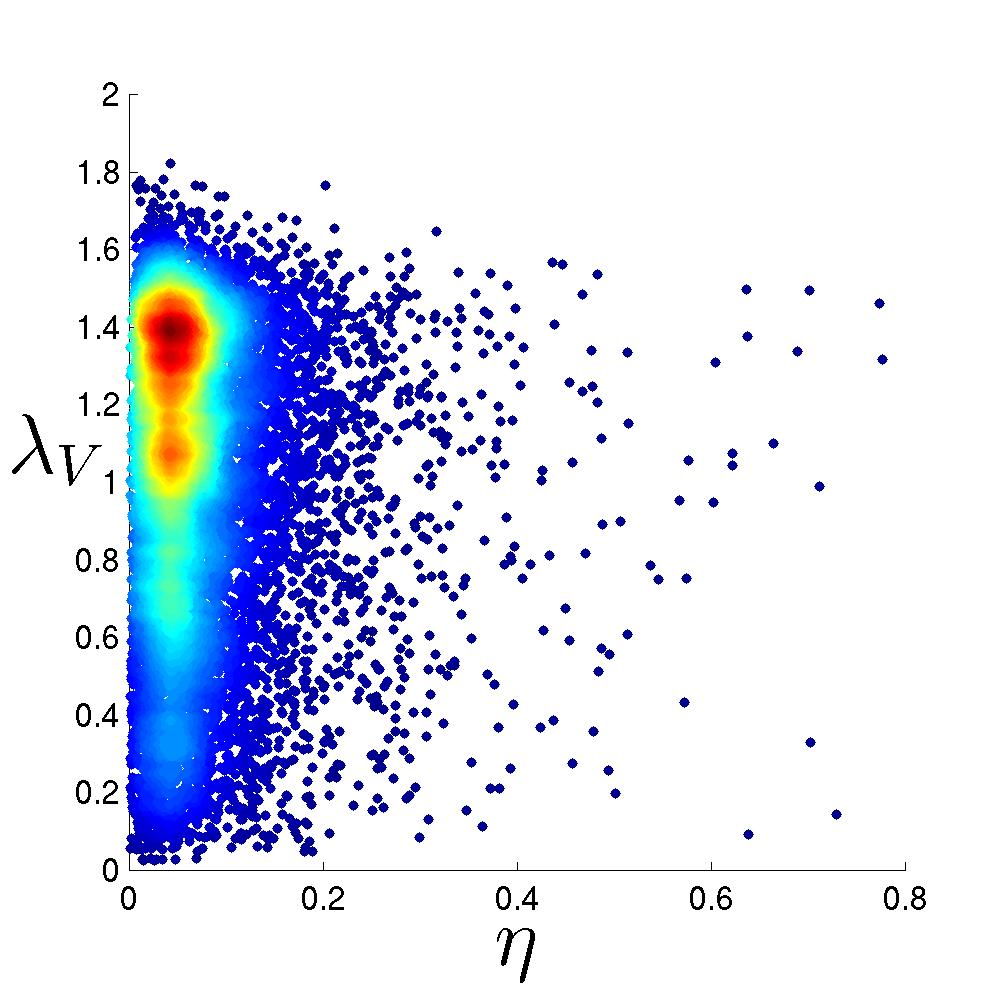}
\includegraphics[width=1.25in]{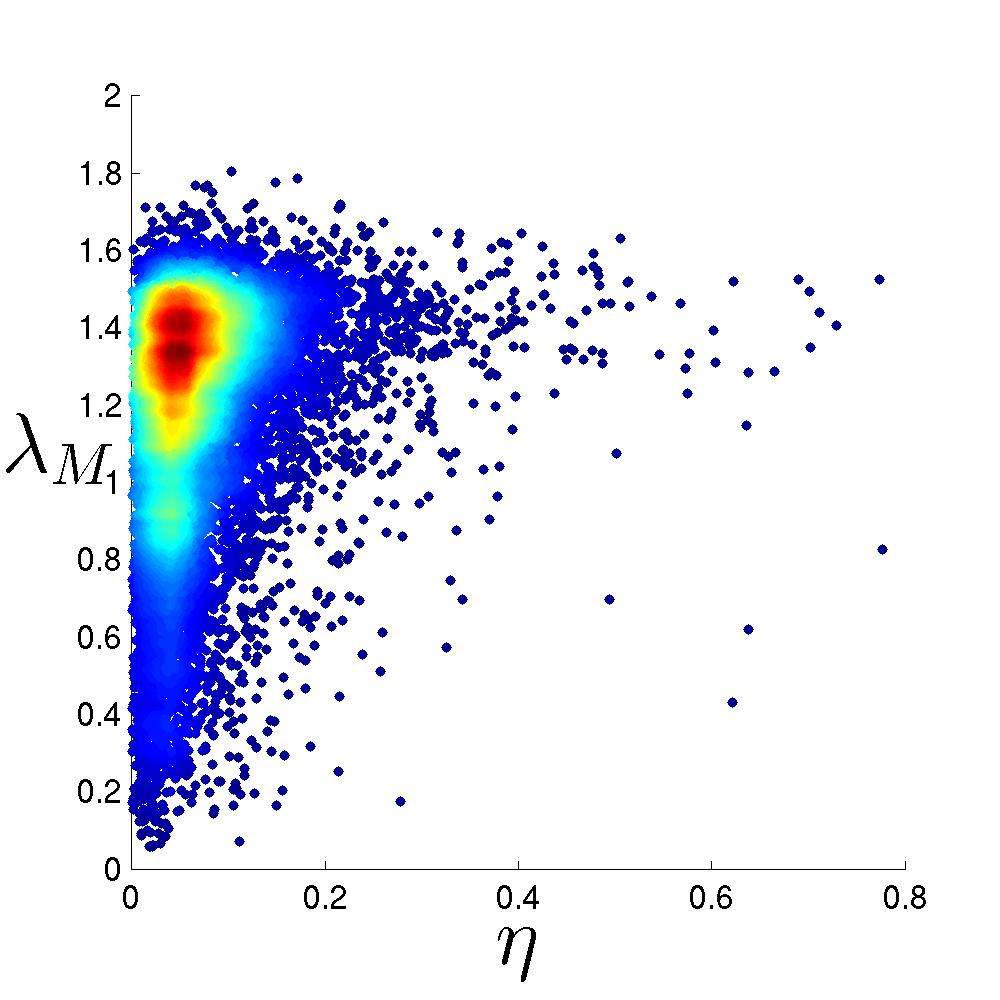}\\
\hskip-0.15in\includegraphics[width=1.25in]{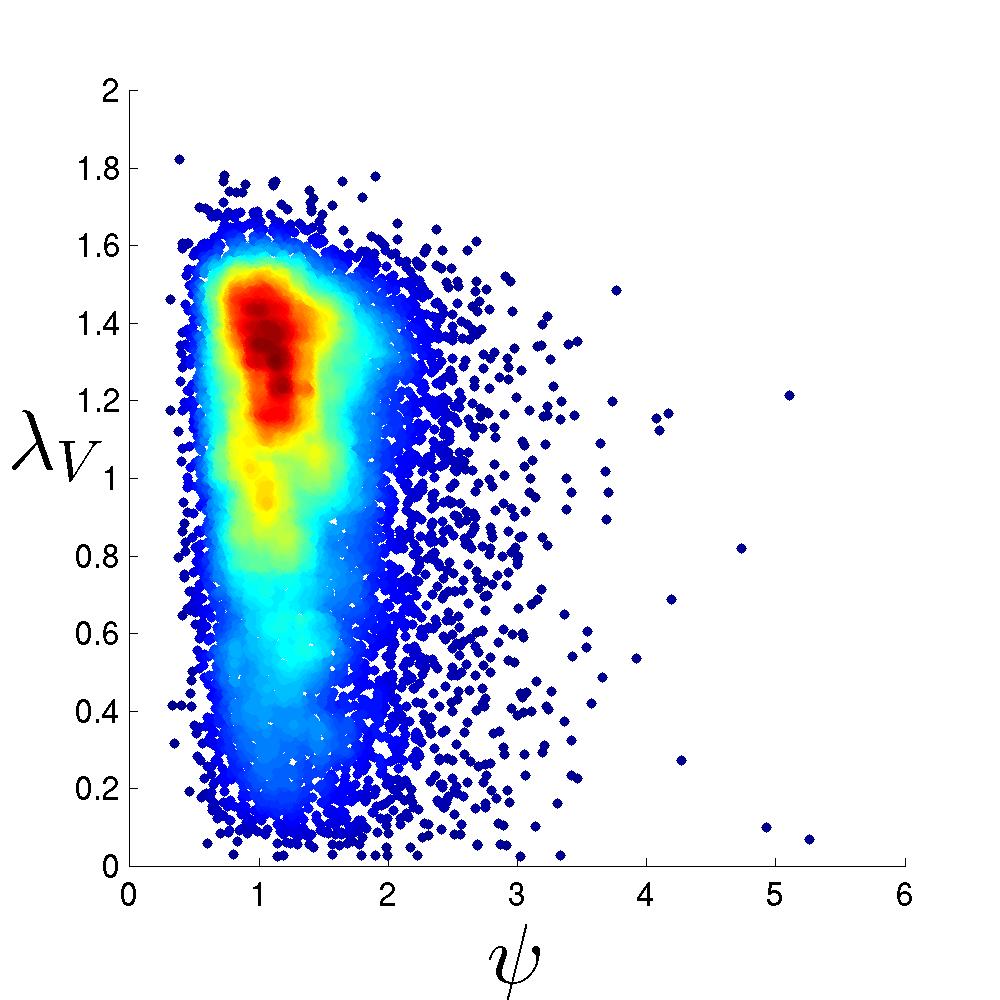}
\includegraphics[width=1.25in]{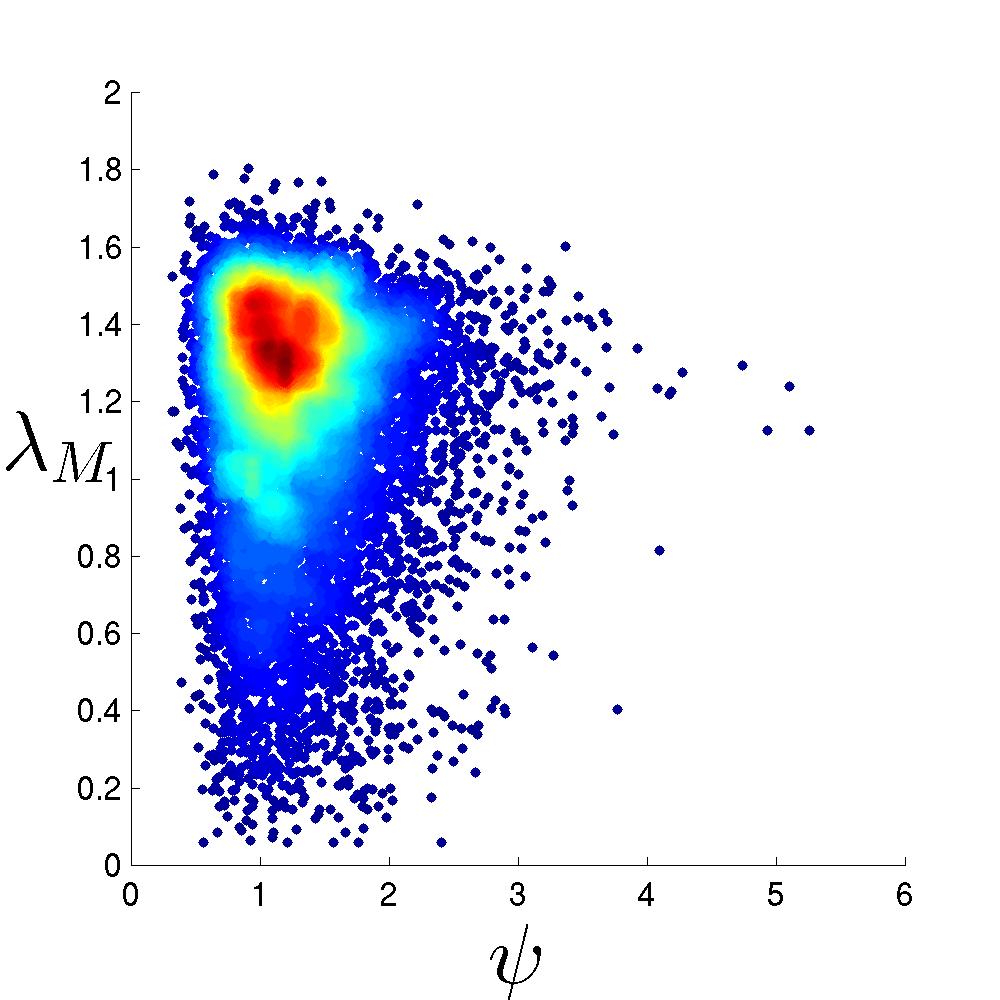}
\includegraphics[width=1.25in]{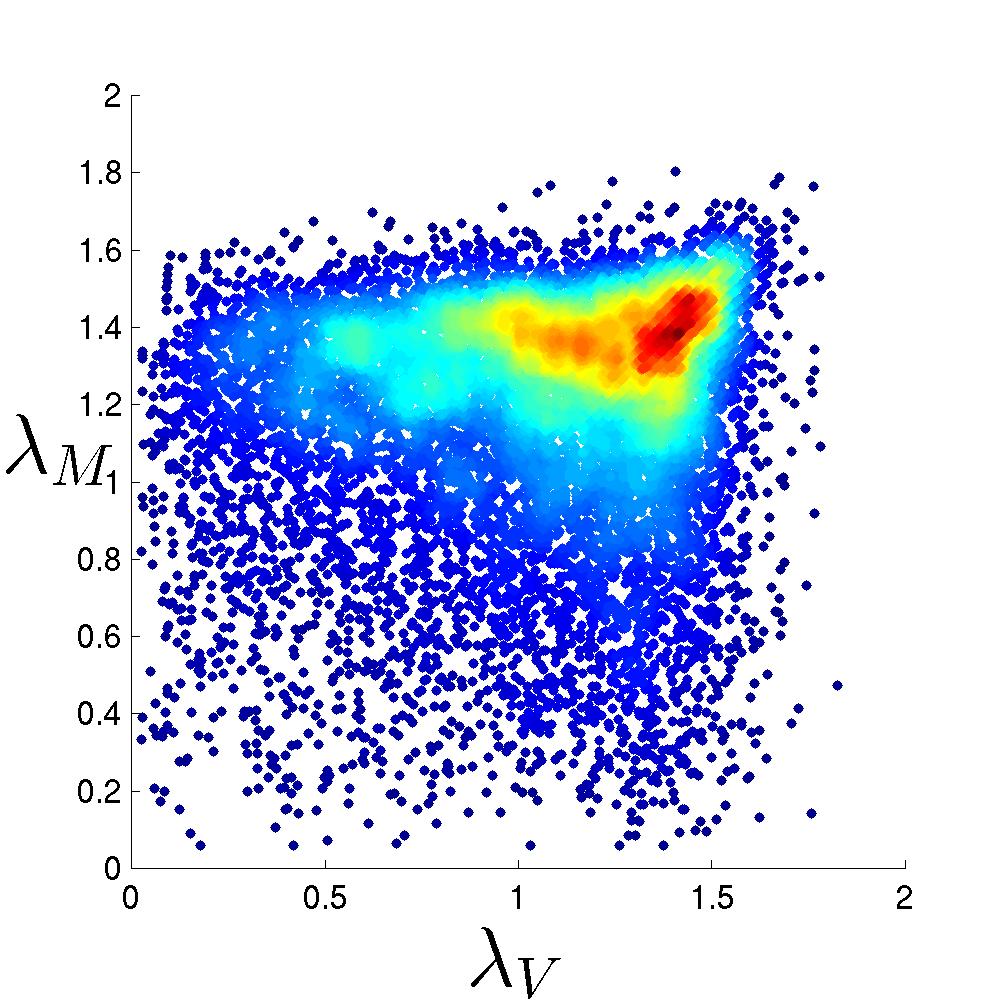}\\
\end{tabular}
\caption{\label{fig:dimensionless_joint} Macaque 400:  bivariate posterior margins for the dimensionless model parameters as annotated.}
\end{figure}

\begin{figure}[ht!]
  \centering
\begin{tabular}{cccc}
\hskip-0.3in\includegraphics[width=1.3in]{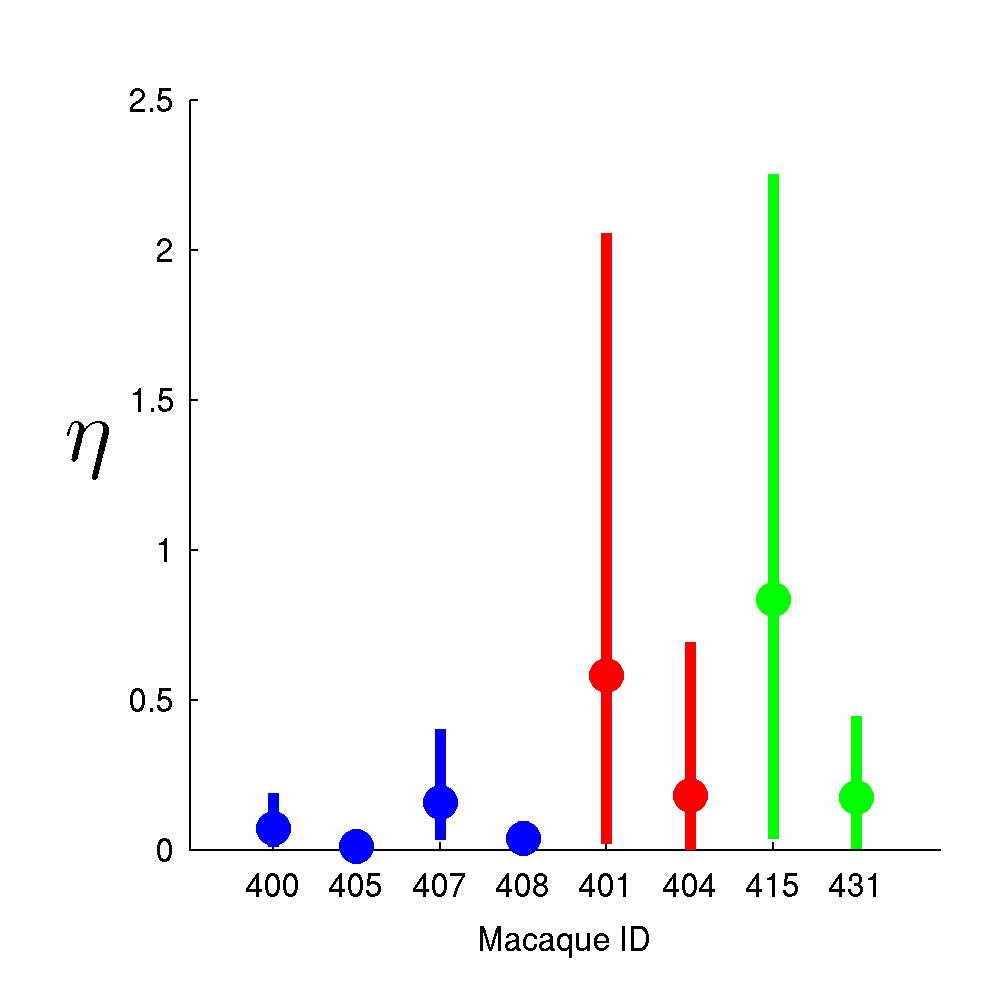}
\includegraphics[width=1.3in]{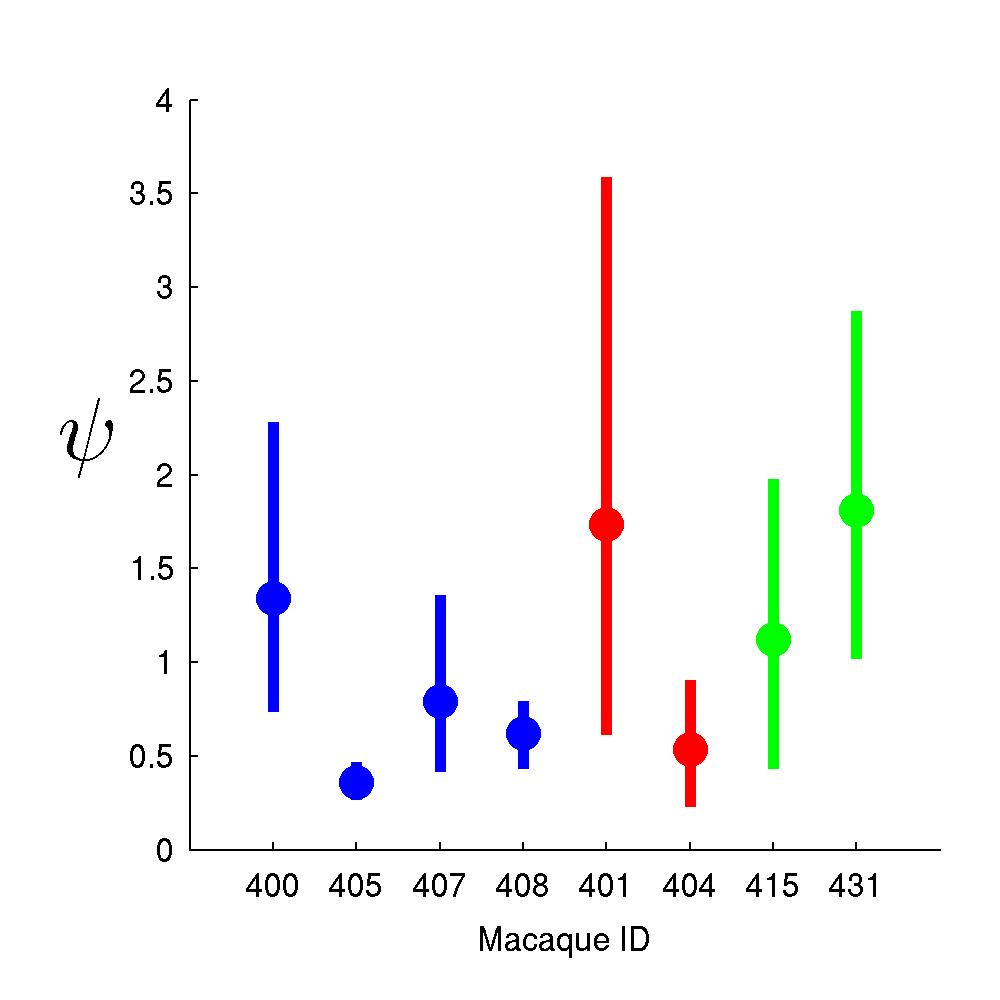}
\includegraphics[width=1.3in]{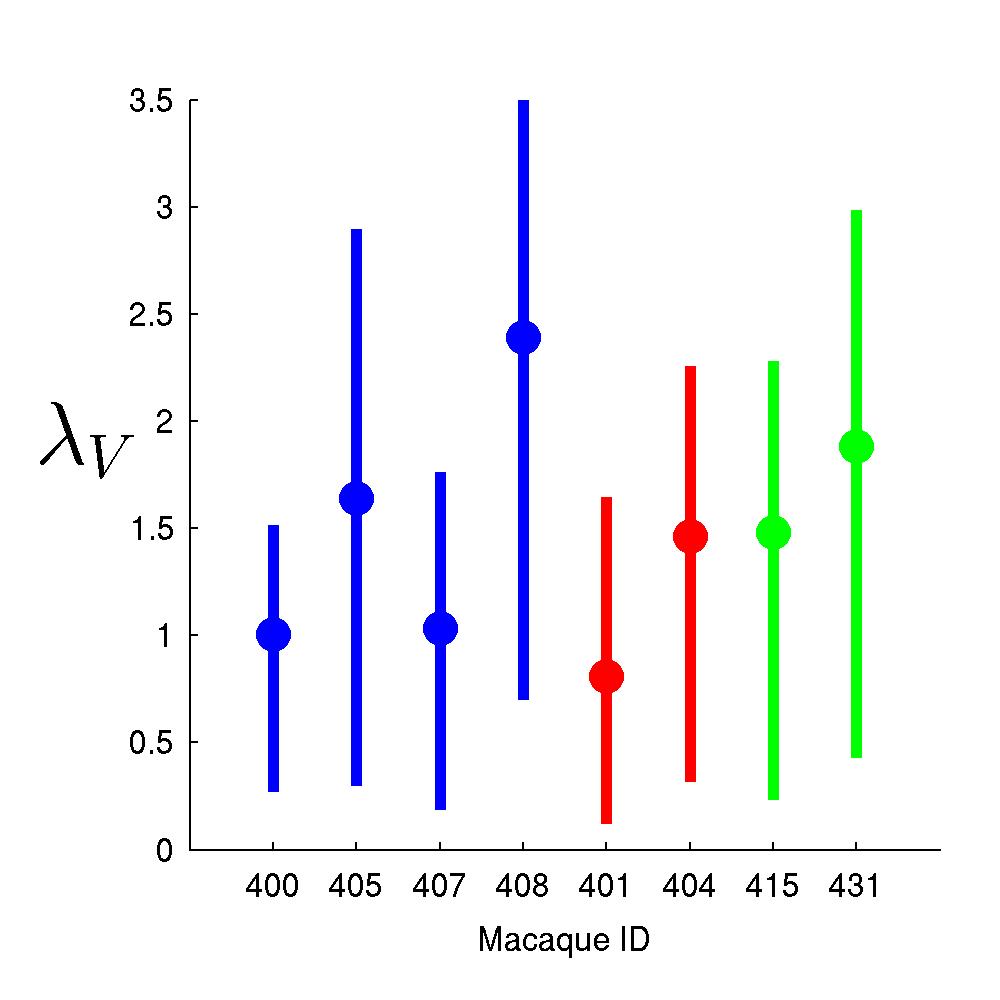}
\includegraphics[width=1.3in]{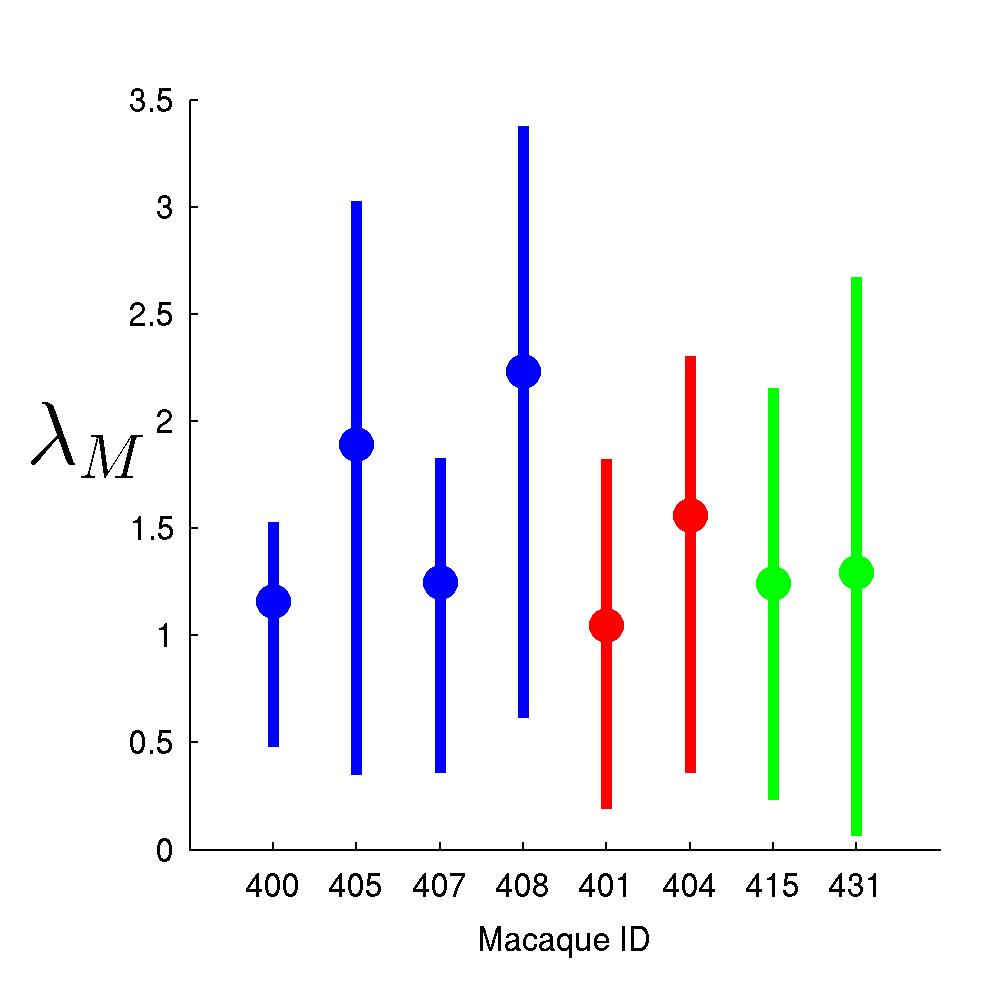}\\
\end{tabular}
\caption{\label{fig:dimensionless_interval} Marginal posterior means and 90\% credible intervals for dimensionless model parameters
on all 8 macaques. Color reflects vaccine immunization route:  intra-nasal/intra-tracheal (blue),
 intra-rectal (red),  intra-vagina (green).}
\end{figure}

The dimensionless form of \eqn{eq:odemodel_dimensionless} is
a \lq\lq canonical skeleton'' of the non-linear model. We can  compare across experiments in terms of
{\em (i)} the   dimensionless  parameters
$\eta  = \alpha\beta /(K_V K_M)$ and $\psi = \rho K_V/\beta$,
and {\em (ii)} the transformed time scale  $ s  = \beta t.$
Also, from Section~\ref{sec:delays},  the implied transformed time delays are
$\lambda_V  =  \beta \tau_V$ and $ \lambda_M  =  \beta \tau_M.$
Figure~\ref{fig:summaries_abc_mcmc} already shows
major differences across macaques in the posteriors for $\beta$, indicating substantial differences in
effective time scales of immune responses across subjects.  For the other dimensionless parameters,
we simply transform the MCMC+ABC posterior samples to approximate
posteriors for $(\eta,\psi,\lambda_V,\lambda_M)$ to
study what has been learned about these  \lq\lq canonical" parameters and how they compare across subjects.
Figures~\ref{fig:dimensionless_marginal},\ref{fig:dimensionless_joint},\ref{fig:dimensionless_interval} show  marginal and bivariate posteriors for macaque \#400, and credible interval summaries for these 4 quantities
on all 8 macaques; the latter shows up evident differences
across macaques, especially for $\eta,\psi.$

\begin{figure}[ht!]
  \centering
 \includegraphics[width=2.0in]{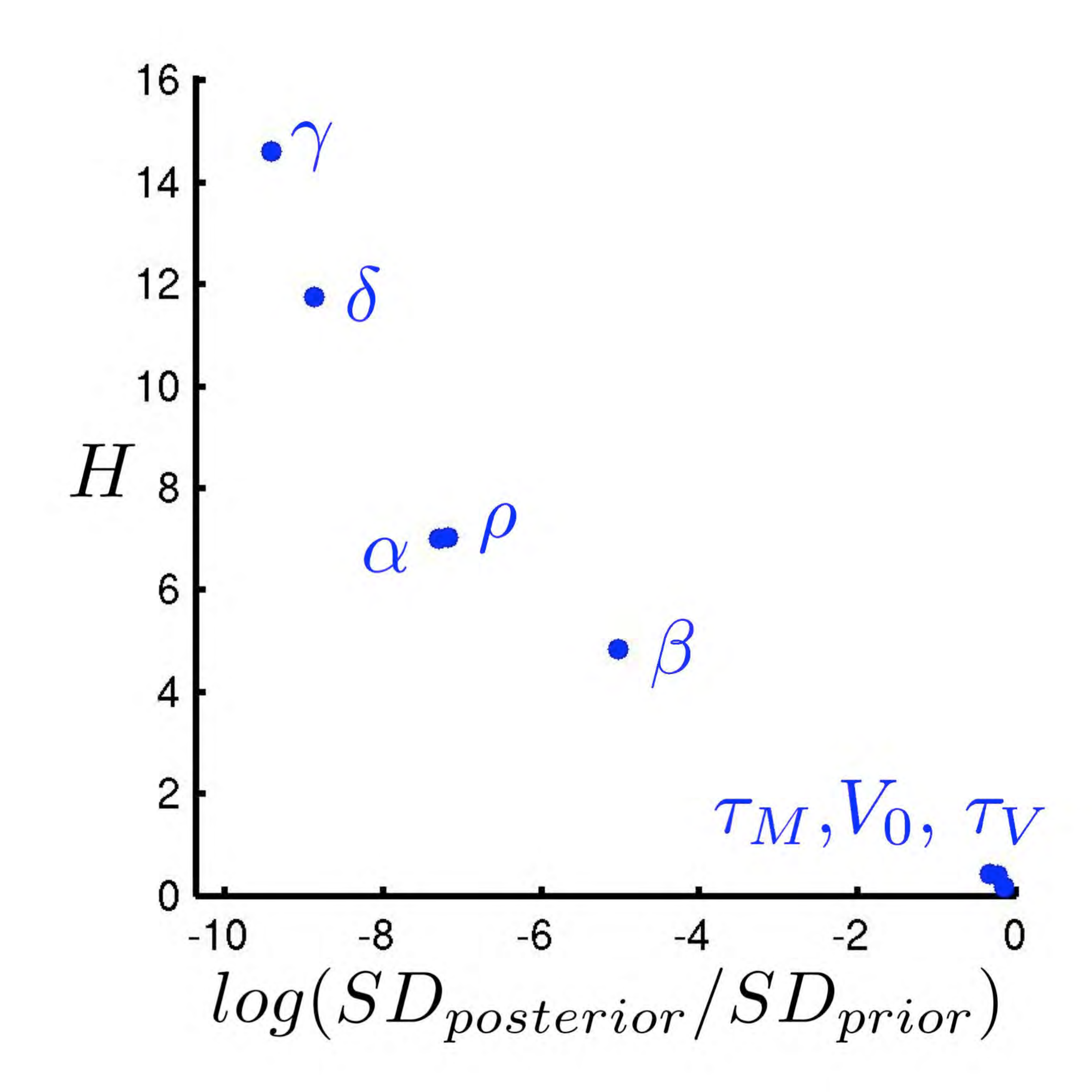}
  \includegraphics[width=2.0in]{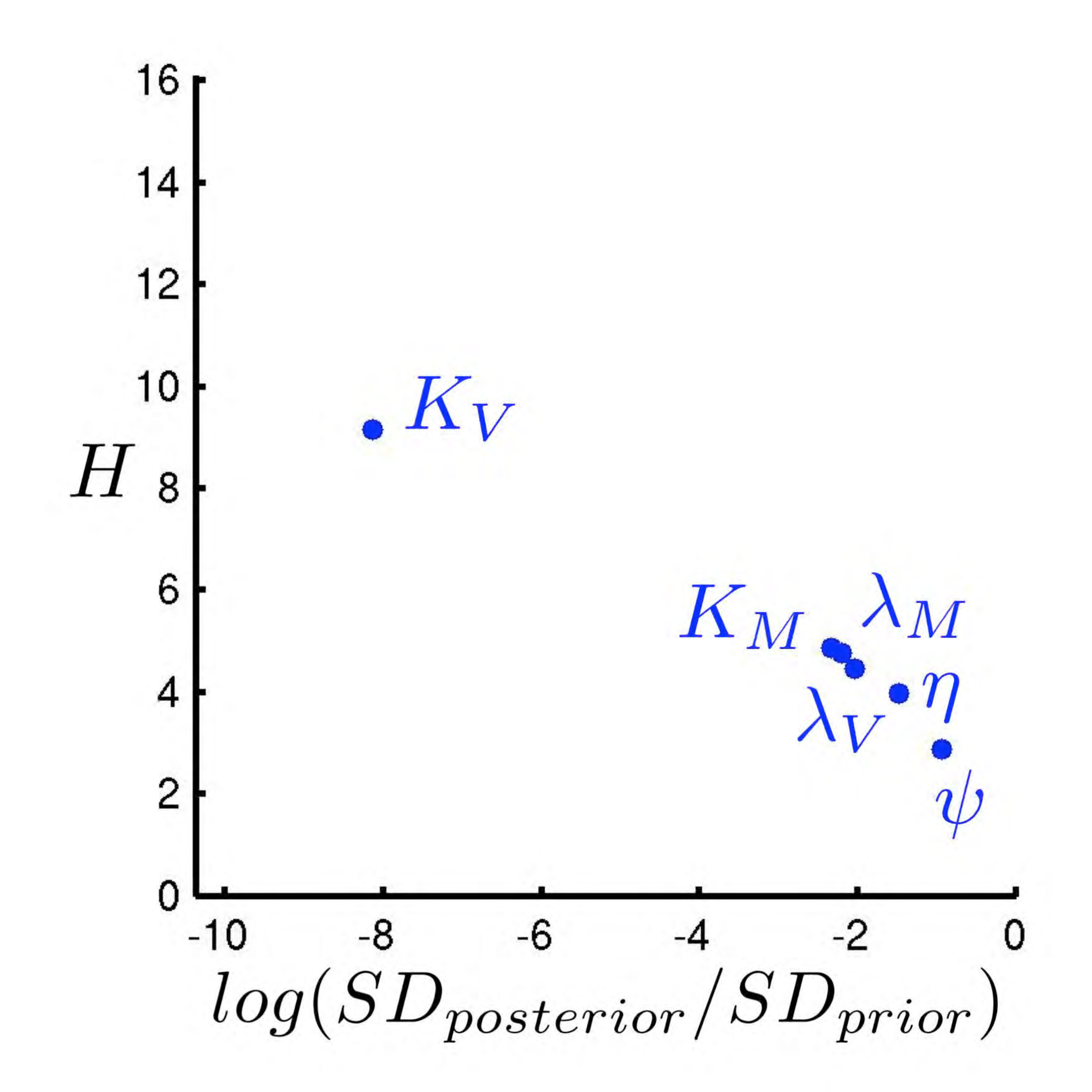} \\
\caption{\label{fig:entroplot_id400} Macaque 400: Learnability summaries.
Posterior:prior relative entropy $H$ against log of posterior to prior standard deviation, in some
cases implicitly based on
parameters following inverse c.d.f. transform to uniform priors as relevant.}
\end{figure}

\subsection{Learnability of Model Parameters}

To quantify and analyze the extent of  {\em learning} about each parameter, we evaluate two indices
based on the marginal posterior samples for each. The first index is the entropy ($H$) of the posterior relative to the prior, computed based on a binned representation of the posterior and prior distributions. The second index is the log ratio of marginal  posterior to prior  standard deviations.  We transformed   parameters having non-uniform priors
(i.e., parameters $K_V, K_M , \eta, \psi, \lambda_V, \lambda_M)$ before computing these   indices;
the transformations were based on inverse prior cdfs so that the indices compare posteriors to {\em uniform}
priors.

 Figure~\ref{fig:entroplot_id400} shows an illustrative example of these ``learnability indices'' for macaque \#400;
those for other macaques are comparable and can be found in the  Supplementary material.
There is limited data-based information about the three parameters $(V_0, \tau_M, \tau_V),$
very much more significant information about parameters $(\gamma,\delta,K_V,\alpha,\rho),$ with the
others in-between.
There is similarly variation in the degrees of \lq\lq learnability" of the dimensionless parameters,
evident in Figure~\ref{fig:entroplot_id400}. While  $\eta,\psi$ appear less learnable than others
for macaque \#400,   their posterior SDs are smaller than prior SDs
by at least a factor of 2, and in the case of $\eta$ it is almost a factor of 5, so that data information on these
two parameters-- earlier noted as reflecting substantial differences across macaques-- is appreciable.

\section{Additional Comments}

The HIV/SIV immune vaccine study exemplifies emerging studies that aim to evaluate and compare  
efficacies of candidate vaccine designs in terms of    resulting dynamic immune
responses. As experimental techniques develop,  we will see larger such experiments--
in human subjects as well as NHPs-- with time series data  at a higher level of time
resolution.  The models developed here represent key aspects of progression of viral and memory cells,
with time delays and stochastic elements that are evidently relevant.  The constrained model form
allows us to identify information in even this very sparse data about some of the model parameters,
including differences across subjects in key rate $(\beta)$ and maximum level ($K_V,K_M)$ parameters
that relate to critical aspects of the response profile. We also capture meaningful levels of between-subject
variability that are likely to be even more relevant in future human studies.

Fitting non-linear dynamic models-- with inherently long series of latent states to be inferred, along with
multiple parameters-- based on sparse time series data is an outstanding research challenge in computational
statistics.  After detailed development and evaluation of a range of MCMC and sequential Monte Carlo techniques,
we have defined a direct coupling of analytic approximation-based MCMC with an ABC strategy that appears
quite effective in Bayesian calibration of our model. Biases inherent in the MCMC analysis based on analytic approximations
can be corrected by a second stage ABC processing of MCMC outputs, while cases with little or no bias remain
relatively unchanged.  Bayesian analysis naturally and gracefully handles the sparsity of data relative to a high-dimensional latent state
and model parameters:  posterior distributions appear similar to priors for parameters that are not learnable,  while
uncertainties in posteriors for other parameters and latent states capture and quantify the extent of information relevant to those
aspects.    Evidently, the overall strategy-- though developed for this applied study-- could be generalized in various ways
as well as used directly in other applications of non-linear dynamic models.

The HIV/SIV vaccination study exhibits this selective learning about model parameters. Marginal posteriors for some parameters
are highly concentrated relative to the priors, whereas for others the posterior is relatively similar to the prior. The inclusion of a time delay is a key model component, and all non-dimensional parameters are learnable. Interestingly, the results raise  suggestions that the dynamics of the immune response depends on the vaccination route. This is not obvious from inspection of the time series data alone, and suggests follow-on experiments.  With additional experimental data coming on stream,  we are now in a position to address future
analyses with perhaps more informative priors that build on the results here.  One other area for further development is
to consider hierarchical models that link parameters  of models in a given immunization route group via a second stage prior, and
even, perhaps, then add a further hierarchical component to link hyper-parameters across route groups. This would enable and
encourage-- when relevant-- information sharing that could be beneficial, especially in contexts like this where data on each subject
is sparse.  Such developments, while beyond the scope of this paper, are under study.

\section*{Acknowledgements}

We gratefully acknowledge the generous sharing of the NHP vaccine trial data by Marjorie Robert-Guroff and Katherine McKinnon at the Immune Biology of Retroviral Infection Section, Vaccine Branch, Center for Cancer Research, National Cancer Institute, and Jean Patterson at the Office of AIDS Research at NIH. We also thank Jacob Frelinger at the Fred Hutchinson Cancer Research Center for helpful discussions during development of the mathematical model.

%\bibliographystyle{chicago}
%\bibliography{../macaques}

\newpage

\begin{thebibliography}{}

\bibitem[\protect\citeauthoryear{Beaumont, Zhang, and Balding}{Beaumont
  et~al.}{2002}]{Beaumont2002}
Beaumont, M., W.~Zhang, and D.~Balding (2002).
\newblock Approximate {B}ayesian computation in population genetics.
\newblock {\em Genetics\/}~{\em 162\/}(4), 2025.

\bibitem[\protect\citeauthoryear{Bonassi, You, and West}{Bonassi
  et~al.}{2011}]{Bonassi2011}
Bonassi, F.~V., L.~You, and M.~West (2011).
\newblock Bayesian learning from marginal data in bionetwork models.
\newblock {\em Statistical Applications in Genetics \& Molecular
  Biology\/}~{\em 10}, Art 49.

\bibitem[\protect\citeauthoryear{Chan, Feng, Ottinger, Foster, West, and
  Kepler}{Chan et~al.}{2008}]{Chan2008}
Chan, C., F.~Feng, J.~Ottinger, D.~Foster, M.~West, and T.~B. Kepler (2008).
\newblock Statistical mixture modelling for cell subtype identification in flow
  cytometry.
\newblock {\em Cytometry, A\/}~{\em 73\/}(693-701), 693--701.

\bibitem[\protect\citeauthoryear{Chan, Lin, Frelinger, Hebert, Gagnon, Landry,
  Sékaly, Enzor, Staats, Weinhold, Jaimes, and West}{Chan
  et~al.}{2010}]{Chan2010}
Chan, C., L.~Lin, J.~Frelinger, V.~Hebert, D.~Gagnon, C.~Landry, R.~P. Sékaly,
  J.~Enzor, J.~Staats, K.~J. Weinhold, M.~Jaimes, and M.~West (2010).
\newblock Optimization of a highly standardized carboxyfluorescein succinimidyl
  ester flow cytometry panel and gating strategy design with discriminative
  information measure evaluation.
\newblock {\em Cytometry A\/}~{\em 77}, 1126--1136.
 

\bibitem[\protect\citeauthoryear{De~Boer}{De~Boer}{2007}]{pmid17202215}
De~Boer, R.~J. (2007, Mar).
\newblock {{U}nderstanding the failure of {C}{D}8+ {T}-cell vaccination against
  simian/human immunodeficiency virus}.
\newblock {\em J. Virol.\/}~{\em 81\/}(6), 2838--2848.

\bibitem[\protect\citeauthoryear{De~Boer, Homann, and Perelson}{De~Boer
  et~al.}{2003}]{DeBoer2003}
De~Boer, R.~J., D.~Homann, and A.~S. Perelson (2003).
\newblock Different dynamics of {CD4+} and {CD8+} {T}  cell responses during and after
  acute lymphocytic choriomeningitis virus infection.
\newblock {\em The Journal of Immunology\/}~{\em 171\/}(8), 3928--3935.

\bibitem[\protect\citeauthoryear{Drovandi and Pettitt}{Drovandi and
  Pettitt}{2011}]{Drovandi2011}
Drovandi, C.~C. and A.~N. Pettitt (2011).
\newblock Estimation of parameters for macroparasite population evolution using
  approximate bayesian computation.
\newblock {\em Biometrics\/}~{\em 67\/}(1), 225--233.

\bibitem[\protect\citeauthoryear{Elemans, al~Basatena, Klatt, Gkekas,
  Silvestri, and Asquith}{Elemans et~al.}{2011}]{elemans2011don}
Elemans, M., N.-K.~S. al~Basatena, N.~R. Klatt, C.~Gkekas, G.~Silvestri, and
  B.~Asquith (2011).
\newblock Why don't cd8+ t cells reduce the lifespan of siv-infected cells in
  vivo?
\newblock {\em PLoS computational biology\/}~{\em 7\/}(9), e1002200.

\bibitem[\protect\citeauthoryear{Evans and Silvestri}{Evans and
  Silvestri}{2013}]{pmid23615116}
Evans, D.~T. and G.~Silvestri (2013, Jul).
\newblock {{N}onhuman primate models in {A}{I}{D}{S} research}.
\newblock {\em Curr Opin HIV AIDS\/}~{\em 8\/}(4), 255--261.

\bibitem[\protect\citeauthoryear{Golightly and Wilkinson}{Golightly and
  Wilkinson}{2011}]{Golightly2011}
Golightly, A. and D.~J. Wilkinson (2011).
\newblock Bayesian parameter inference for stochastic biochemical network
  models using particle {M}arkov chain {M}onte {C}arlo.
\newblock {\em Interface Focus\/}~{\em 1\/}(6), 807--820.

\bibitem[\protect\citeauthoryear{Johnston}{Johnston}{2000}]{johnston2000role}
Johnston, M.~I. (2000).
\newblock The role of nonhuman primate models in aids vaccine development.
\newblock {\em Molecular Medicine Today\/}~{\em 6\/}(7), 267--270.

\bibitem[\protect\citeauthoryear{Lin, Chan, Hadrup, Froesig, Wang, and
  West}{Lin et~al.}{2013}]{Lin2012}
Lin, L., C.~Chan, S.~R. Hadrup, T.~M. Froesig, Q.~Wang, and M.~West (2013).
\newblock Hierarchical {B}ayesian mixture modelling for antigen-specific {T}-cell
  subtyping in combinatorially encoded flow cytometry studies.
\newblock {\em Statistical Applications in Genetics and Molecular
  Biology\/}~{\em 12}, 309--331.

\bibitem[\protect\citeauthoryear{Liu and West}{Liu and West}{2001}]{Liu2001}
Liu, J. and M.~West (2001).
\newblock Combined parameter and state estimation in simulation-based
  filtering.
\newblock In A.~Doucet, J.~D. Freitas, and N.~Gordon (Eds.), {\em Sequential
  Monte Carlo Methods in Practice}, pp.\  197--217. New York: Springer-Verlag.

\bibitem[\protect\citeauthoryear{Morgan, Marthas, Miller, Duerr, Cheng-Mayer,
  Desrosiers, Flores, Haigwood, Hu, Johnson, et~al.}{Morgan
  et~al.}{2008}]{morgan2008use}
Morgan, C., M.~Marthas, C.~Miller, A.~Duerr, C.~Cheng-Mayer, R.~Desrosiers,
  J.~Flores, N.~Haigwood, S.-L. Hu, R.~P. Johnson, et~al. (2008).
\newblock The use of nonhuman primate models in hiv vaccine development.
\newblock {\em PLoS medicine\/}~{\em 5\/}(8), e173.

\bibitem[\protect\citeauthoryear{Niemi and West}{Niemi and
  West}{2010}]{Niemi2010}
Niemi, J.~B. and M.~West (2010).
\newblock Adaptive mixture modelling metropolis methods for {B}ayesian analysis
  of non-linear state-space models.
\newblock {\em Journal of Computational and Graphical Statistics\/}~{\em 19},
  260--280.
\newblock PMC2887612.

\bibitem[\protect\citeauthoryear{Patterson, Kuate, Daltabuit-Test, Li, Xiao,
  McKinnon, DiPasquale, Cristillo, Venzon, Haase, and Robert-Guroff}{Patterson
  et~al.}{2012}]{pmid22441384}
Patterson, L.~J., S.~Kuate, M.~Daltabuit-Test, Q.~Li, P.~Xiao, K.~McKinnon,
  J.~DiPasquale, A.~Cristillo, D.~Venzon, A.~Haase, and M.~Robert-Guroff (2012,
  May).
\newblock {{R}eplicating adenovirus-simian immunodeficiency virus ({S}{I}{V})
  vectors efficiently prime {S}{I}{V}-specific systemic and mucosal immune
  responses by targeting myeloid dendritic cells and persisting in rectal
  macrophages, regardless of immunization route}.
\newblock {\em Clin. Vaccine Immunol.\/}~{\em 19\/}(5), 629--637.

\bibitem[\protect\citeauthoryear{Patterson and Robert-Guroff}{Patterson and
  Robert-Guroff}{2008}]{pmid18694354}
Patterson, L.~J. and M.~Robert-Guroff (2008, Sep).
\newblock {{R}eplicating adenovirus vector prime/protein boost strategies for
  {H}{I}{V} vaccine development}.
\newblock {\em Expert Opin Biol Ther\/}~{\em 8\/}(9), 1347--1363.

\bibitem[\protect\citeauthoryear{Prado and West}{Prado and
  West}{2010}]{Prado2010}
Prado, R. and M.~West (2010).
\newblock {\em Time Series: Modelling, Computation \& Inference}.
\newblock Chapman \& Hall/CRC Press.

\bibitem[\protect\citeauthoryear{Pritchard, Seielstad, Perez-Lezaun, and
  Feldman}{Pritchard et~al.}{1999}]{Pritchard1999}
Pritchard, J., M.~Seielstad, A.~Perez-Lezaun, and M.~Feldman (1999).
\newblock Population growth of human {Y} chromosomes: a study of {Y} chromosome
  microsatellites.
\newblock {\em Molecular Biology and Evolution\/}~{\em 16\/}(12), 1791.

\bibitem[\protect\citeauthoryear{Scherer and McLean}{Scherer and
  McLean}{2002}]{scherer2002mathematical}
Scherer, A. and A.~McLean (2002).
\newblock Mathematical models of vaccination.
\newblock {\em British Medical Bulletin\/}~{\em 62\/}(1), 187--199.

\bibitem[\protect\citeauthoryear{Toni, Welch, Strelkowa, Ipsen, and
  Stumpf}{Toni et~al.}{2009}]{toni2009}
Toni, T., D.~Welch, N.~Strelkowa, A.~Ipsen, and M.~Stumpf (2009).
\newblock Approximate {B}ayesian computation scheme for parameter inference and
  model selection in dynamical systems.
\newblock {\em Journal of the Royal Society Interface\/}~{\em 6\/}(31),
  187--202.

\bibitem[\protect\citeauthoryear{West}{West}{1992}]{West1992b}
West, M. (1992).
\newblock Modelling with mixtures (with discussion).
\newblock In J.~M. Bernardo, J.~O. Berger, A.~P. Dawid, and A.~F.~M. Smith
  (Eds.), {\em Bayesian Statistics 4}, pp.\  503--524. Oxford University Press.

\bibitem[\protect\citeauthoryear{West}{West}{1993a}]{West1993b}
West, M. (1993a).
\newblock Approximating posterior distributions by mixtures.
\newblock {\em Journal of the Royal Statistical Society: Series B (Statistical
  Methology)\/}~{\em 54}, 553--568.

\bibitem[\protect\citeauthoryear{West}{West}{1993b}]{West1993a}
West, M. (1993b).
\newblock Mixture models, {M}onte {C}arlo, {B}ayesian updating and dynamic
  models.
\newblock {\em Computing Science and Statistics\/}~{\em 24}, 325--333.

\bibitem[\protect\citeauthoryear{West and Harrison}{West and
  Harrison}{1997}]{West1997}
West, M. and P.~J. Harrison (1997).
\newblock {\em Bayesian Forecasting \& Dynamic Models\/} (2nd ed.).
\newblock Springer Verlag.

\bibitem[\protect\citeauthoryear{Wilkinson}{Wilkinson}{2006}]{Wilk:stoc:2006}
Wilkinson, D. (2006).
\newblock {\em Stochastic Modelling for Systems Biology}.
\newblock London: Chapman \& Hall/CRC.

\bibitem[\protect\citeauthoryear{Wodarz}{Wodarz}{2008}]{pmid18586297}
Wodarz, D. (2008, Sep).
\newblock {{I}mmunity and protection by live attenuated {H}{I}{V}/{S}{I}{V}
  vaccines}.
\newblock {\em Virology\/}~{\em 378\/}(2), 299--305.

\end{thebibliography}

\newpage
  
\section*{Supplementary Material: Additional Figures} 

This supplement includes the additional graphical summaries of posteriors from the
analysis of all 8 macaques, together with the raw time series data. 

\vfill
\begin{figure}[htbp!]
  \centering
 \includegraphics[width=2.0in]{plot_data_400}
 \includegraphics[width=2.0in]{plot_data_401} \\
 \includegraphics[width=2.0in]{plot_data_404}
 \includegraphics[width=2.0in]{plot_data_405}
  \includegraphics[width=2.0in]{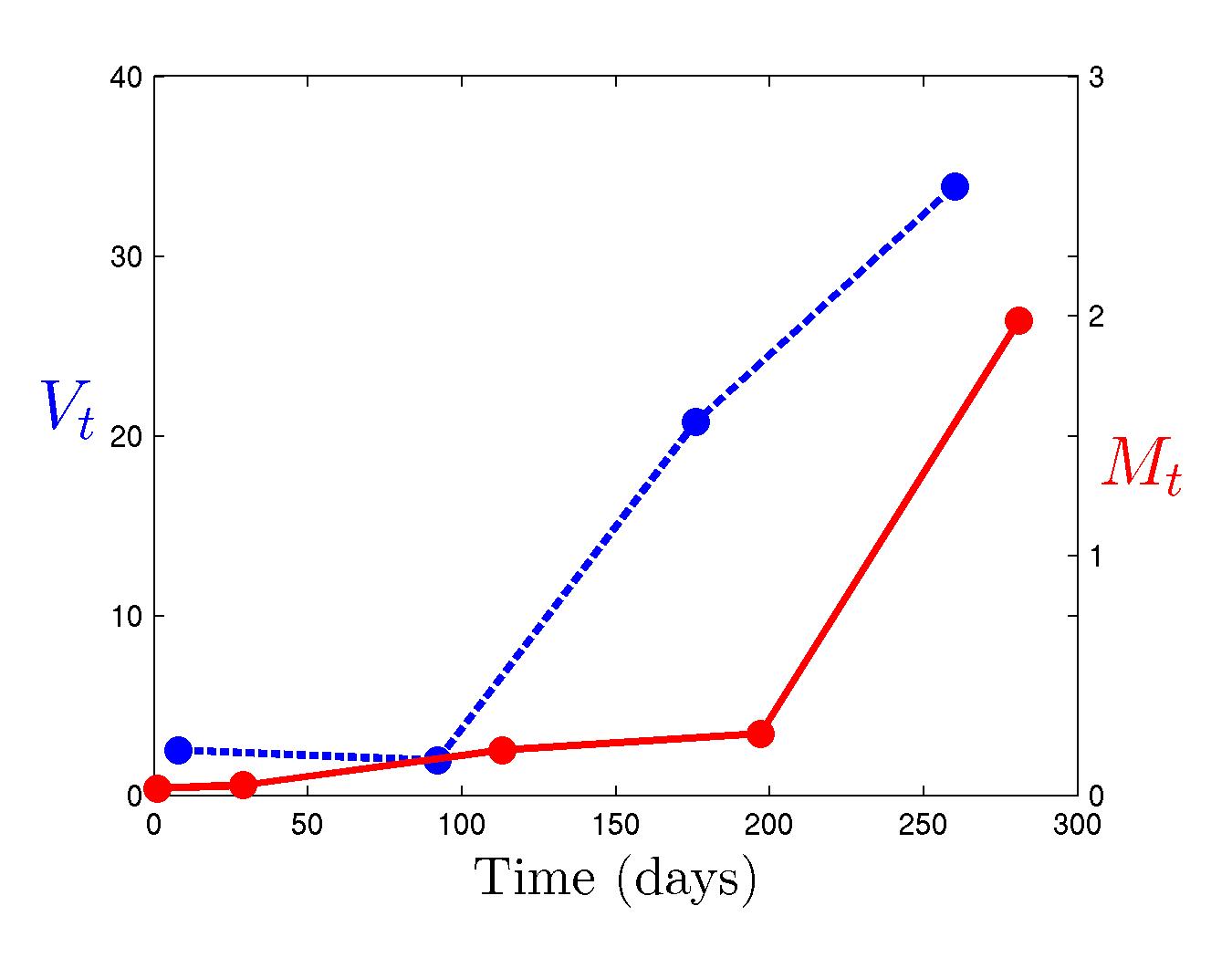}
 \includegraphics[width=2.0in]{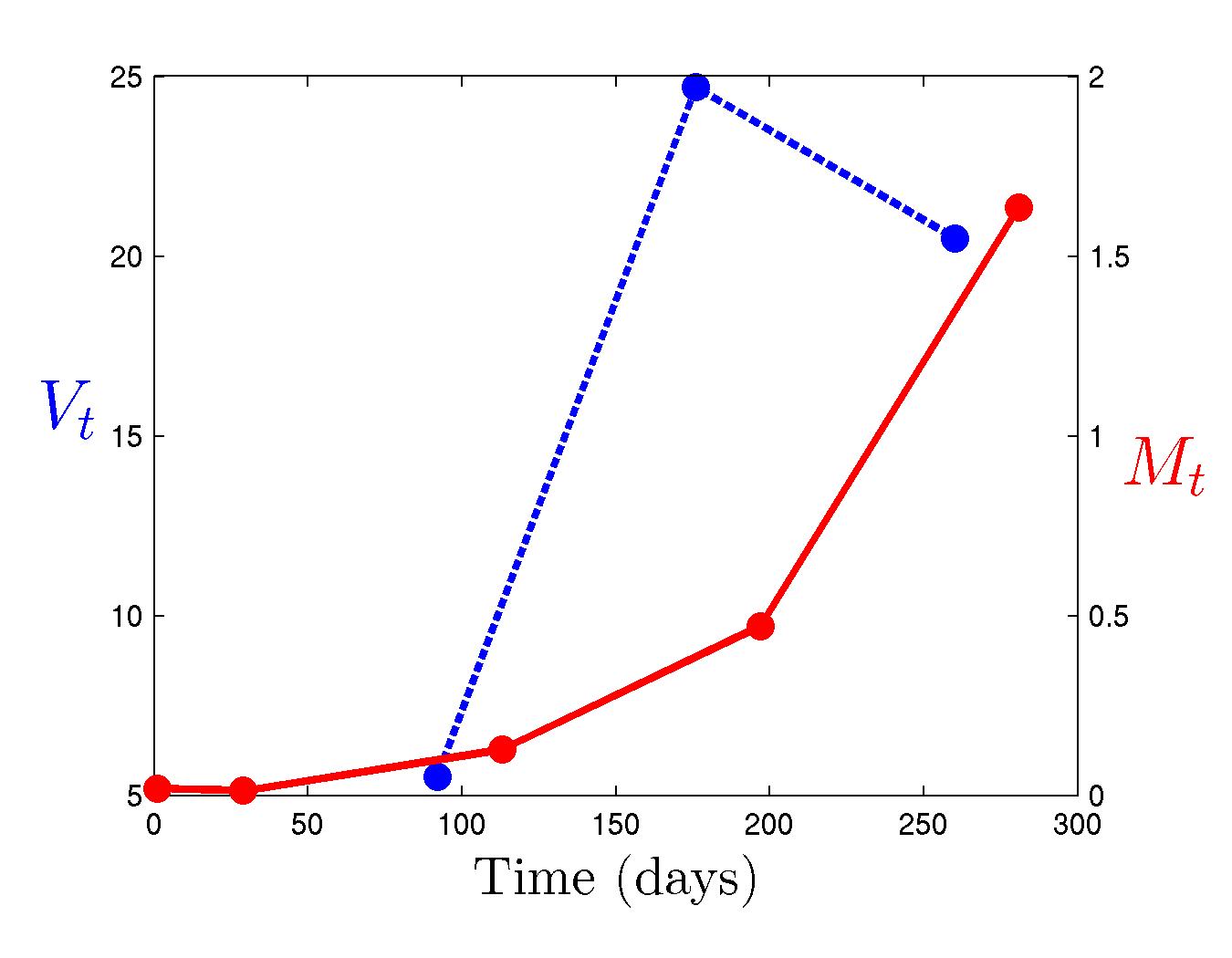} \\
 \includegraphics[width=2.0in]{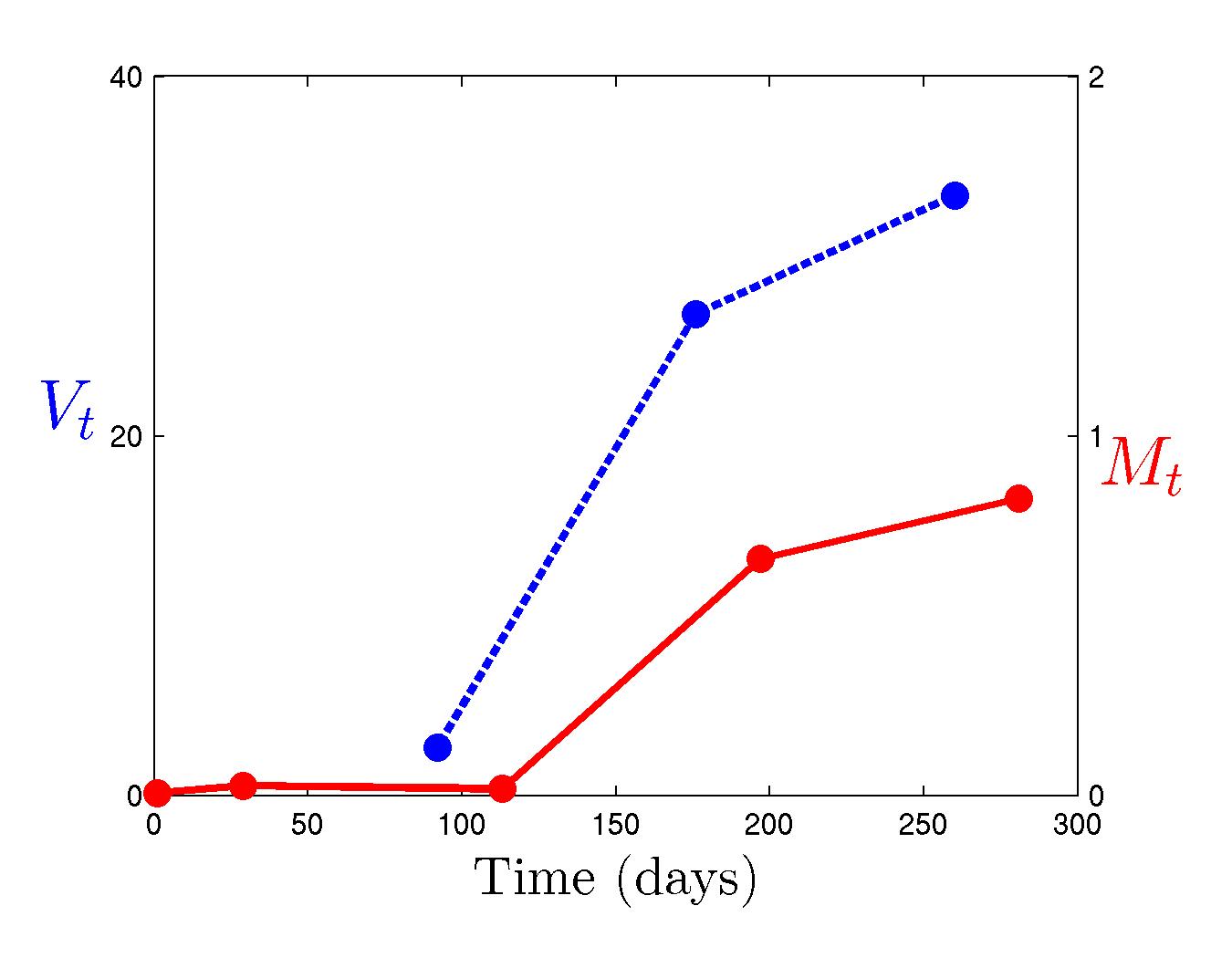}
 \includegraphics[width=2.0in]{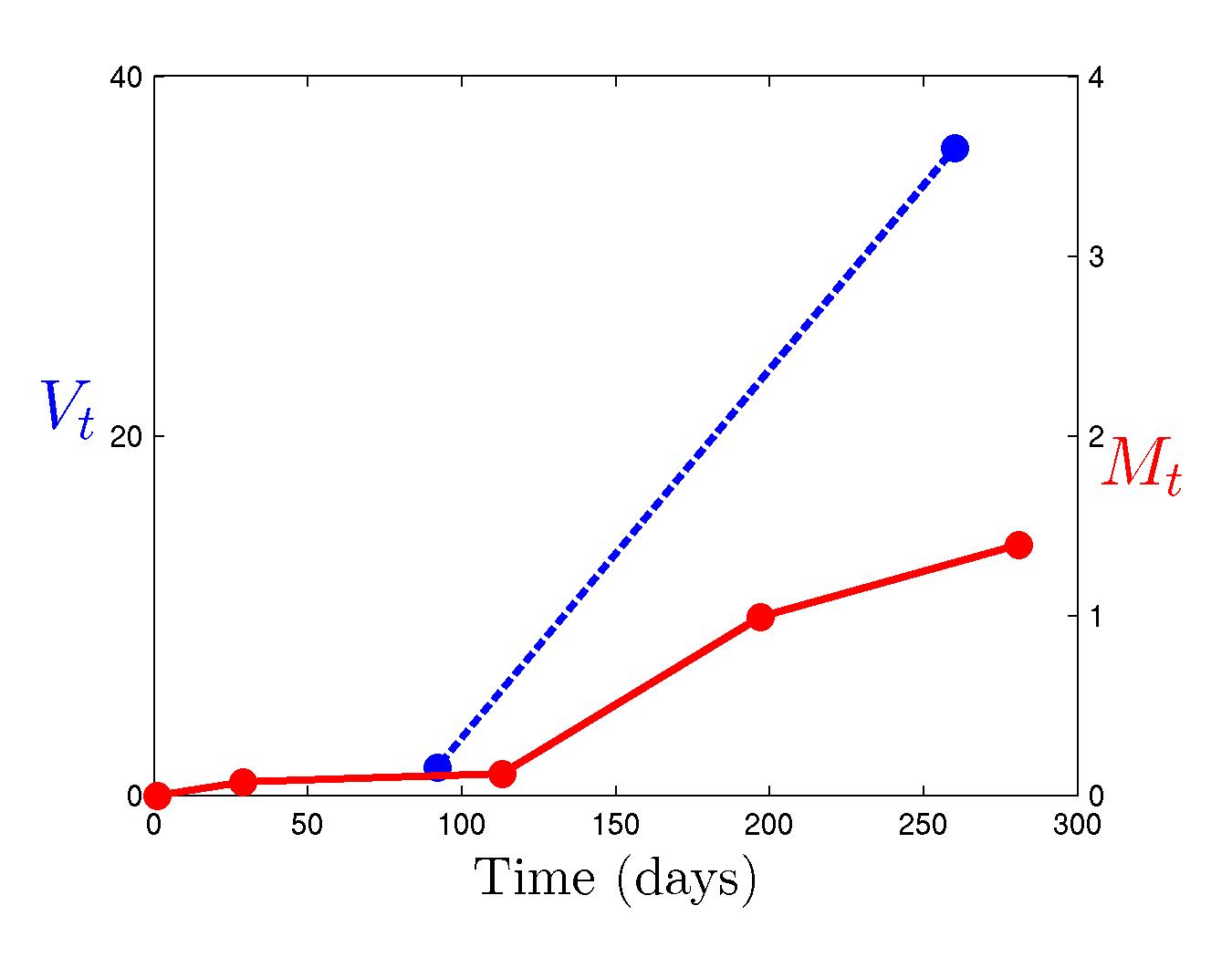}
\caption{Observed data  $V_t$ (blue) and $M_t$ (red) for all 8
macaques (subject IDs 400, 401, 404, 405, 407, 408, 415 and 431, respectively, from top left running left-to-right then down the rows).}
\end{figure}\vfill

\newpage

\begin{figure}[ht]
  \centering
 \includegraphics[width=2.30in]{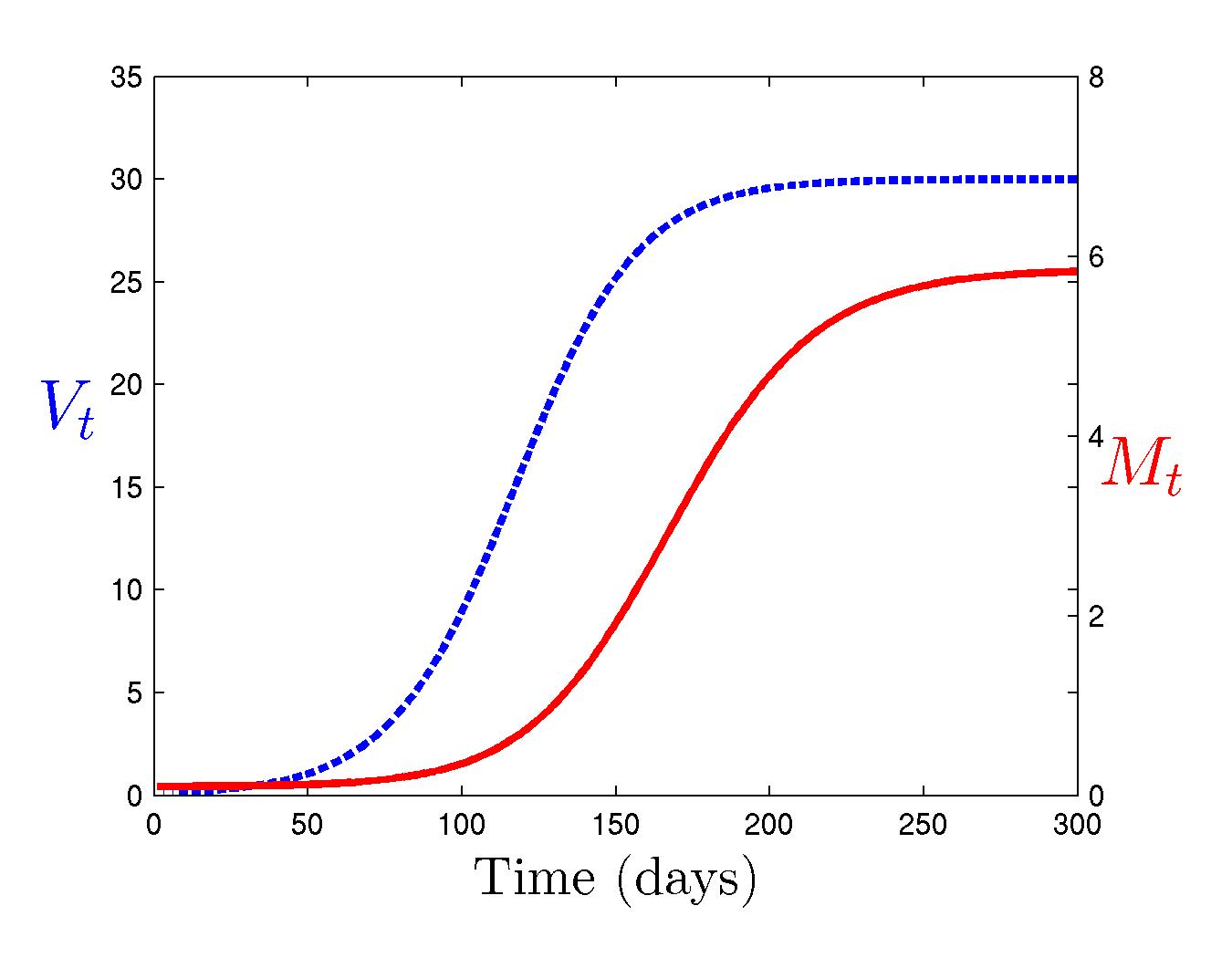}
 \includegraphics[width=2.30in]{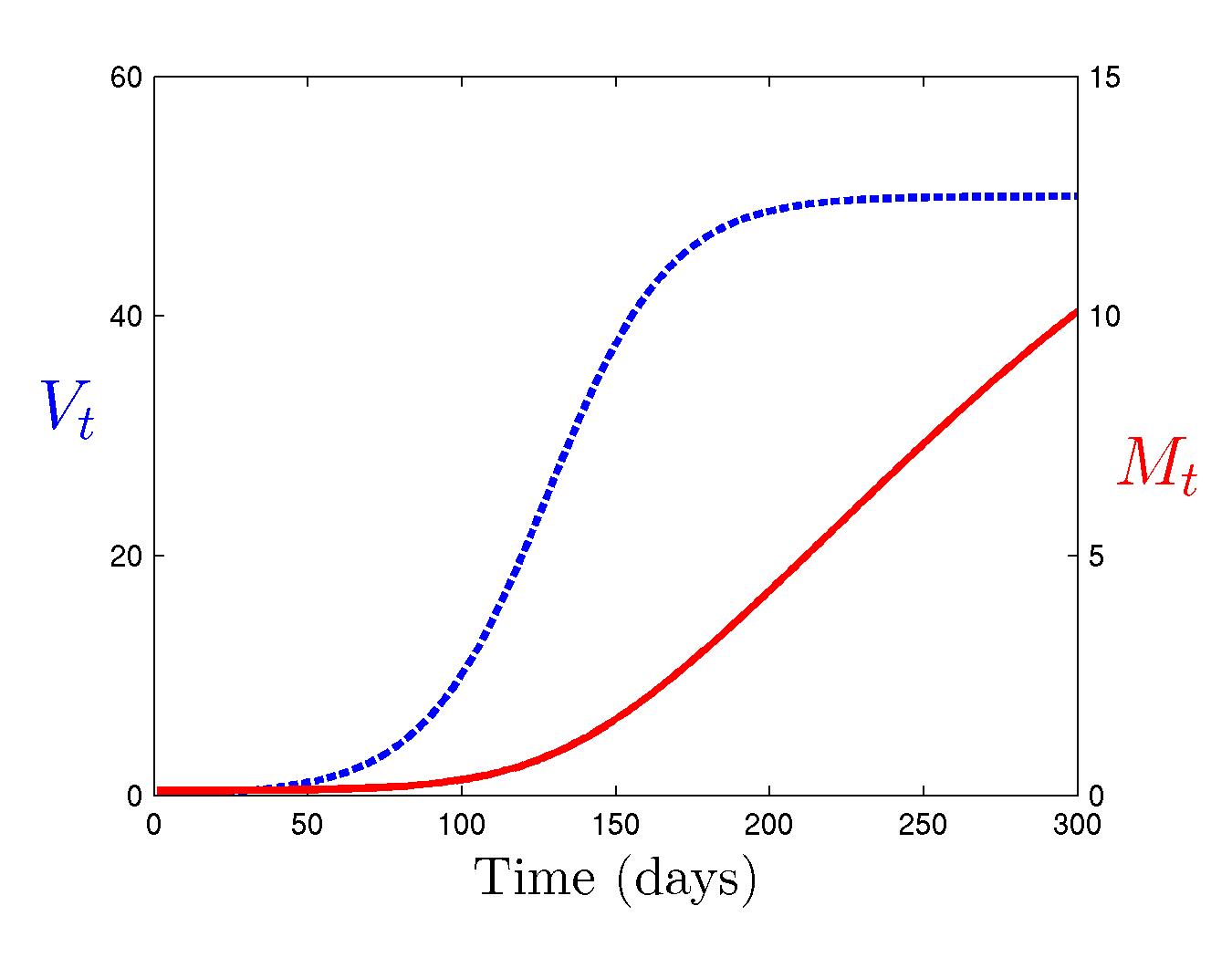} \\
 \includegraphics[width=2.30in]{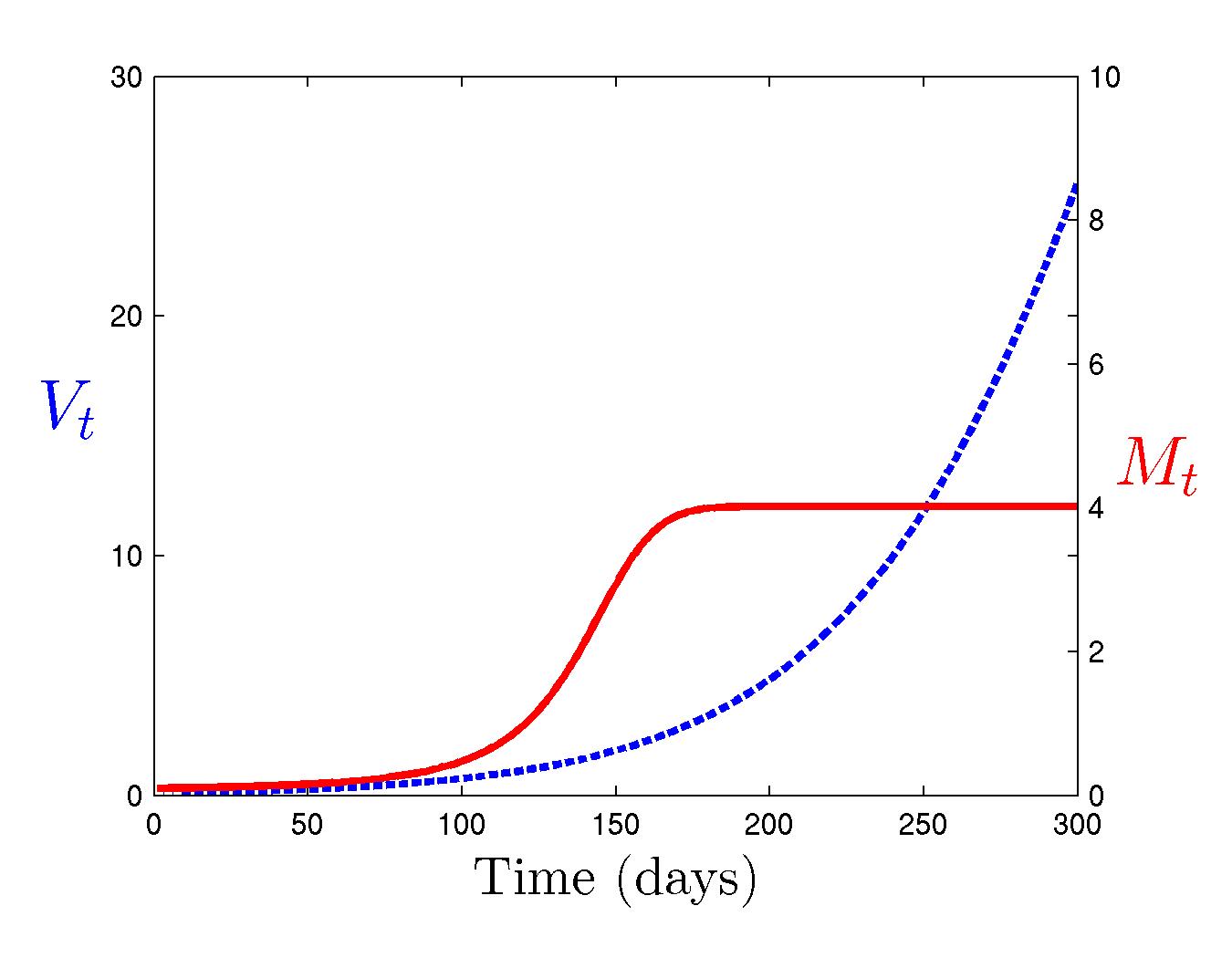}
 \includegraphics[width=2.30in]{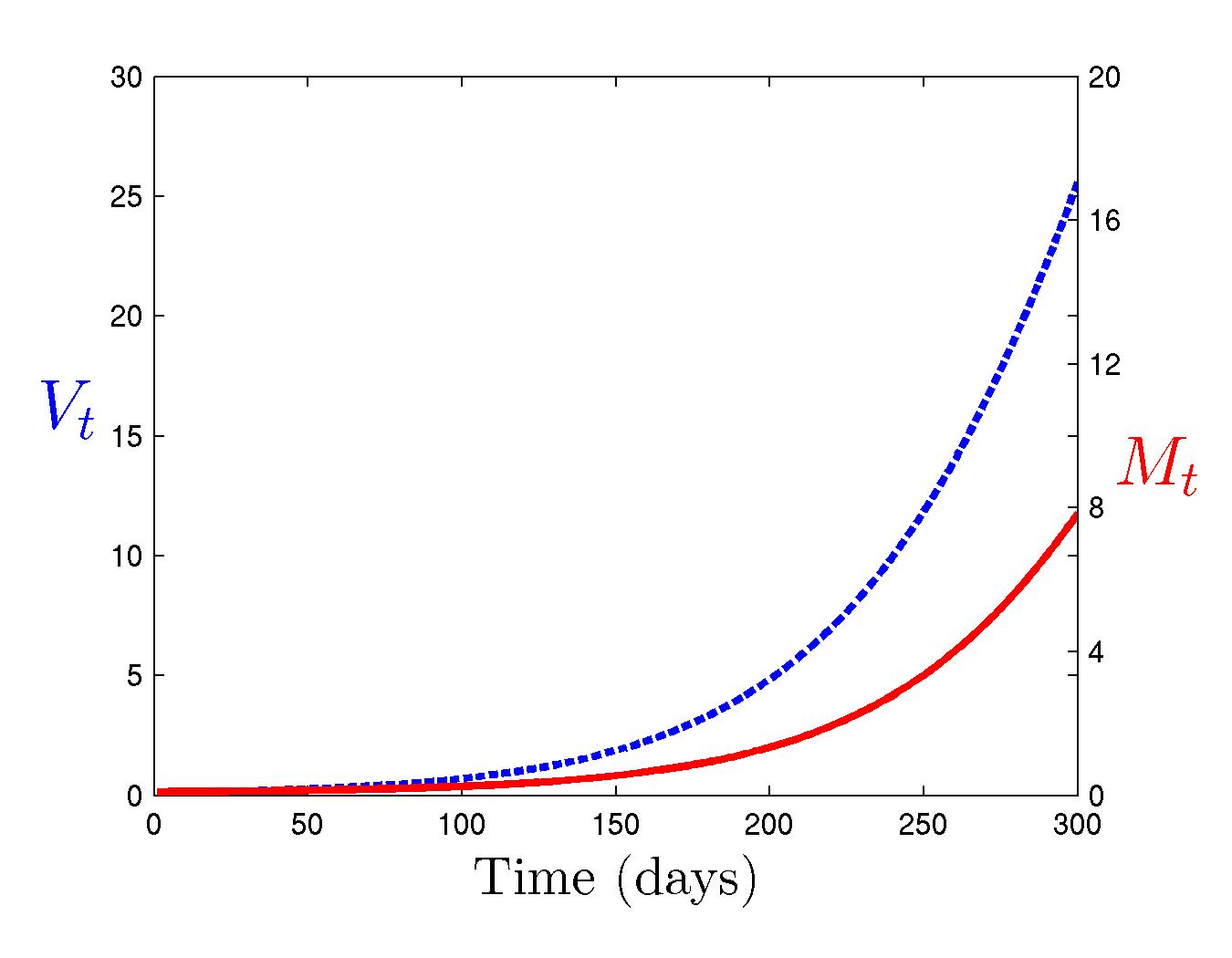}
\caption{\label{fig:Splots_ode_abc_mcmc} Synthetic trajectories of of $V_t$ (blue) and $M_t$ (red) from the initial ODE model based on forward integration, using $(V_0, M_0)=(0.5, 0.5)$ as initial values. The four outcomes of the model are based on the parameters$\{\beta, K_V, \alpha, \rho, K_M \}$ given by  (0.05, 30, 0.001, 0.001, 5), (0.05, 50, 0.001, 0.0001, 10), (0.02, 80, 0.001, 0.05, 4), and (0.02, 80, 0.005, 0.0001, 6).}
\end{figure}

\newpage

\begin{figure}[hb!]
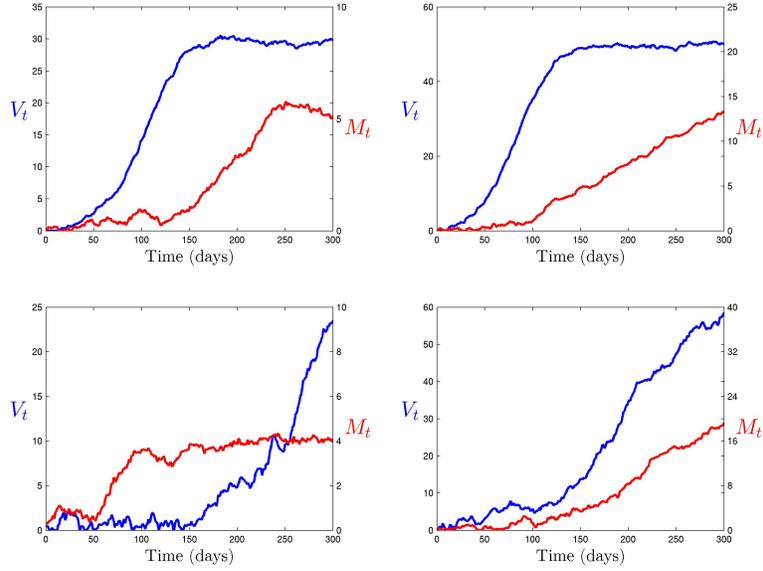

  \centering
 \includegraphics[width=2.0in]{plot_sde_1}
 \includegraphics[width=2.0in]{plot_sde_2} \\
 \includegraphics[width=2.0in]{plot_sde_3}
 \includegraphics[width=2.0in]{plot_sde_4}
\caption{\label{fig:Splots_sde_abc_mcmc} Synthetic trajectories of $V_t$ (blue) and $M_t$ (red) from 
the discrete-time, non-linear stochastic model with time delays. These start at 
 $(V_0, M_0)=(0.5, 0.5)$ and use 
model parameters chosen to be consistent with posteriors from the macaque study.
The four outcomes (left to right, top down) are based on the parameters $\{\beta, \delta, \alpha, \rho, \gamma, \tau_V, \tau_M, \kappa^2_V, \kappa^2_M\}$ given by  (0.05, 0.05/30, 0.001, 0.001, 0.001/5, 5, 30, 0.025, 0.005), (0.05, 0.05/50, 0.001, 0.0001, 0.0001/10, 20, 5, 0.05, 0.01), (0.02, 0.02/80, 0.001, 0.05, 0.05/4, 10, 30, 0.1, 0.01), and (0.02, 0.02/80, 0.005, 0.0001, 0.0001/6, 20, 20, 0.2, 0.05).}
\end{figure}

\newpage
\begin{figure}[ht!]
  \centering
\begin{tabular}{cccc}
\hskip-0.3in\includegraphics[width=1.3in]{sum_group_V0}
\includegraphics[width=1.3in]{sum_group_beta}
\includegraphics[width=1.3in]{sum_group_alpha}
\includegraphics[width=1.3in]{sum_group_rho}\\
\hskip-0.3in\includegraphics[width=1.3in]{sum_group_tauV}
\includegraphics[width=1.3in]{sum_group_tauM}
\includegraphics[width=1.3in]{sum_group_KV}
\includegraphics[width=1.3in]{sum_group_KM}\\
\hskip-0.3in\includegraphics[width=1.3in]{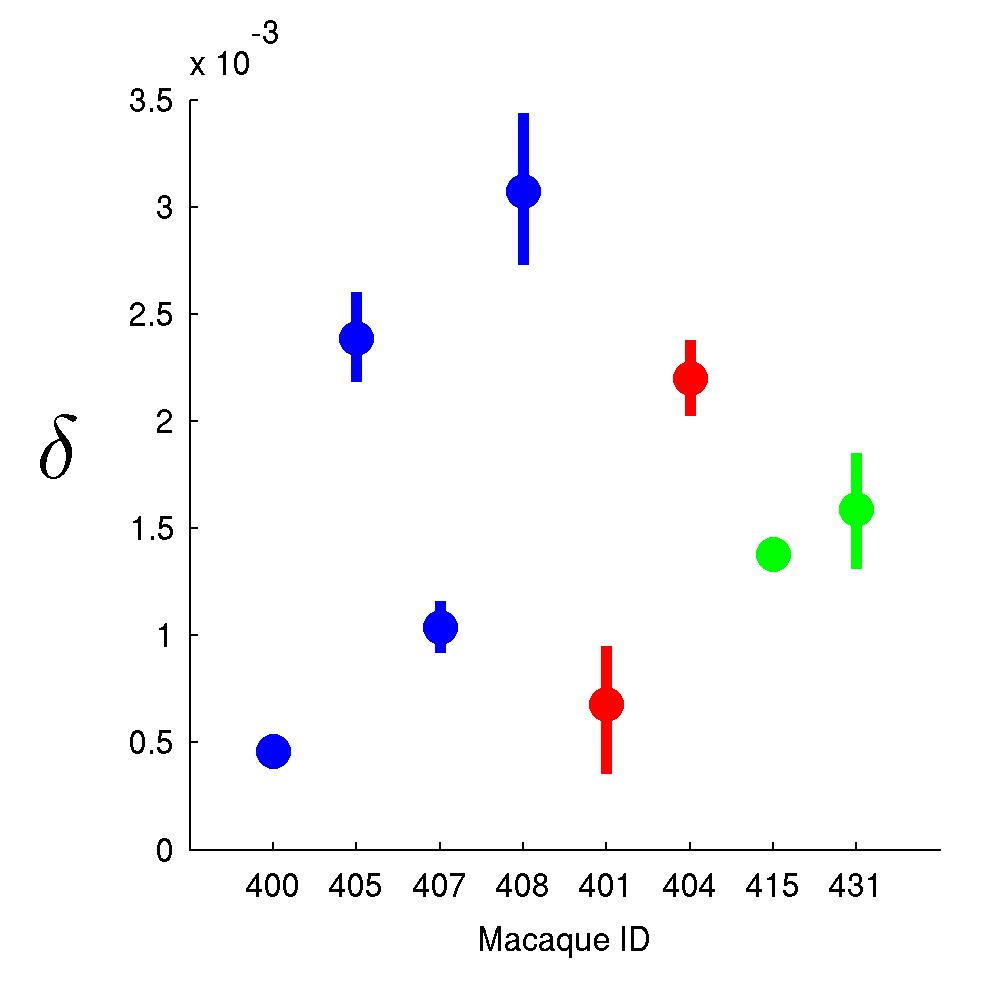}
\includegraphics[width=1.3in]{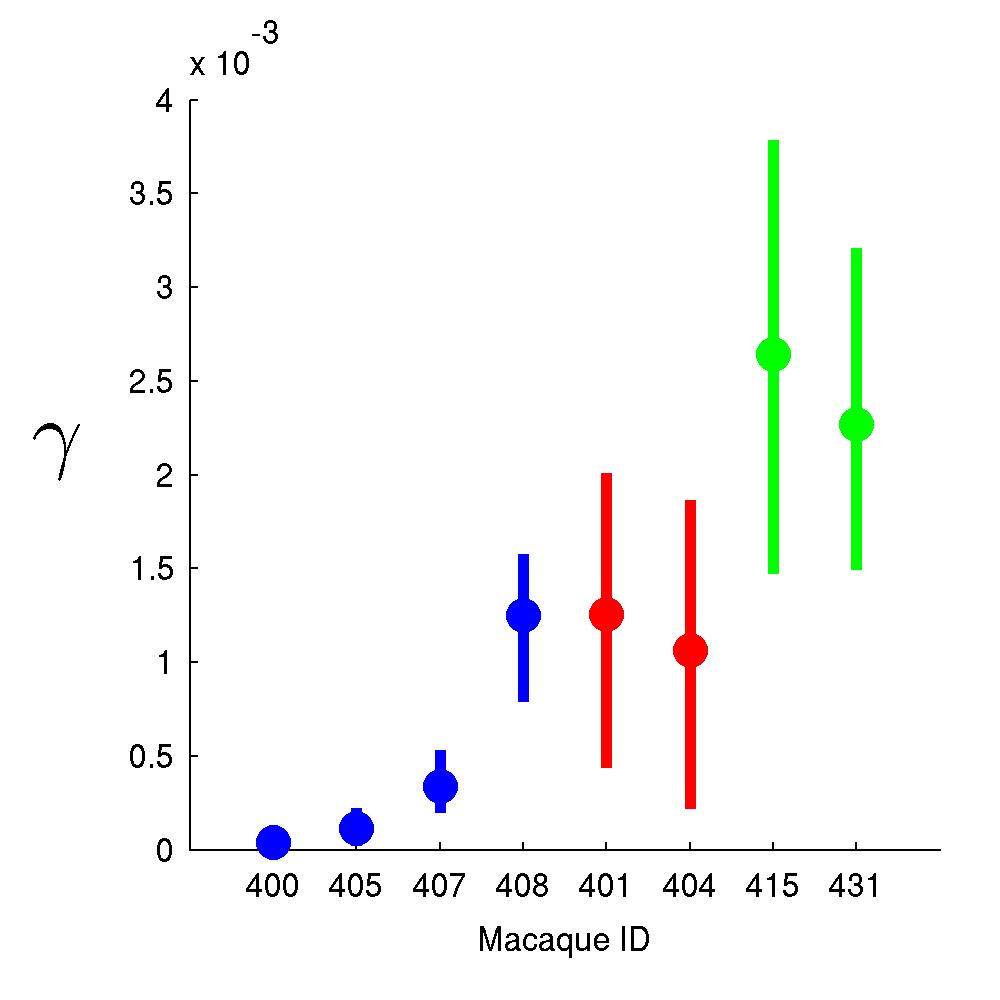}
\includegraphics[width=1.3in]{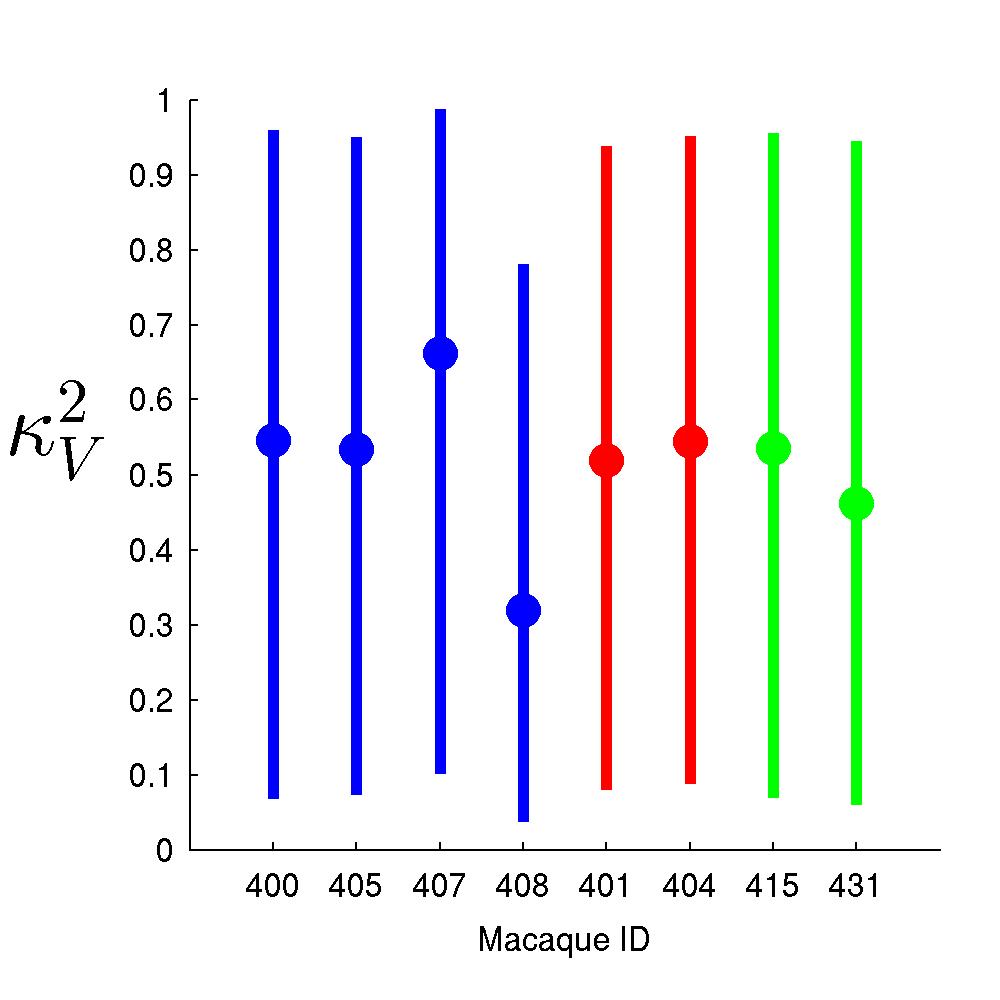}
\includegraphics[width=1.3in]{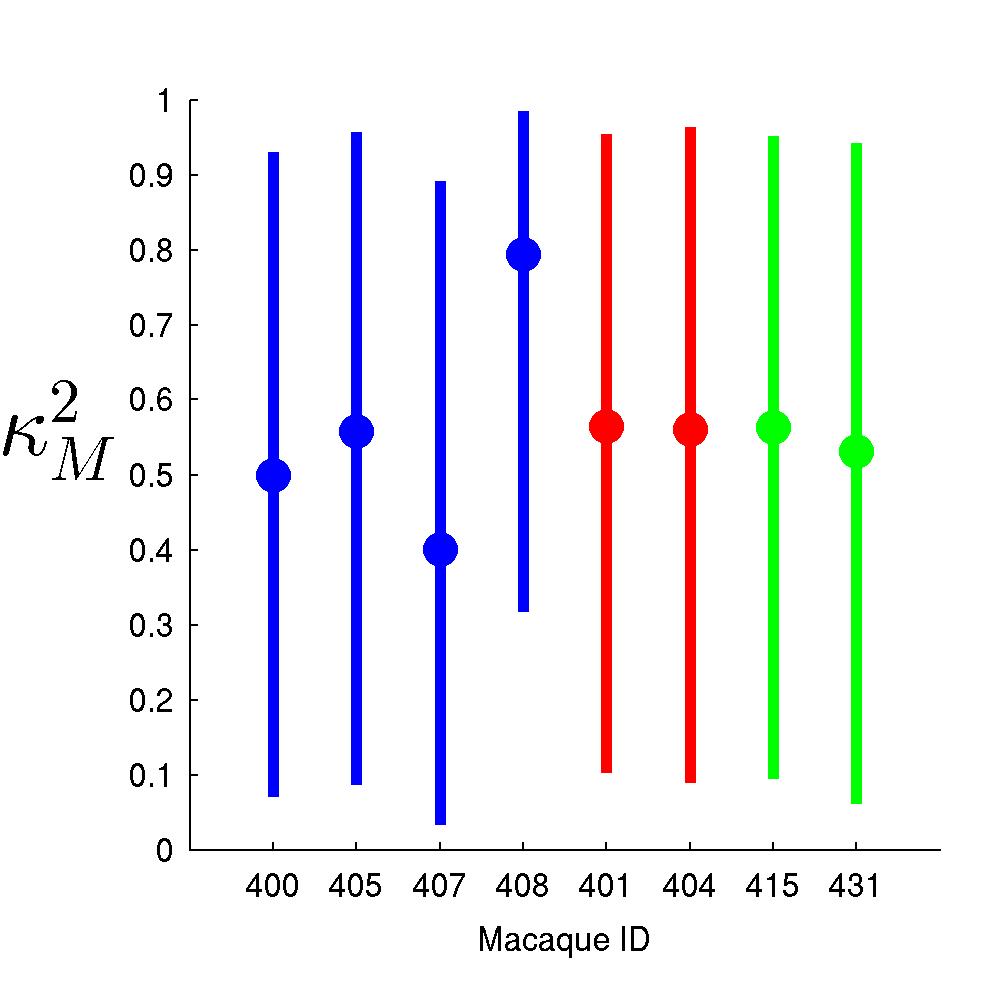}\\
\hskip-0.3in\includegraphics[width=1.3in]{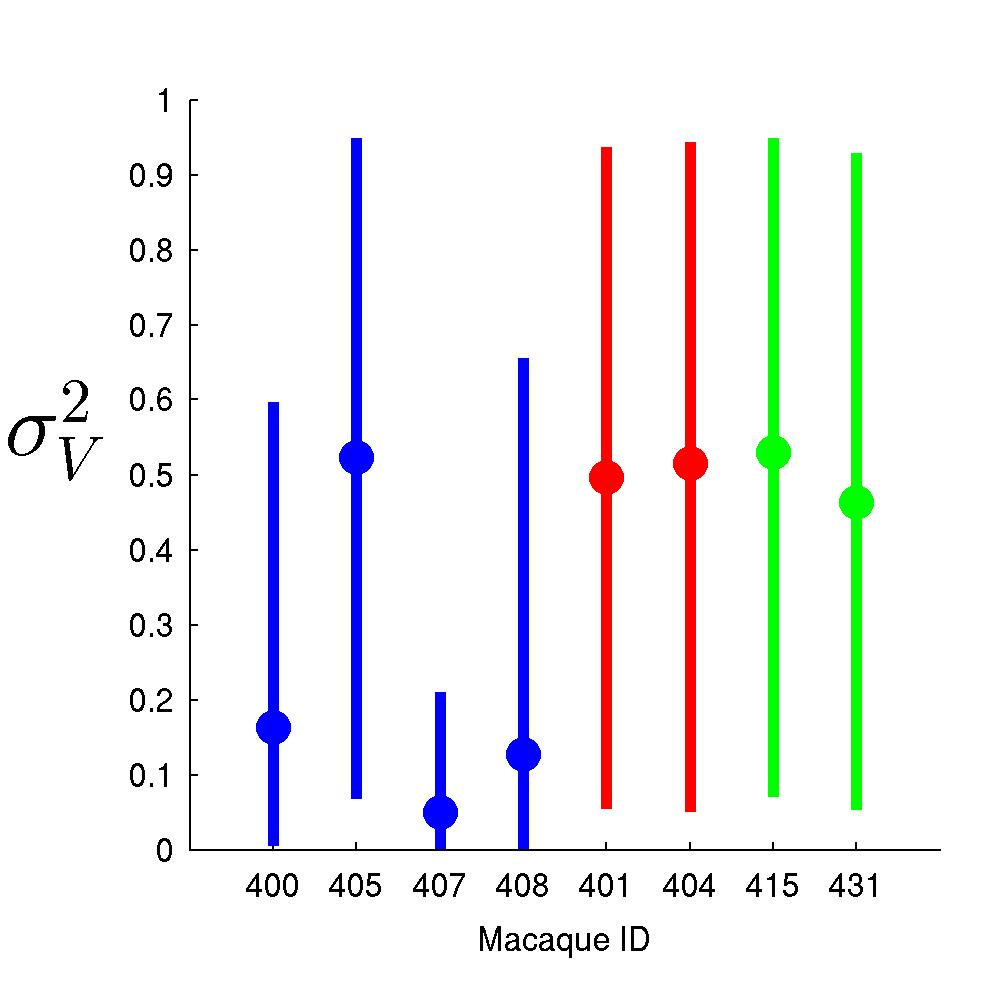}
\includegraphics[width=1.3in]{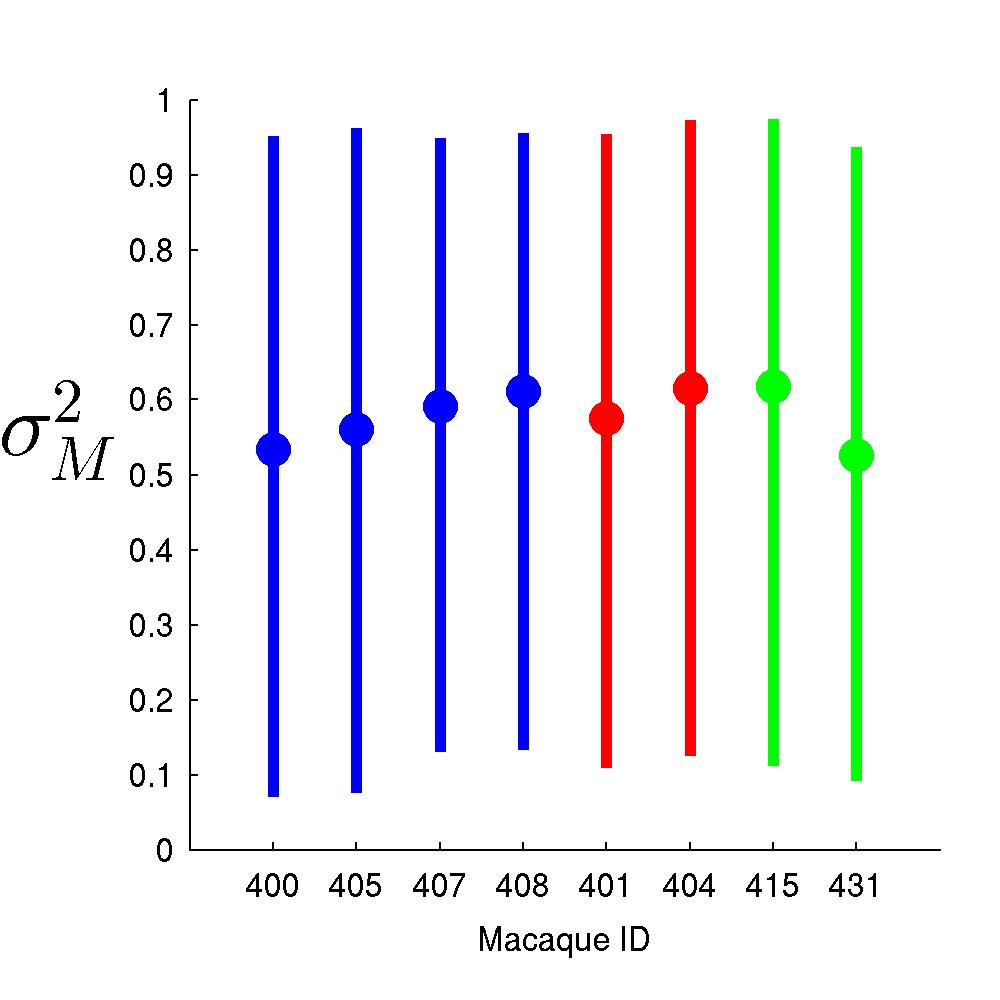}
\end{tabular}
\caption{\label{fig:Ssummaries_abc_mcmc} Marginal posterior means and 90\% credible intervals for the full set of model parameters
on all 8 macaques. Color reflects vaccine immunization route:  intra-nasal/intra-tracheal (blue),
 intra-rectal (red),  intra-vagina (green).}
\end{figure}

\newpage

\begin{figure}[ht!]
  \centering
\includegraphics[width=2.2in]{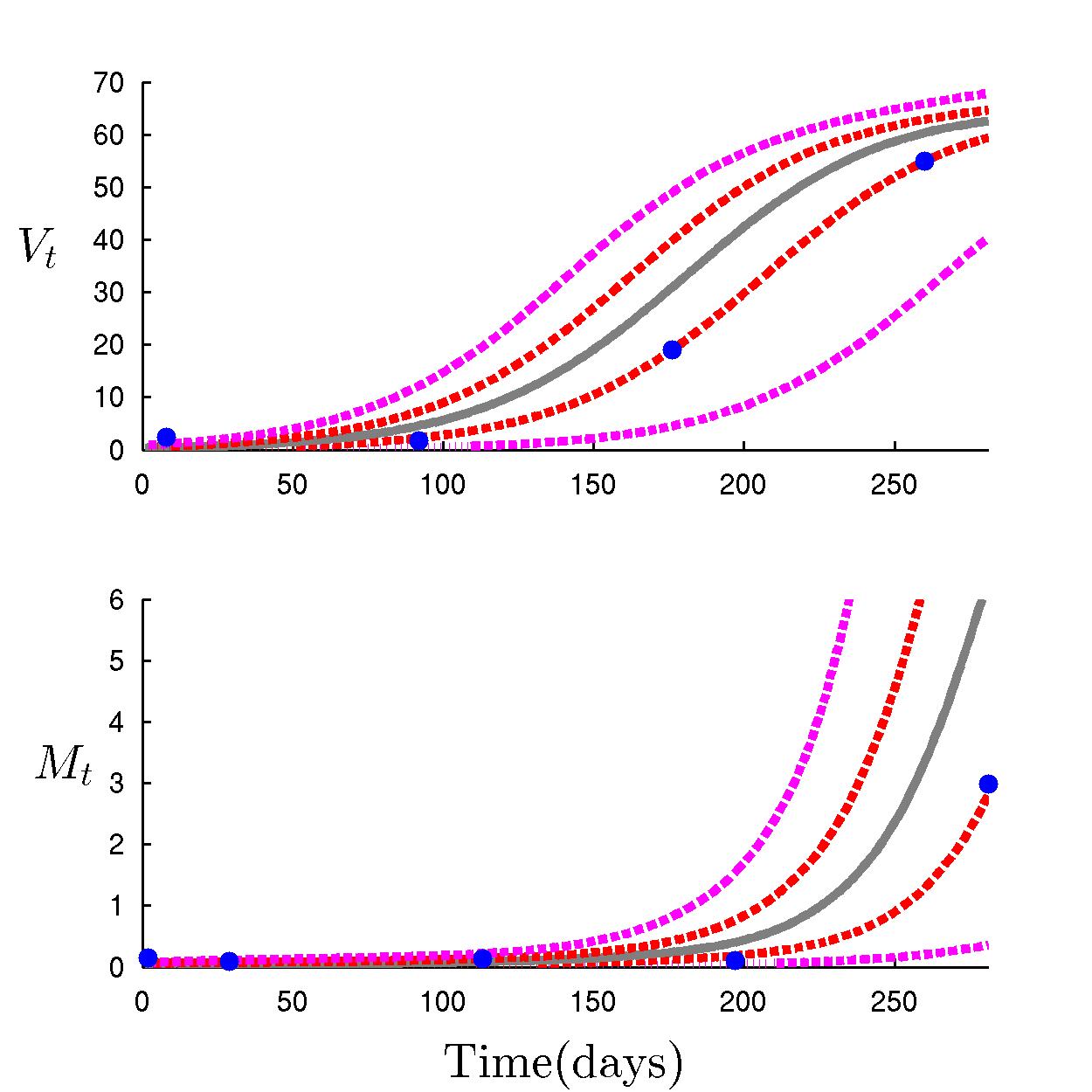}
\includegraphics[width=2.2in]{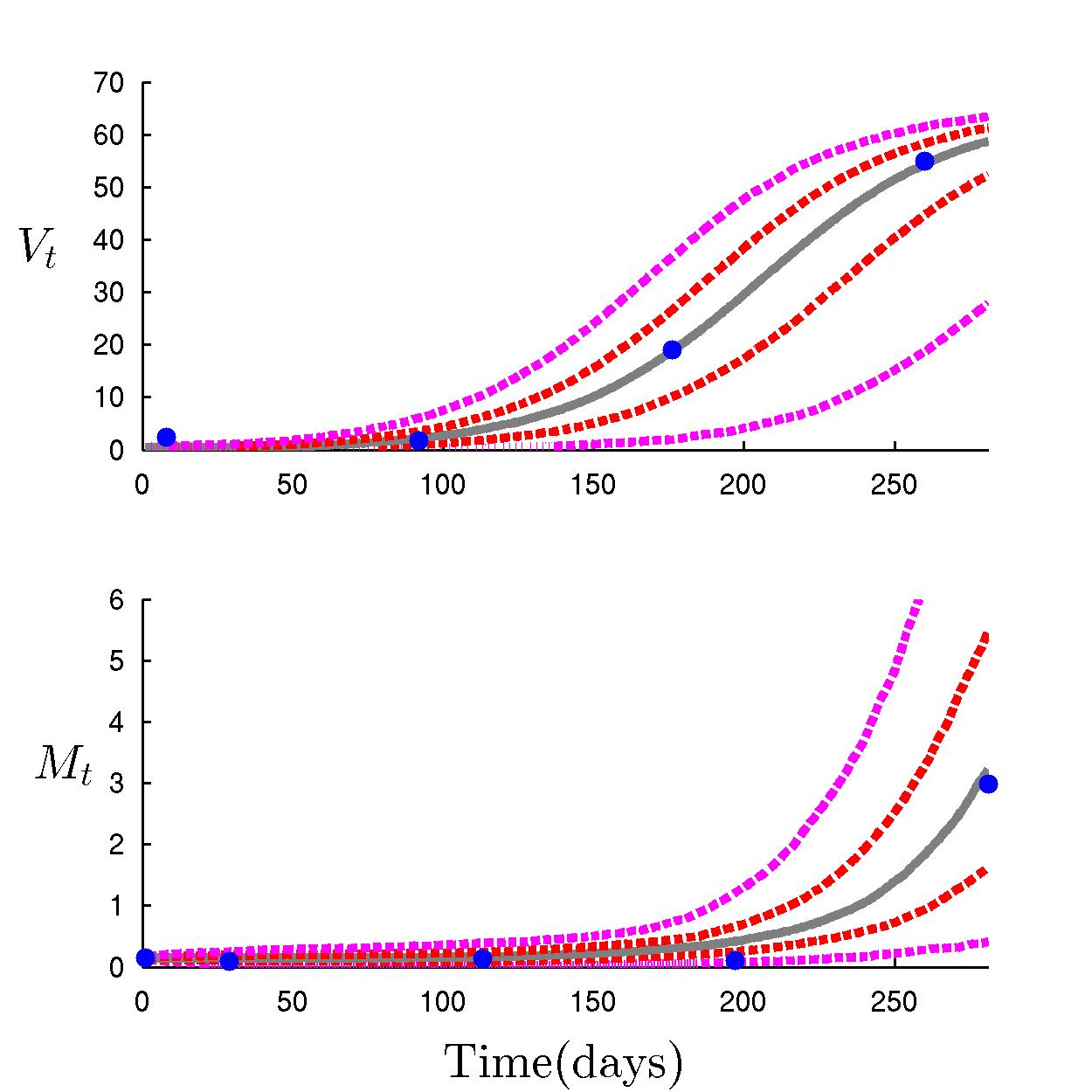} \\
\includegraphics[width=2.2in]{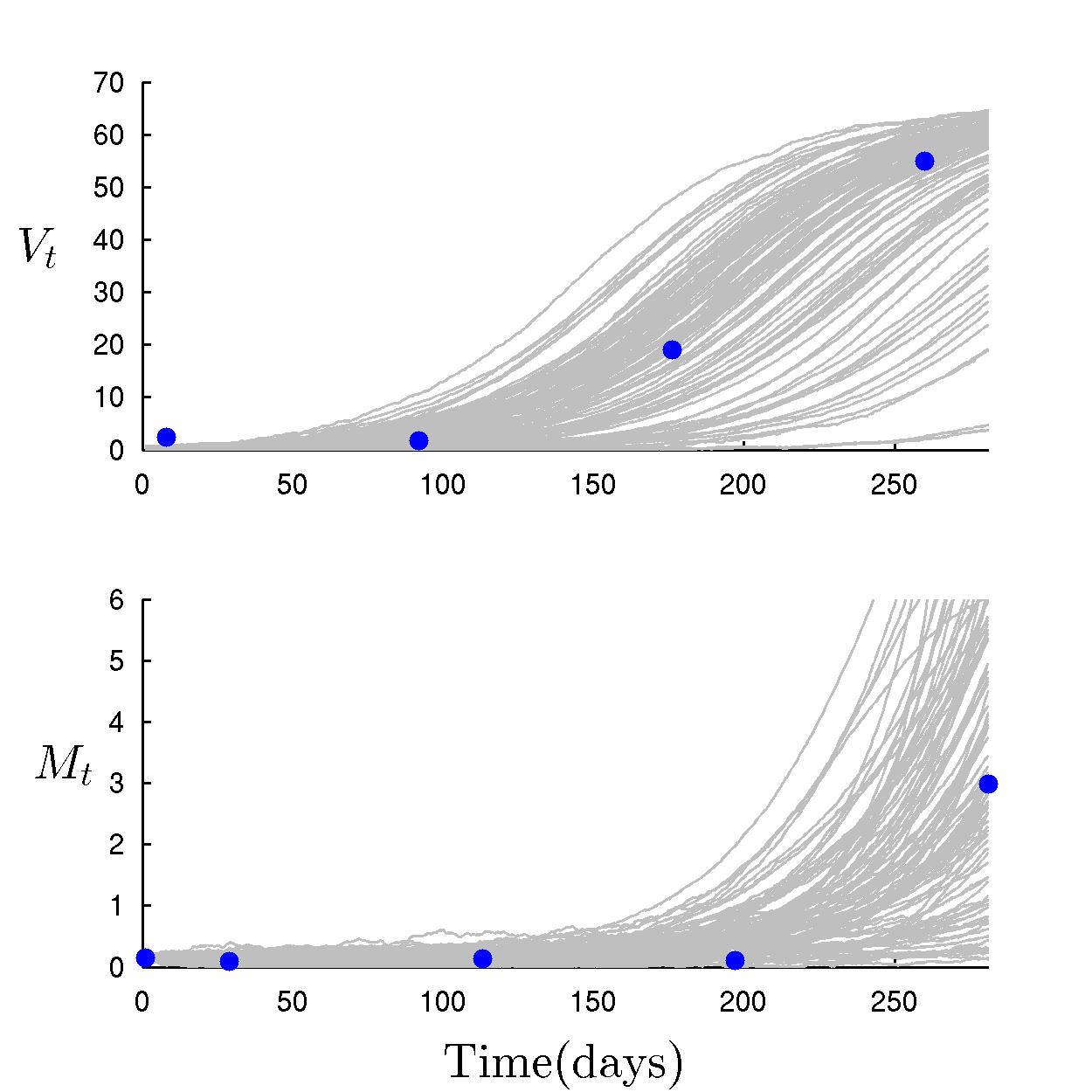}\\
\caption{\label{fig:Spredplots_id400} Posterior predictions for macaque \#400:
{\em Top left:}  Observed data (blue);
50\% (red), 95\% (magenta) credible bands and median (gray) of posterior predictions for the latent states
using only MCMC analysis.
{\em Top right:} Same format, but based on MCMC+ABC analysis.
{\em Lower center:}
100 sample trajectories from the posterior predictive distribution of the latent states from MCMC+ABC analysis.}
\end{figure}

\newpage

\begin{figure}[ht!]
  \centering
\includegraphics[width=2.2in]{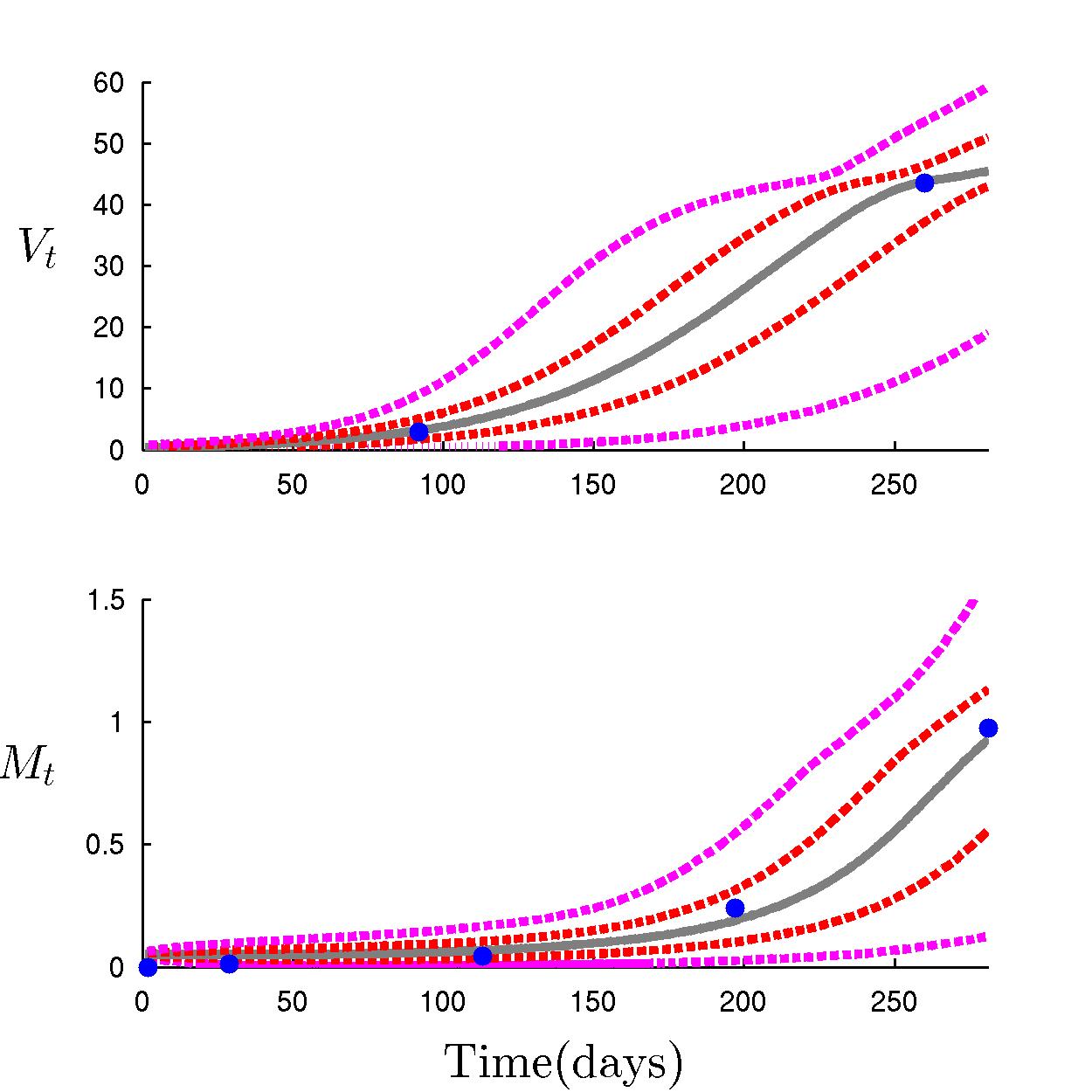}
\includegraphics[width=2.2in]{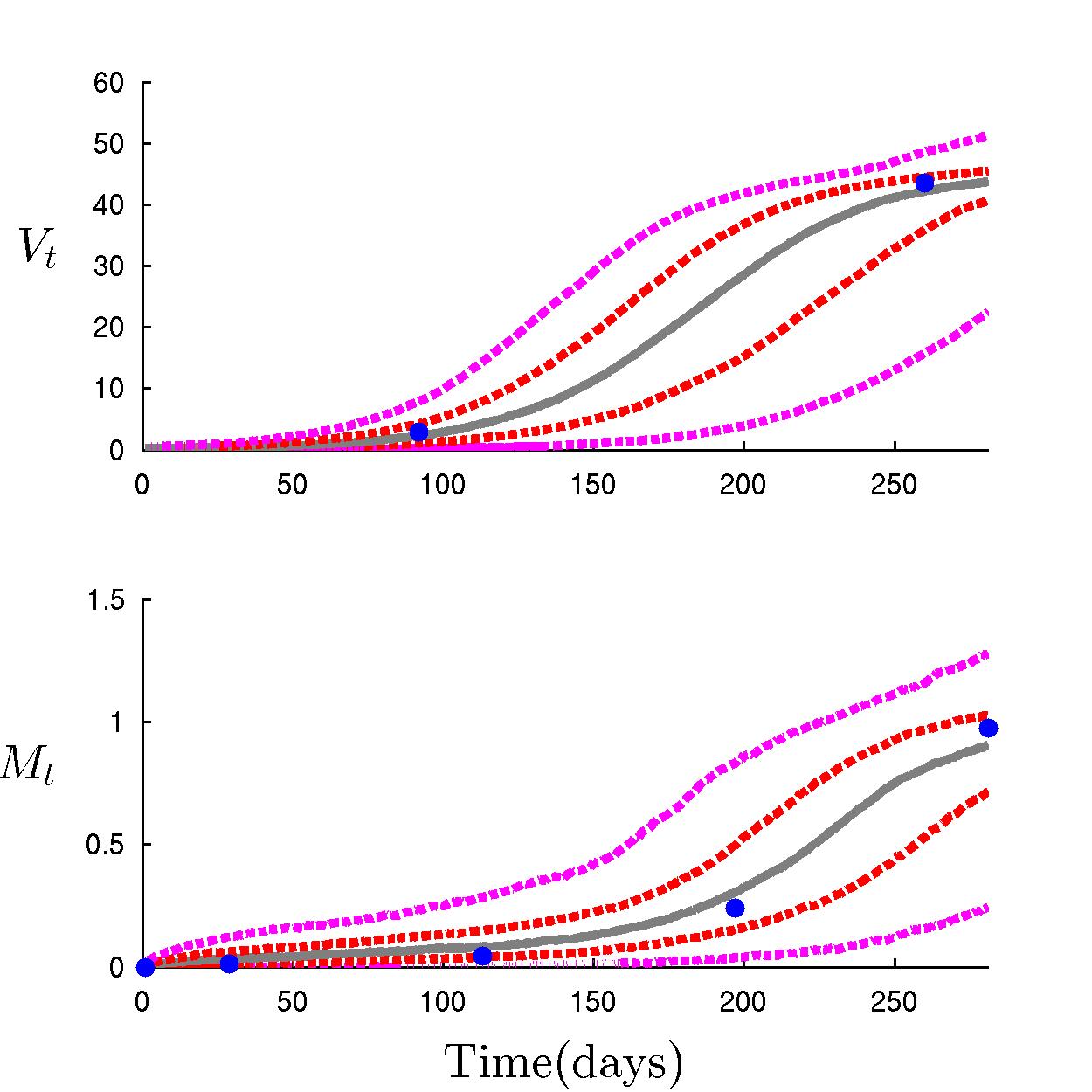} \\
\includegraphics[width=2.2in]{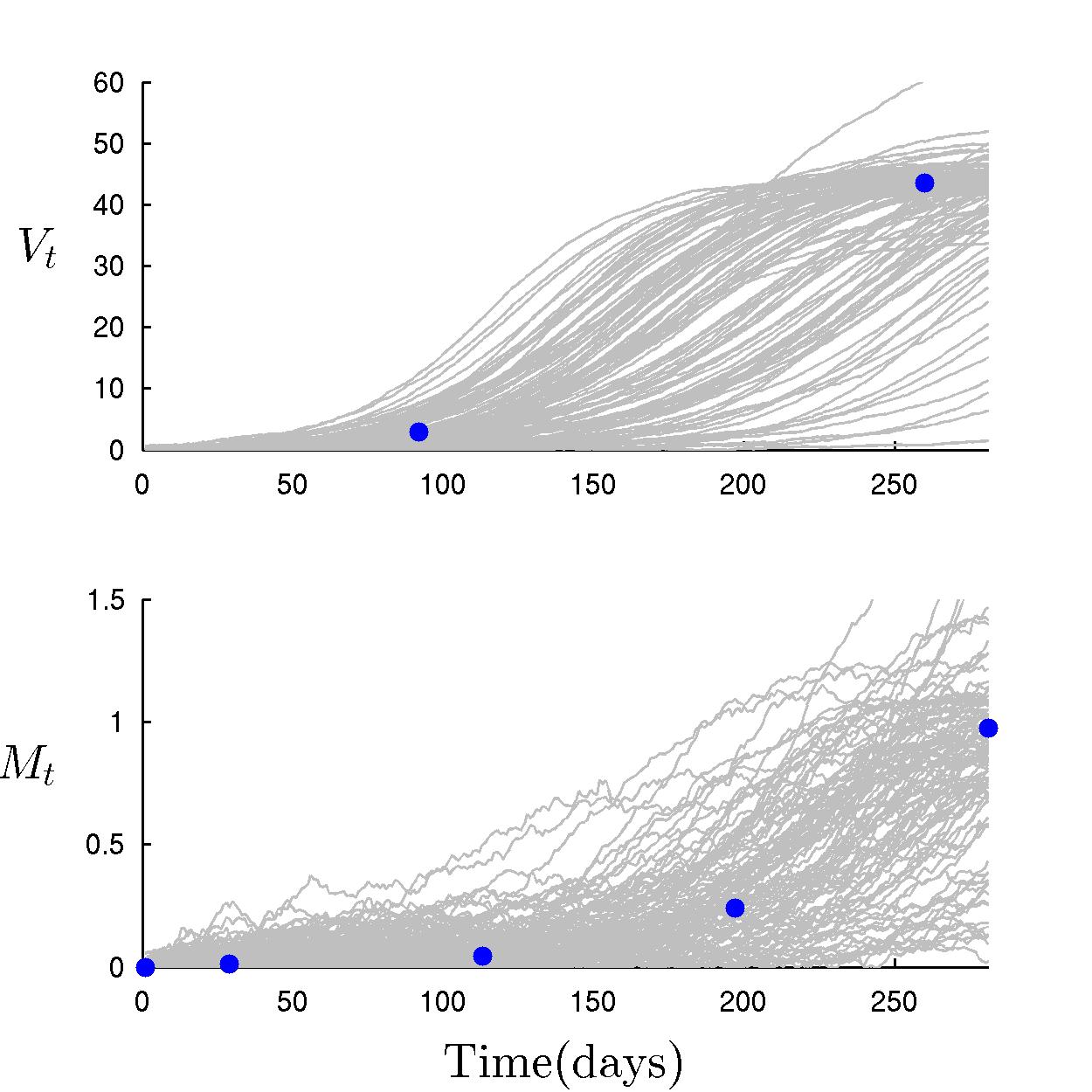}\\
\caption{\label{fig:Spredplots_id400} Posterior predictions for macaque \#401:
{\em Top left:}  Observed data (blue);
50\% (red), 95\% (magenta) credible bands and median (gray) of posterior predictions for the latent states
using only MCMC analysis.
{\em Top right:} Same format, but based on MCMC+ABC analysis.
{\em Lower center:}
100 sample trajectories from the posterior predictive distribution of the latent states from MCMC+ABC analysis.}
\end{figure}

\newpage

\begin{figure}[ht!]
  \centering
\includegraphics[width=2.2in]{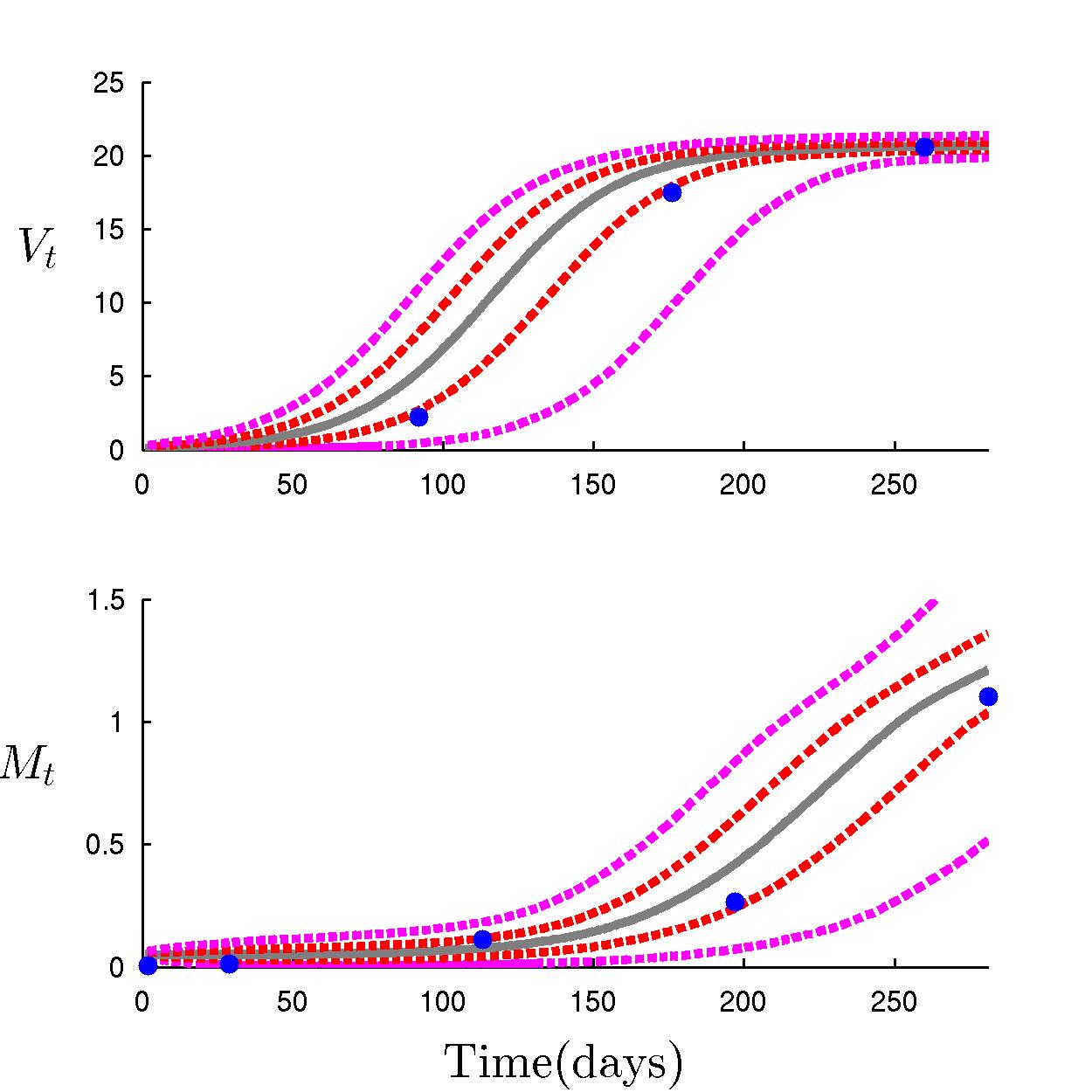}
\includegraphics[width=2.2in]{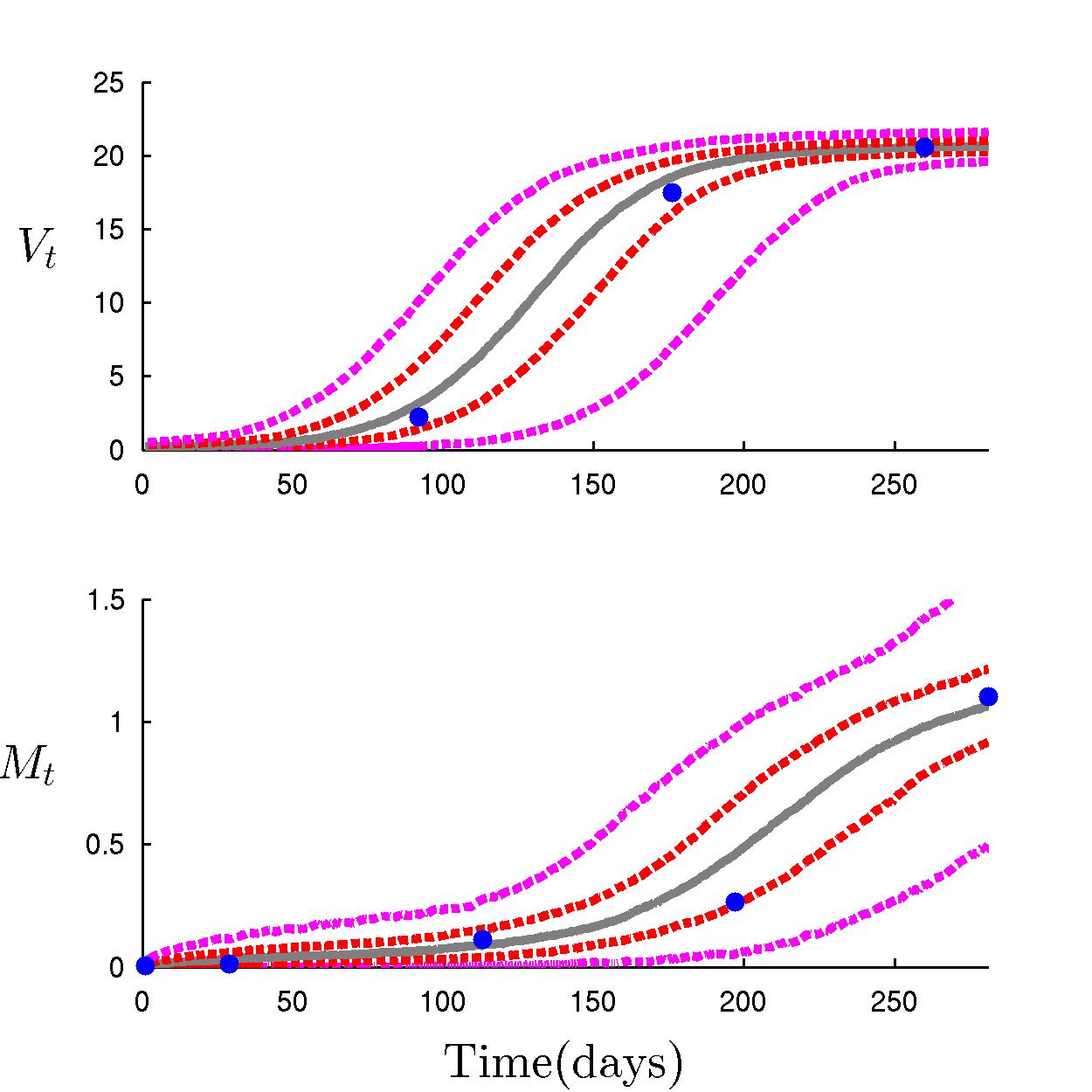} \\
\includegraphics[width=2.2in]{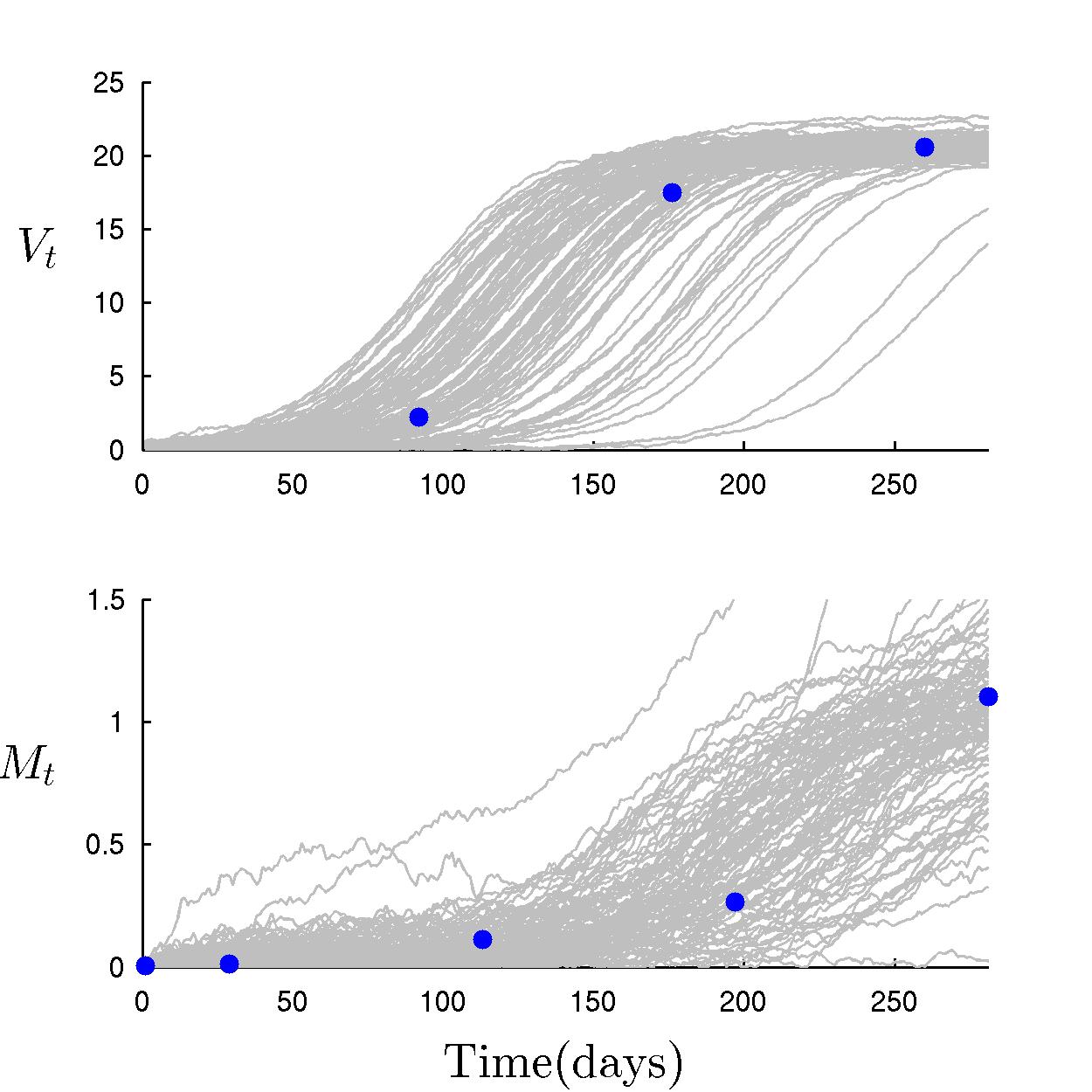}\\
\caption{\label{fig:Spredplots_id400} Posterior predictions for macaque \#404:
{\em Top left:}  Observed data (blue);
50\% (red), 95\% (magenta) credible bands and median (gray) of posterior predictions for the latent states
using only MCMC analysis.
{\em Top right:} Same format, but based on MCMC+ABC analysis.
{\em Lower center:}
100 sample trajectories from the posterior predictive distribution of the latent states from MCMC+ABC analysis.}
\end{figure}

\newpage

\begin{figure}[ht!]
  \centering
\includegraphics[width=2.2in]{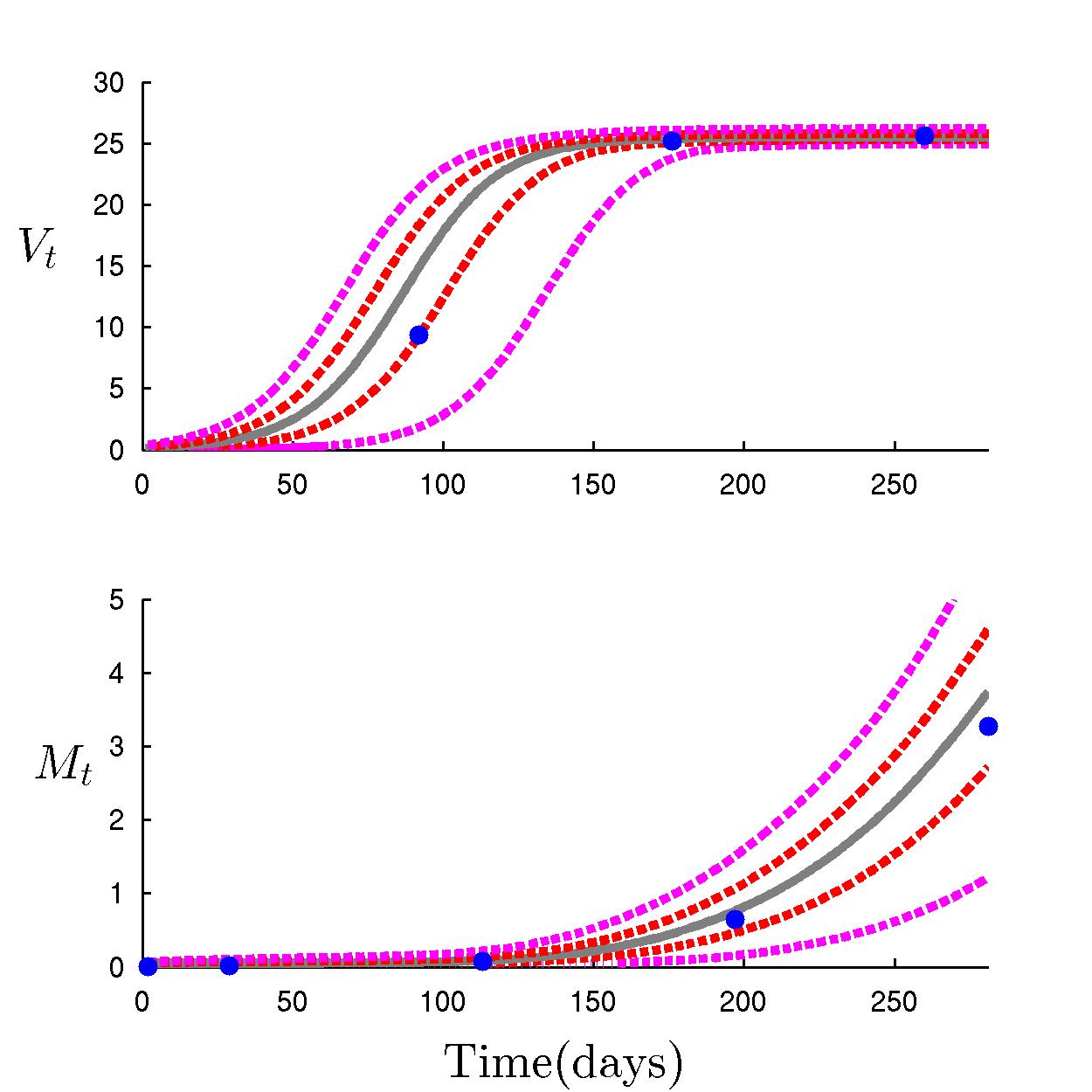}
\includegraphics[width=2.2in]{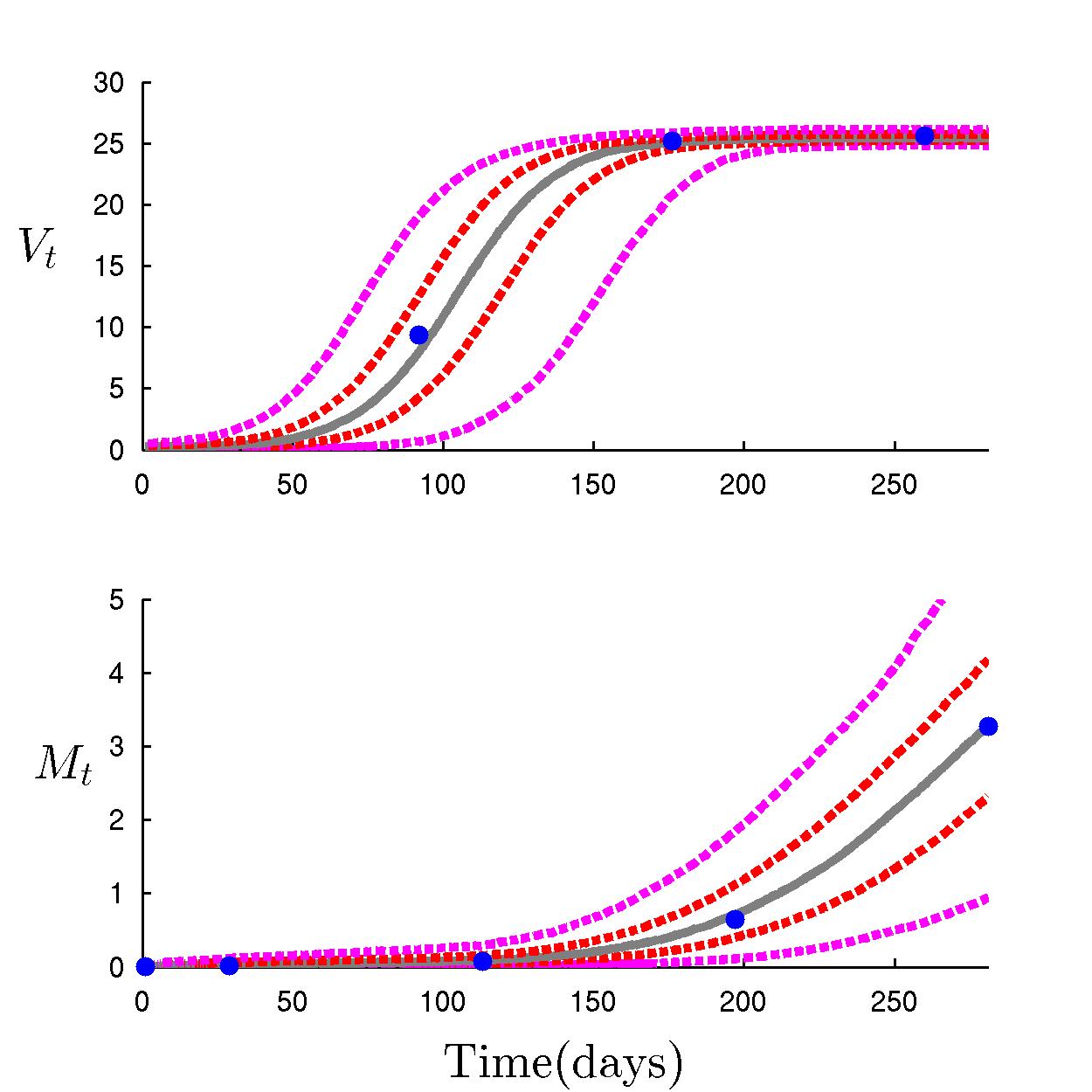} \\
\includegraphics[width=2.2in]{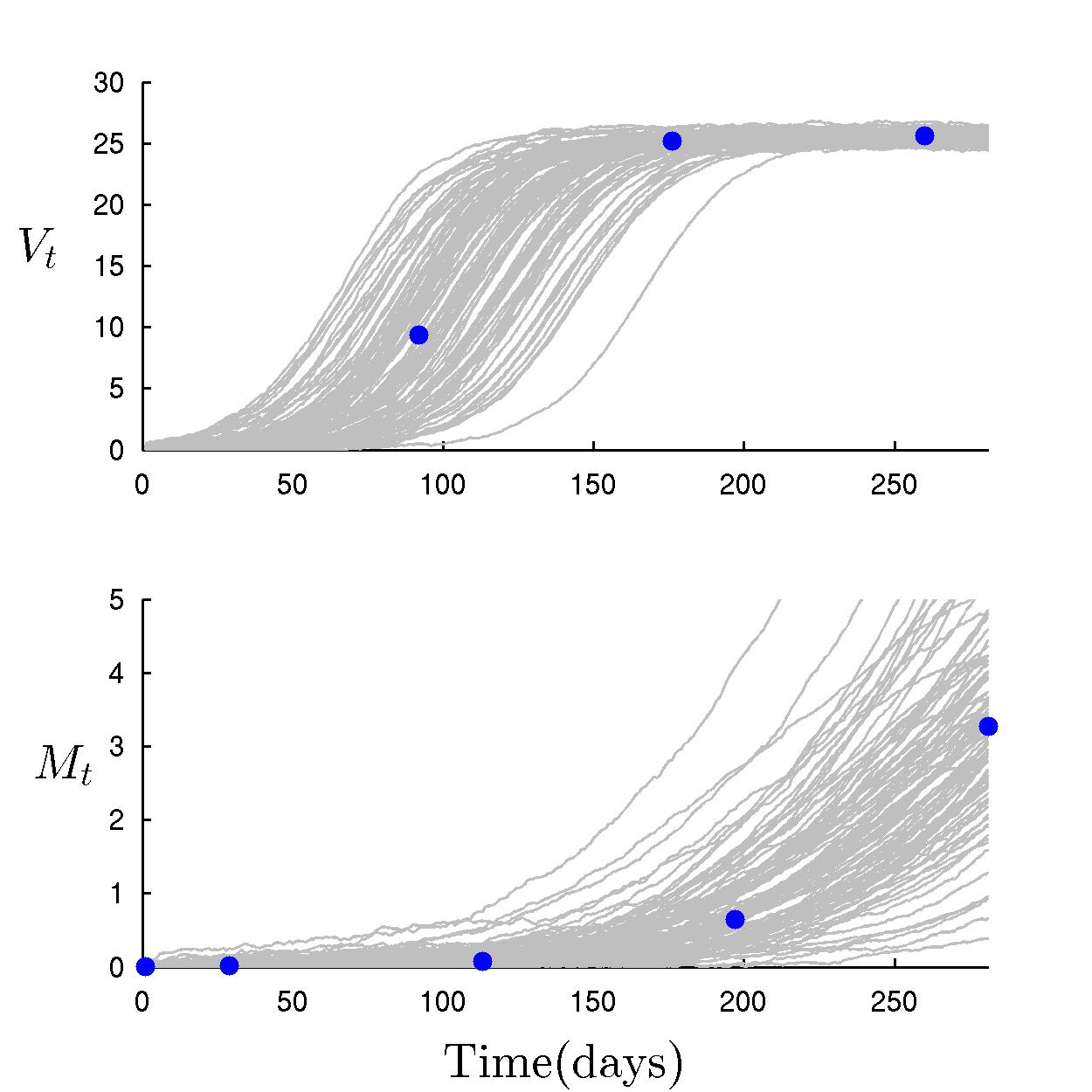}\\
\caption{\label{fig:Spredplots_id400} Posterior predictions for macaque \#405:
{\em Top left:}  Observed data (blue);
50\% (red), 95\% (magenta) credible bands and median (gray) of posterior predictions for the latent states
using only MCMC analysis.
{\em Top right:} Same format, but based on MCMC+ABC analysis.
{\em Lower center:}
100 sample trajectories from the posterior predictive distribution of the latent states from MCMC+ABC analysis.}
\end{figure}

\newpage

\begin{figure}[ht!]
  \centering
\includegraphics[width=2.2in]{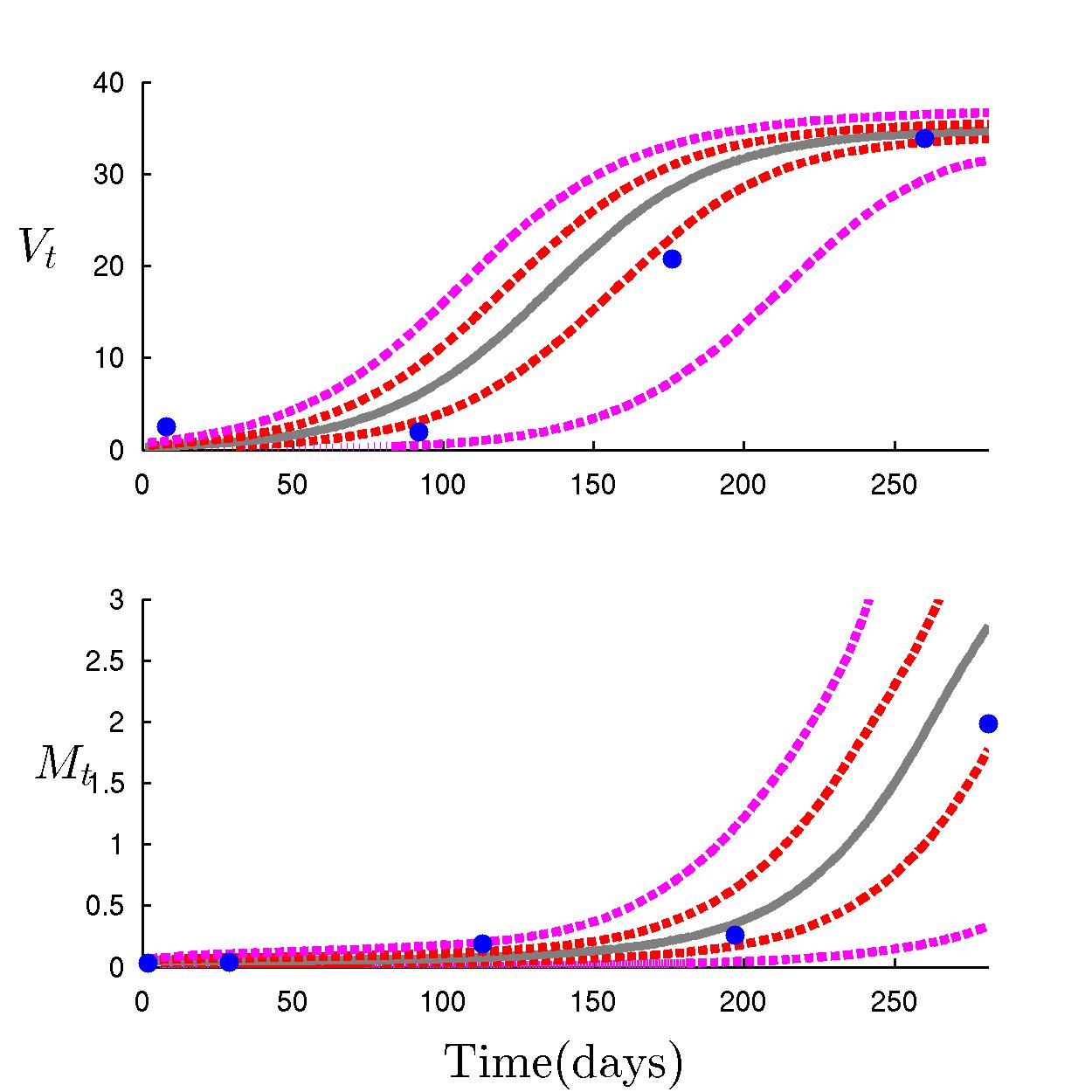}
\includegraphics[width=2.2in]{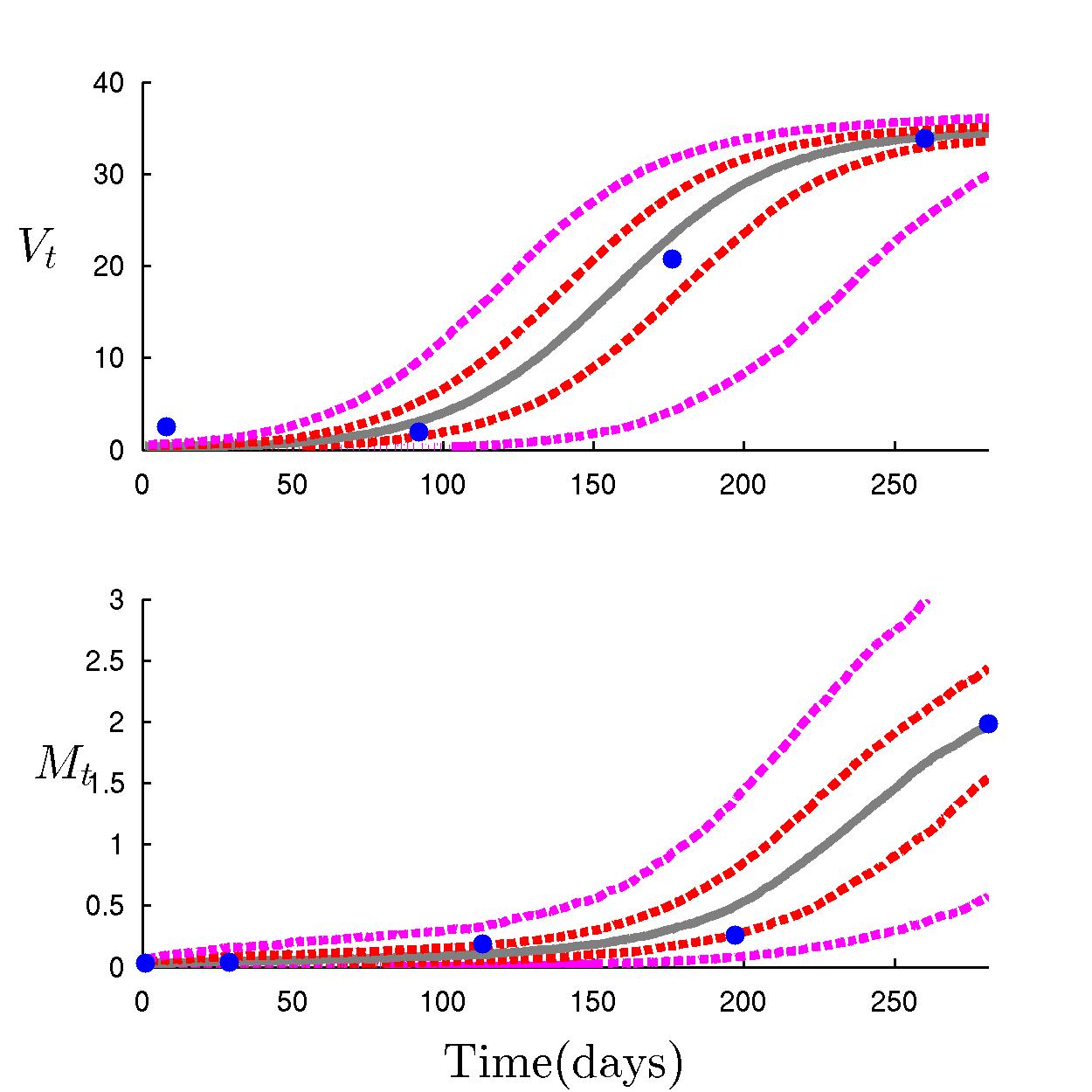} \\
\includegraphics[width=2.2in]{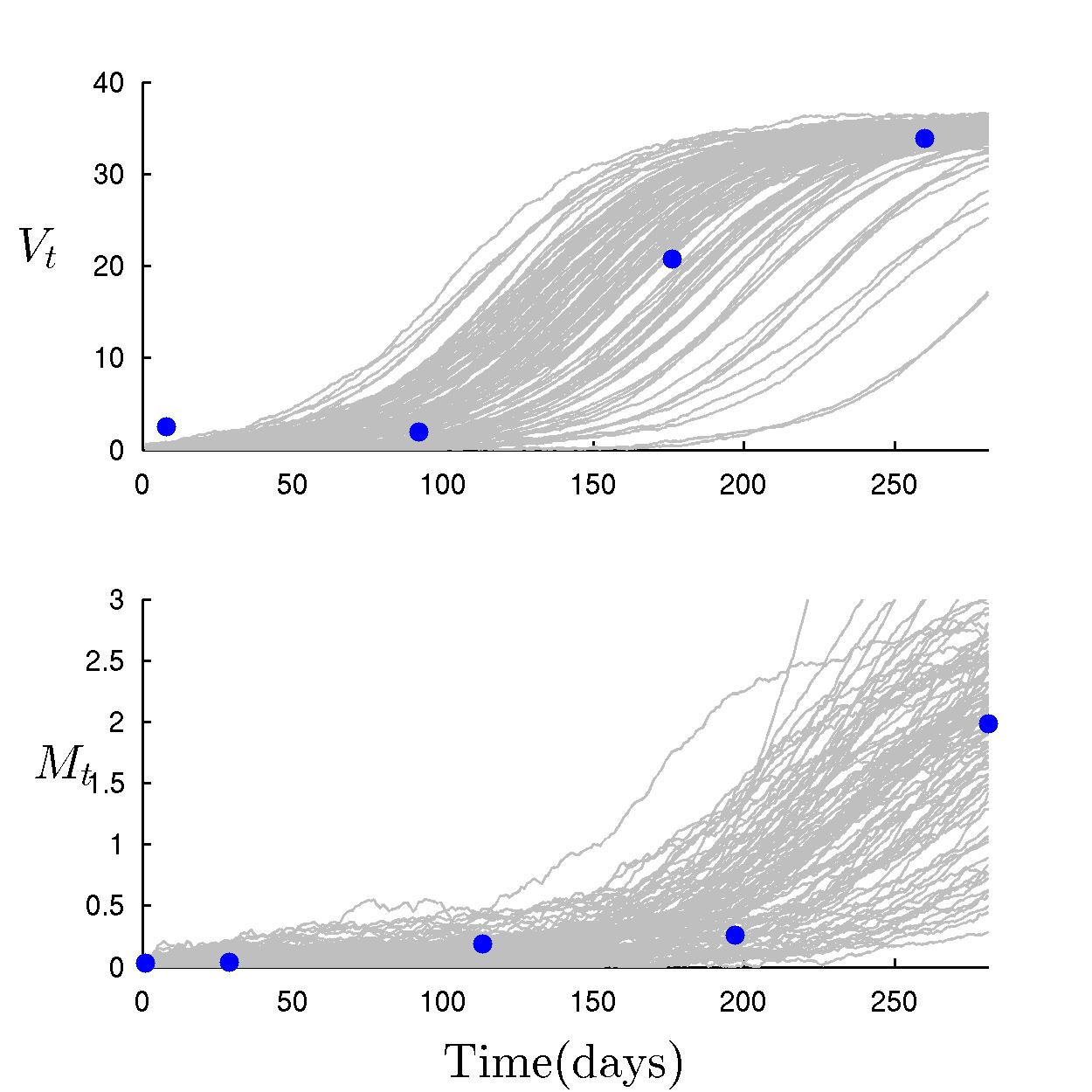}\\
\caption{\label{fig:Spredplots_id400} Posterior predictions for macaque \#407:
{\em Top left:}  Observed data (blue);
50\% (red), 95\% (magenta) credible bands and median (gray) of posterior predictions for the latent states
using only MCMC analysis.
{\em Top right:} Same format, but based on MCMC+ABC analysis.
{\em Lower center:}
100 sample trajectories from the posterior predictive distribution of the latent states from MCMC+ABC analysis.}
\end{figure}

\newpage

\begin{figure}[ht!]
  \centering
\includegraphics[width=2.2in]{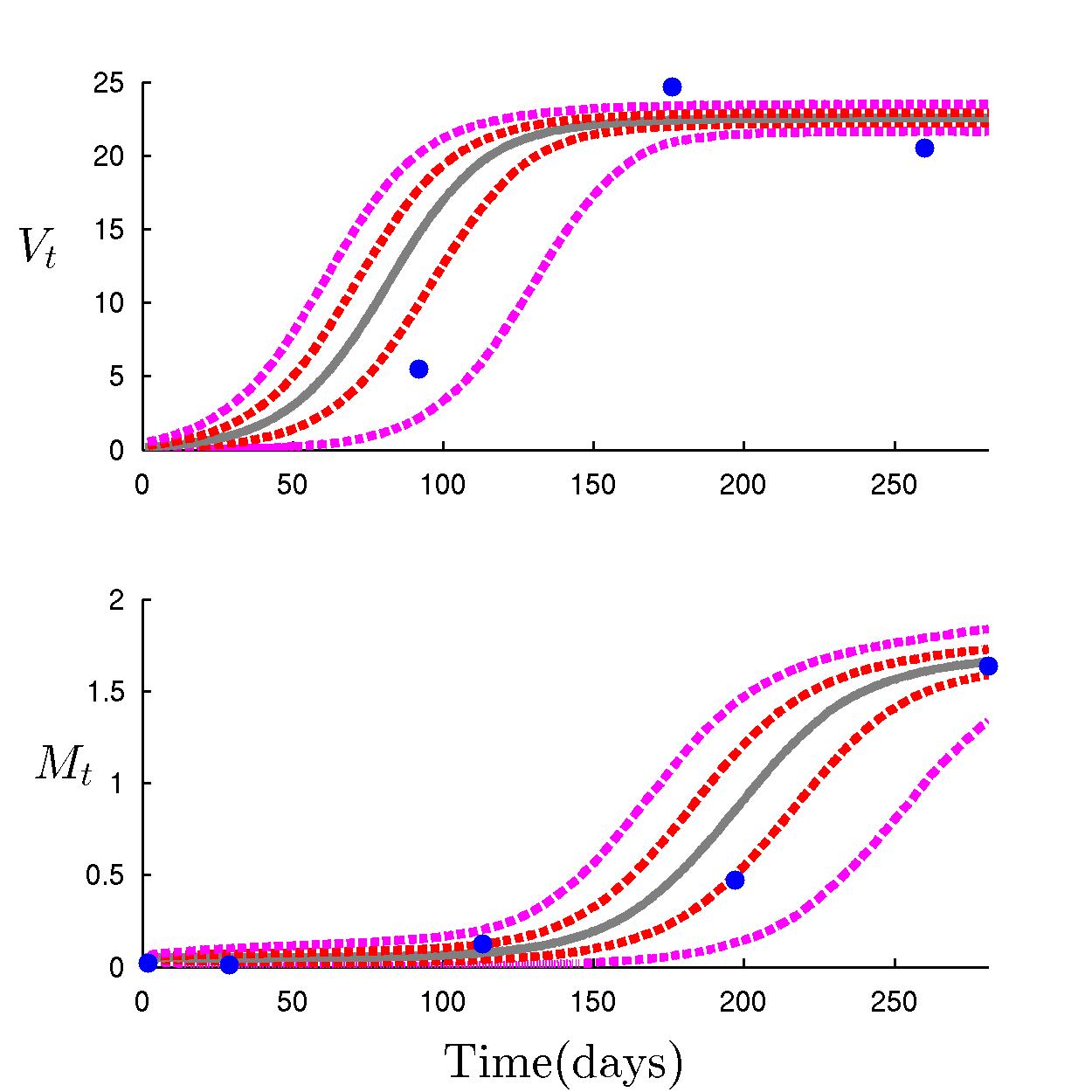}
\includegraphics[width=2.2in]{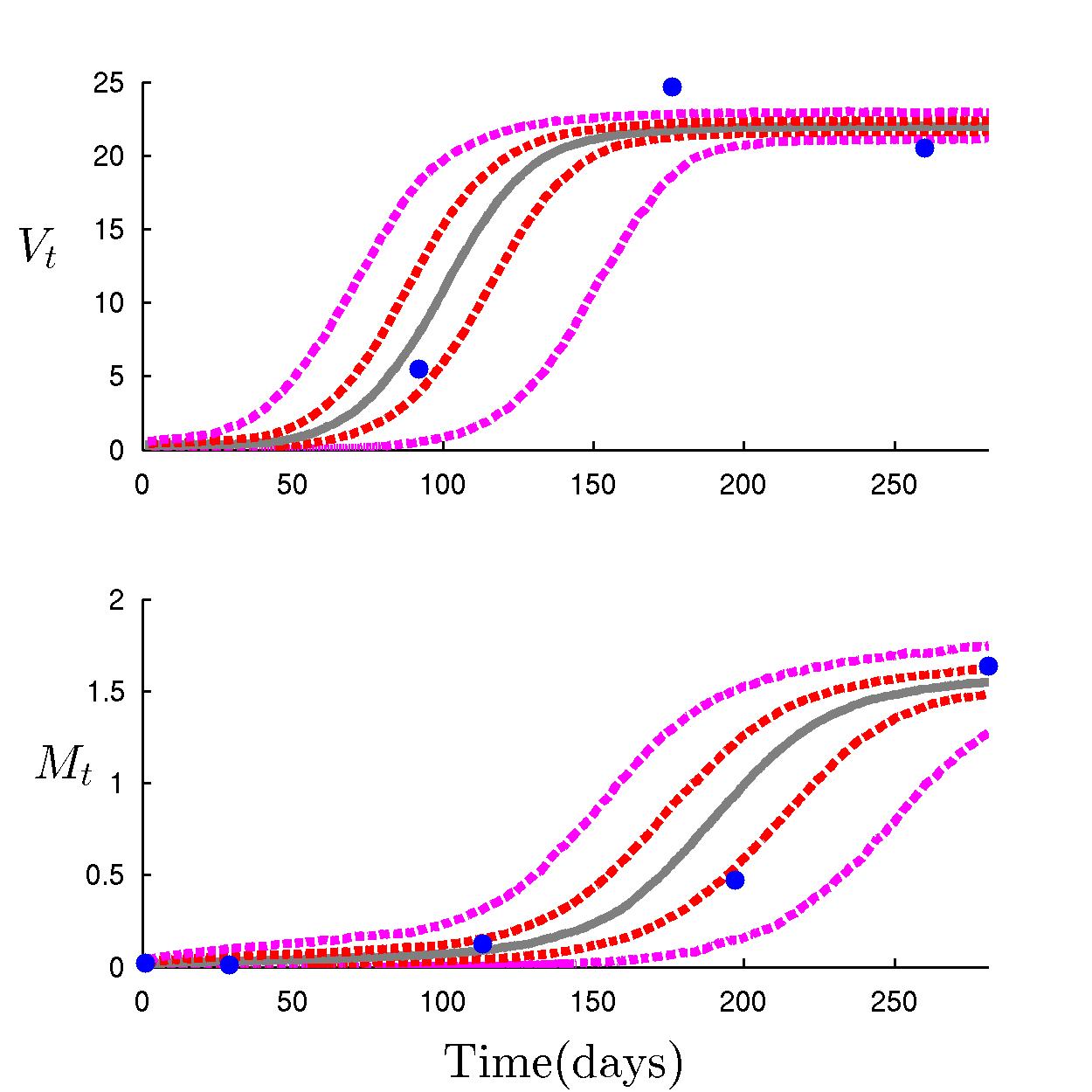} \\
\includegraphics[width=2.2in]{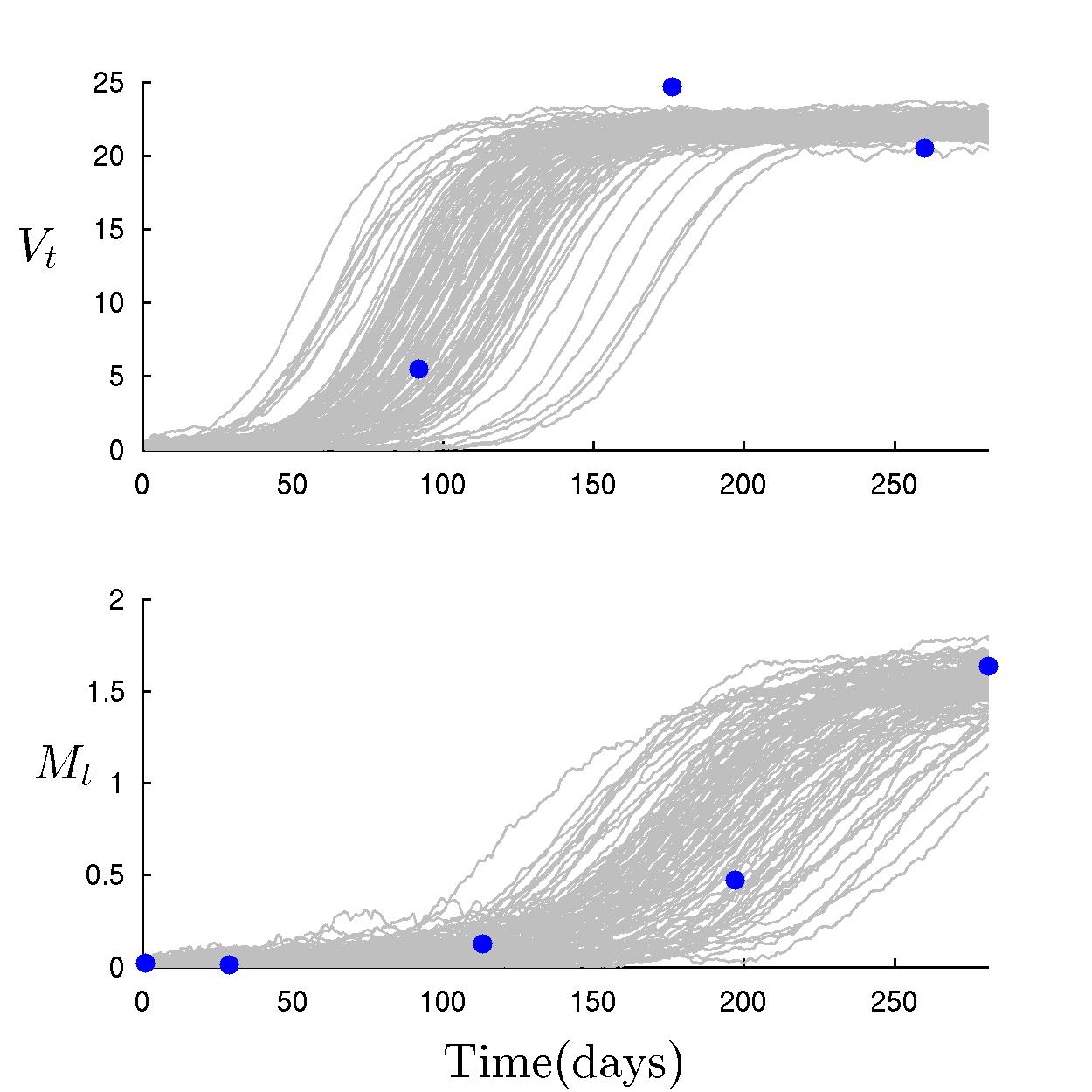}\\
\caption{\label{fig:Spredplots_id400} Posterior predictions for macaque \#408:
{\em Top left:}  Observed data (blue);
50\% (red), 95\% (magenta) credible bands and median (gray) of posterior predictions for the latent states
using only MCMC analysis.
{\em Top right:} Same format, but based on MCMC+ABC analysis.
{\em Lower center:}
100 sample trajectories from the posterior predictive distribution of the latent states from MCMC+ABC analysis.}
\end{figure}

\newpage

\begin{figure}[ht!]
  \centering
\includegraphics[width=2.2in]{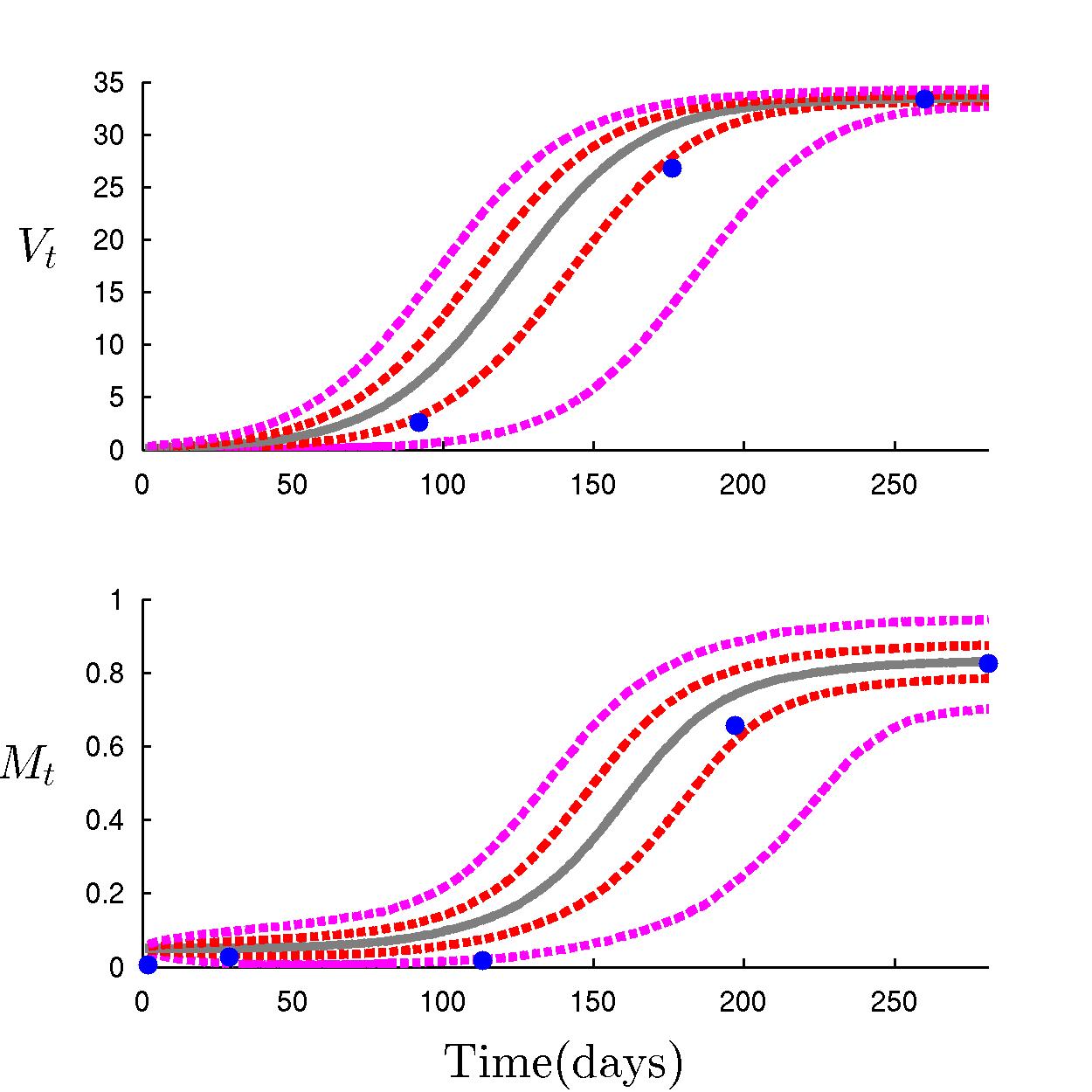}
\includegraphics[width=2.2in]{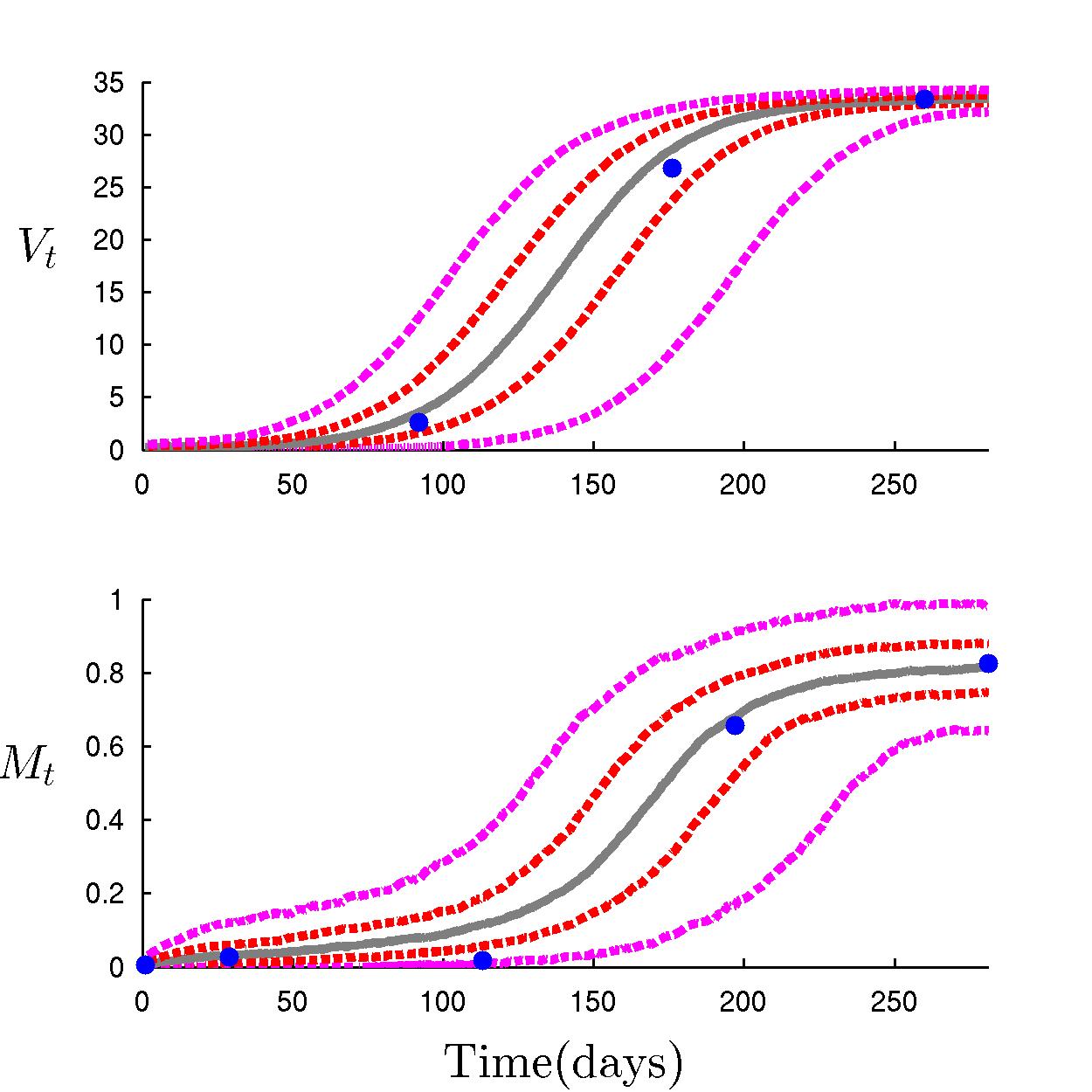} \\
\includegraphics[width=2.2in]{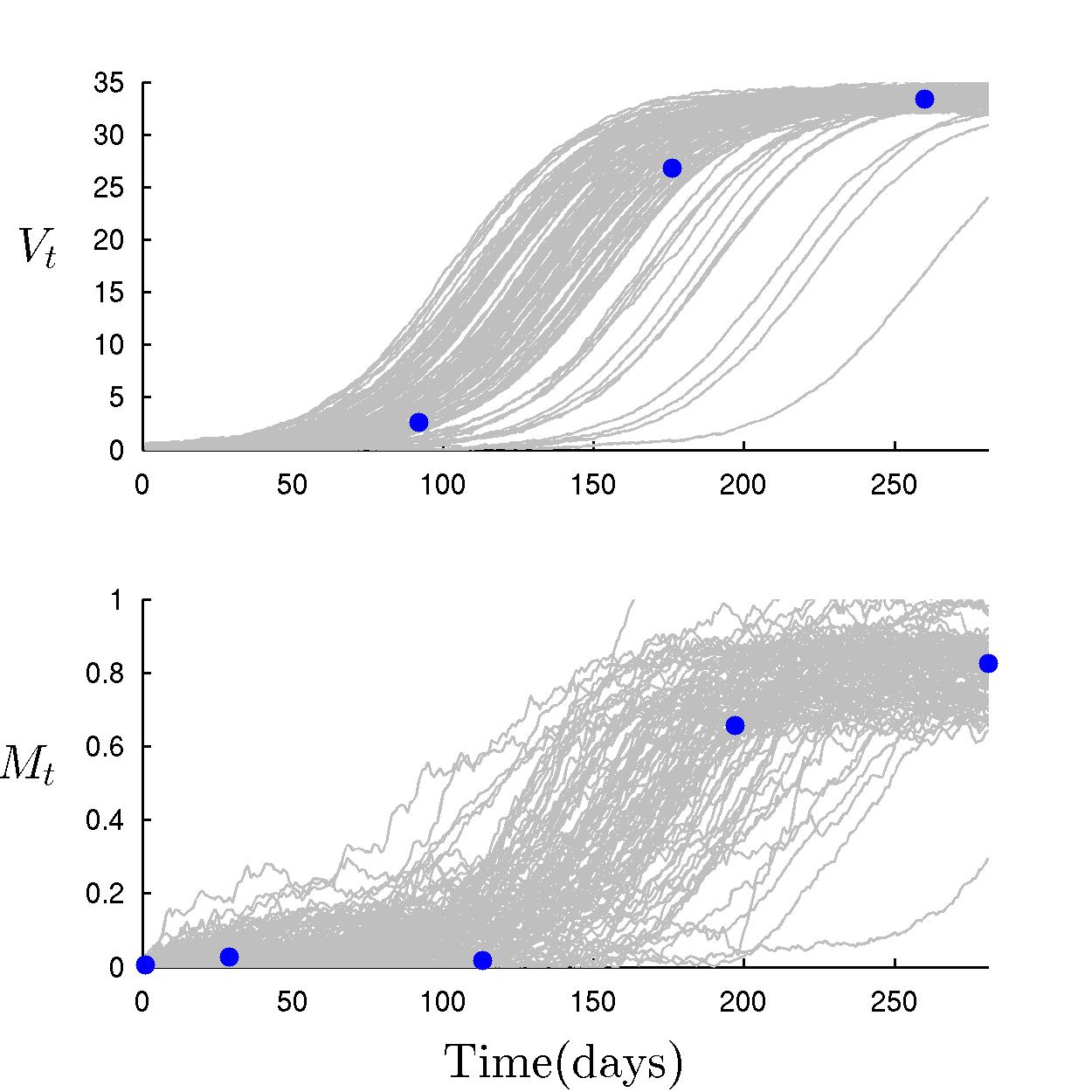}\\
\caption{\label{fig:Spredplots_id400} Posterior predictions for macaque \#415:
{\em Top left:}  Observed data (blue);
50\% (red), 95\% (magenta) credible bands and median (gray) of posterior predictions for the latent states
using only MCMC analysis.
{\em Top right:} Same format, but based on MCMC+ABC analysis.
{\em Lower center:}
100 sample trajectories from the posterior predictive distribution of the latent states from MCMC+ABC analysis.}
\end{figure}

\newpage

\begin{figure}[ht!]
  \centering
\includegraphics[width=2.2in]{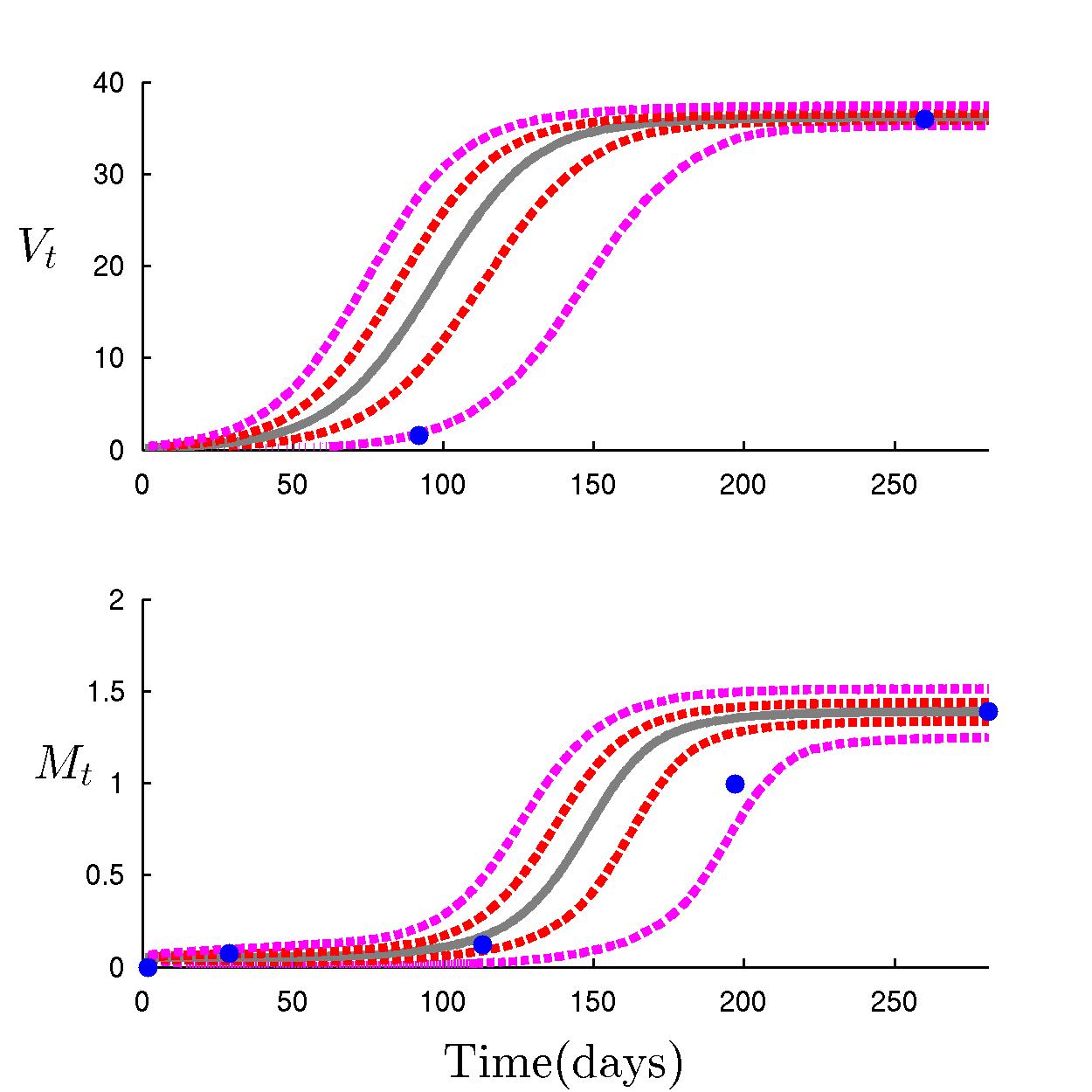}
\includegraphics[width=2.2in]{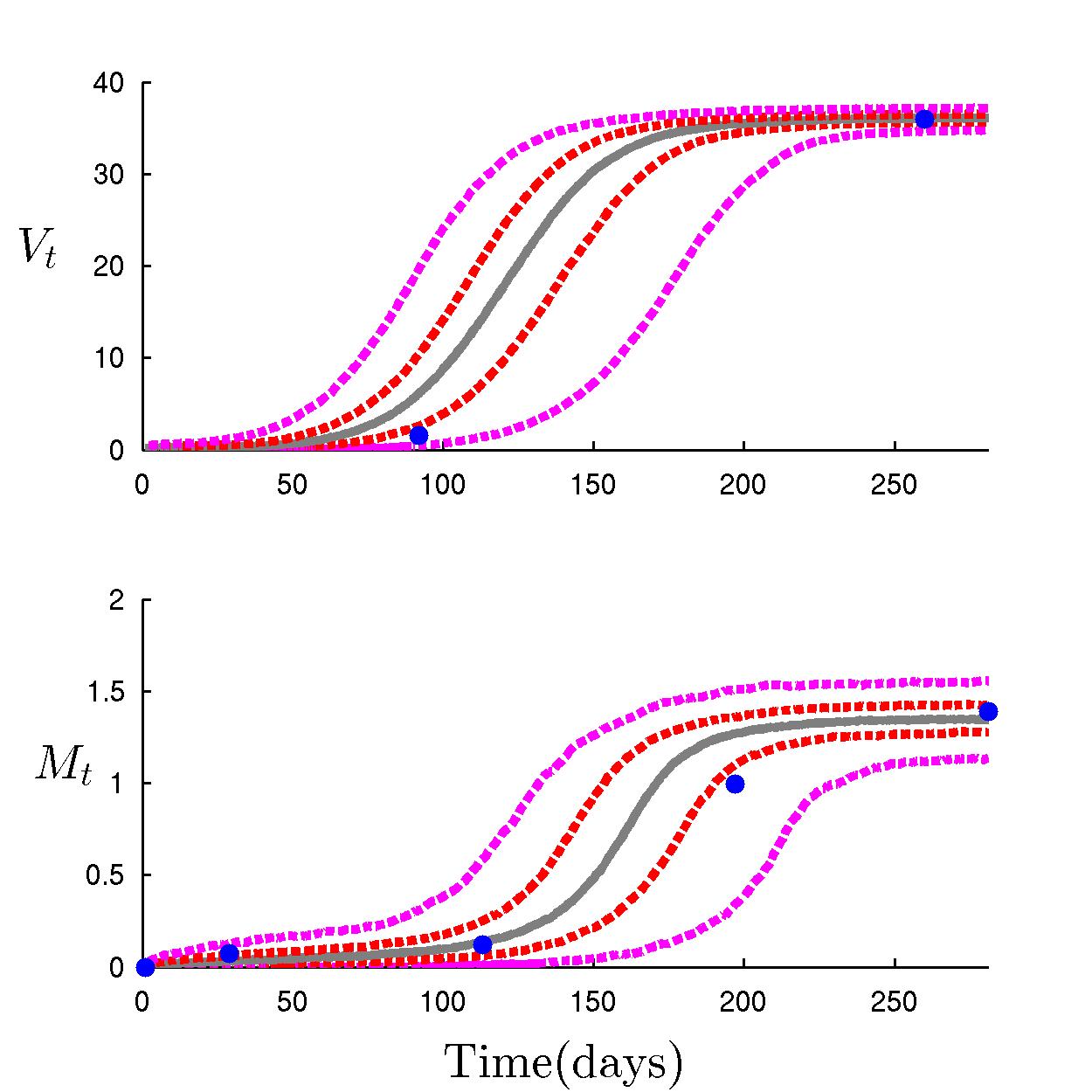} \\
\includegraphics[width=2.2in]{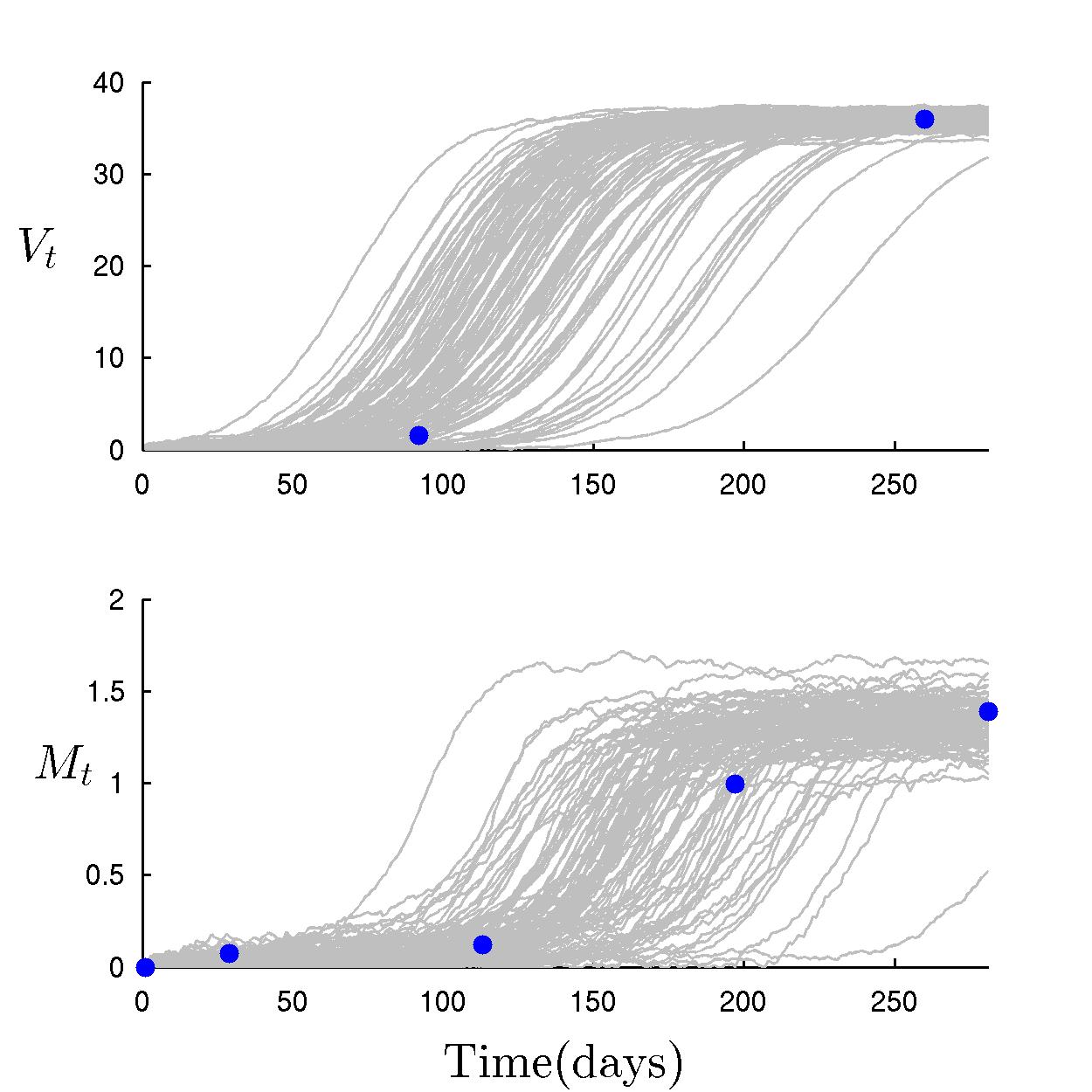}\\
\caption{\label{fig:Spredplots_id400} Posterior predictions for macaque \#431:
{\em Top left:}  Observed data (blue);
50\% (red), 95\% (magenta) credible bands and median (gray) of posterior predictions for the latent states
using only MCMC analysis.
{\em Top right:} Same format, but based on MCMC+ABC analysis.
{\em Lower center:}
100 sample trajectories from the posterior predictive distribution of the latent states from MCMC+ABC analysis.}
\end{figure}

\newpage
\begin{figure}[ht!]
  \centering
 \includegraphics[width=2.0in]{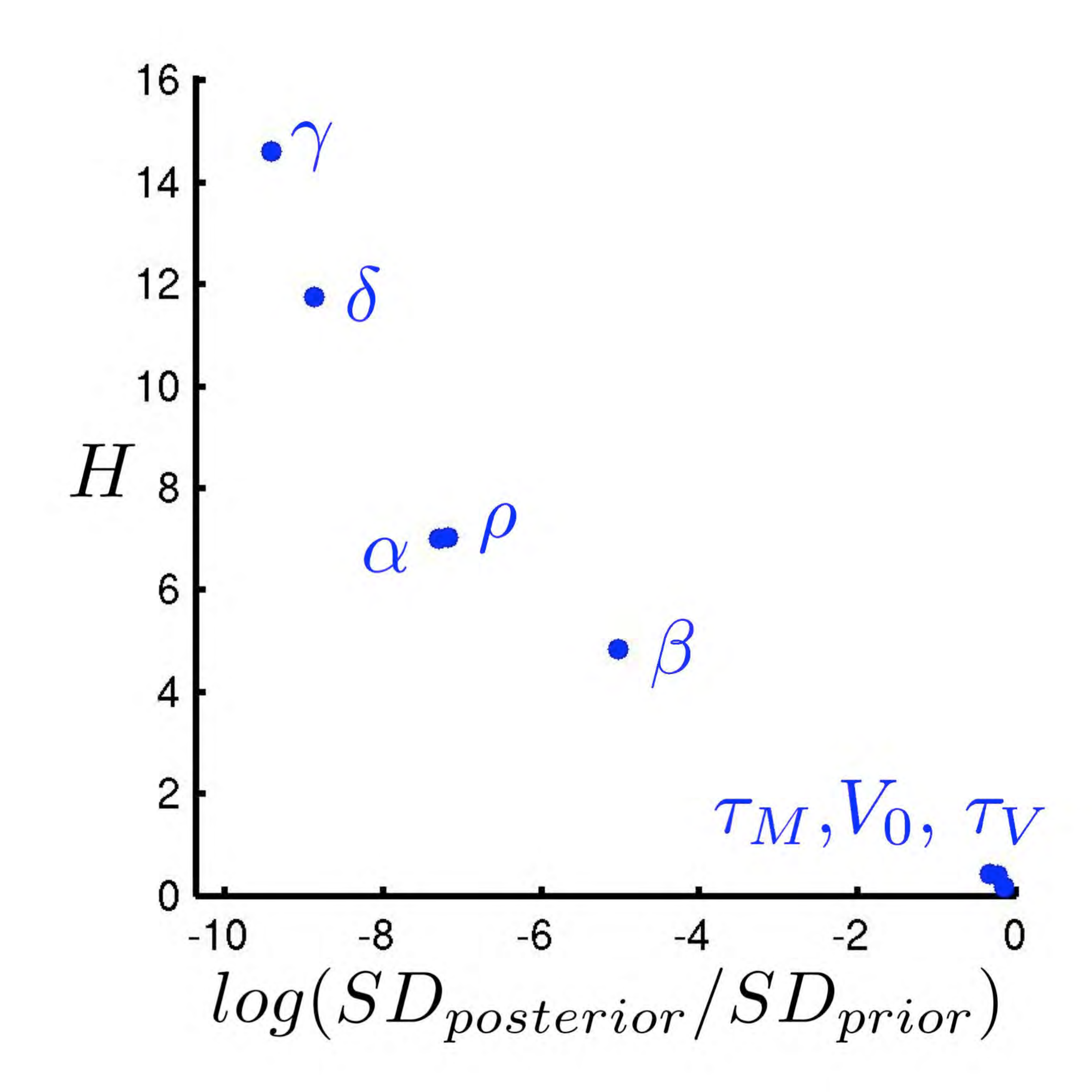}
  \includegraphics[width=2.0in]{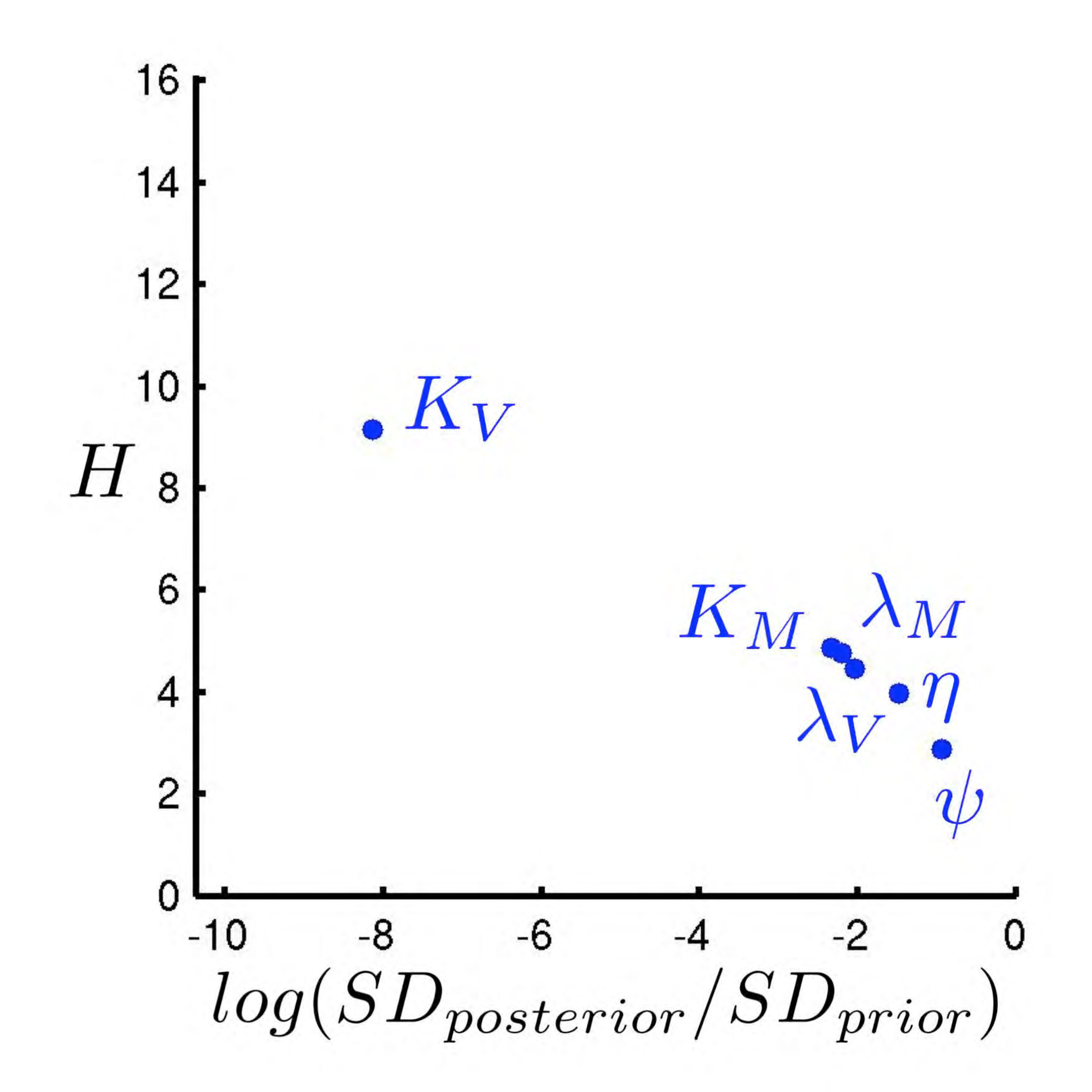} \\
\caption{Macaque 400: Learnability summaries.
Posterior:prior relative entropy $H$ against log of posterior to prior standard deviation, in some
cases implicitly based on
parameters following inverse c.d.f. transform to uniform priors as relevant.}
\end{figure}

\begin{figure}[ht!]
  \centering
 \includegraphics[width=2.0in]{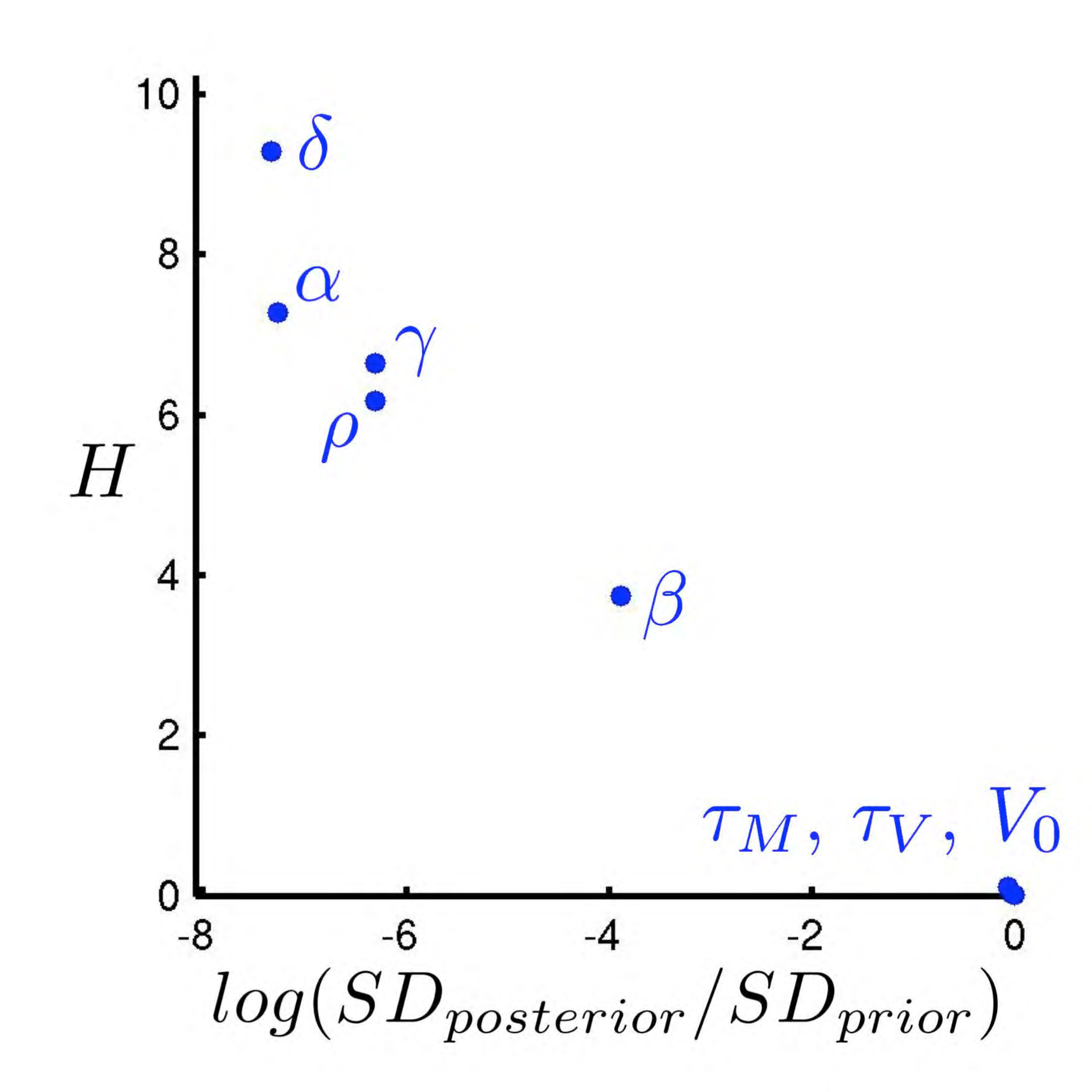}
  \includegraphics[width=2.0in]{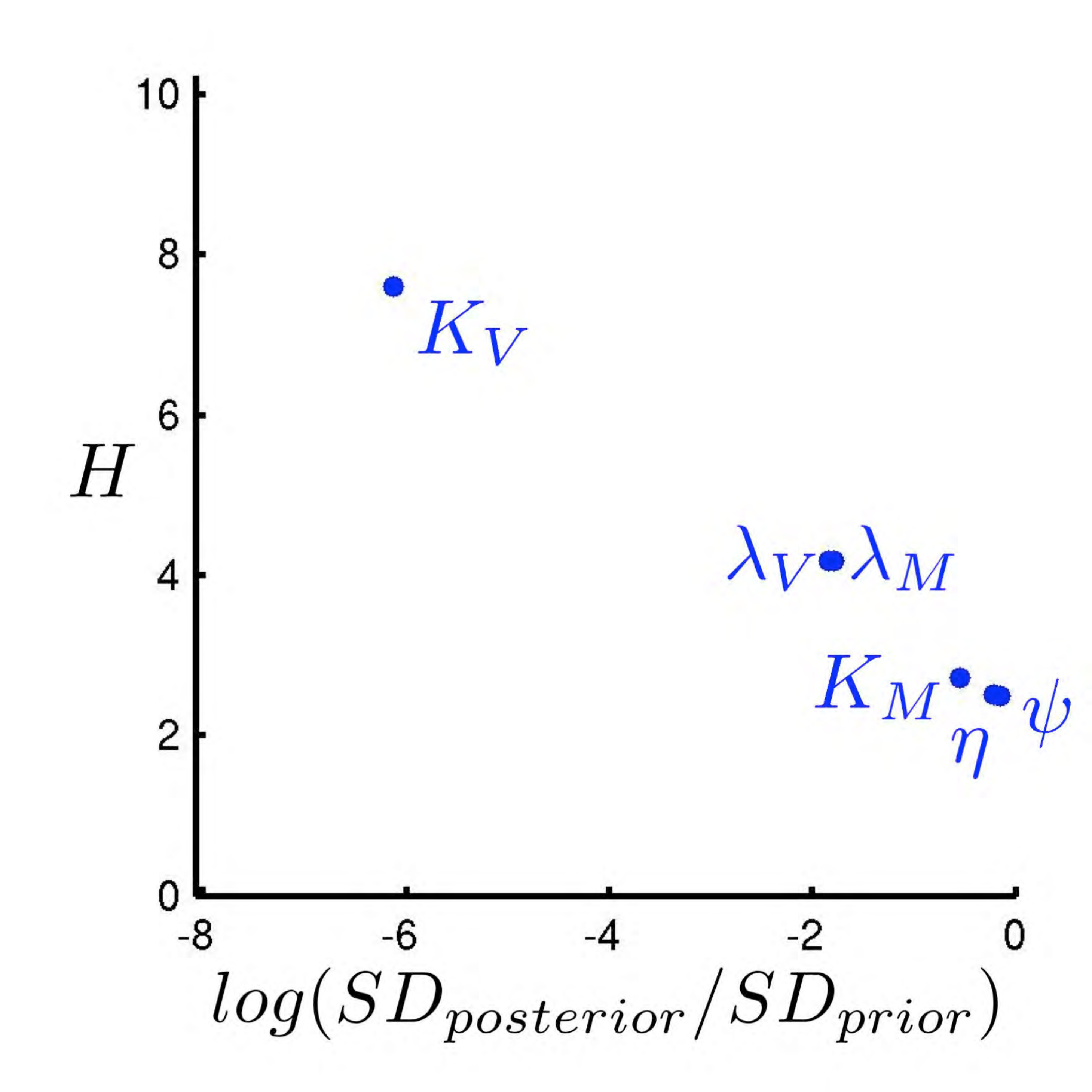} \\
\caption{Macaque 400: Learnability summaries.
Posterior:prior relative entropy $H$ against log of posterior to prior standard deviation, in some
cases implicitly based on
parameters following inverse c.d.f. transform to uniform priors as relevant.}
\end{figure}

\newpage
\begin{figure}[ht!]
  \centering
 \includegraphics[width=2.0in]{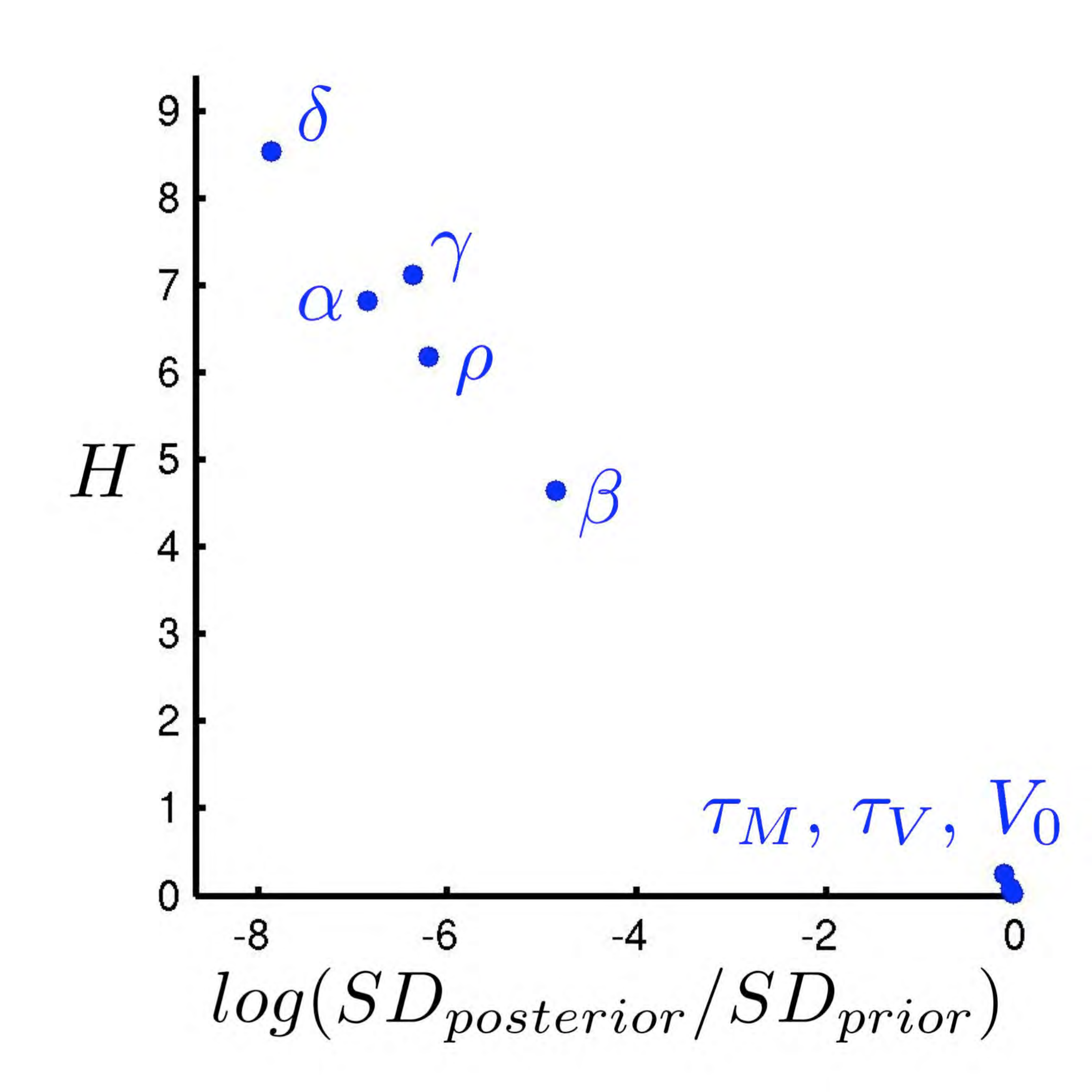}
  \includegraphics[width=2.0in]{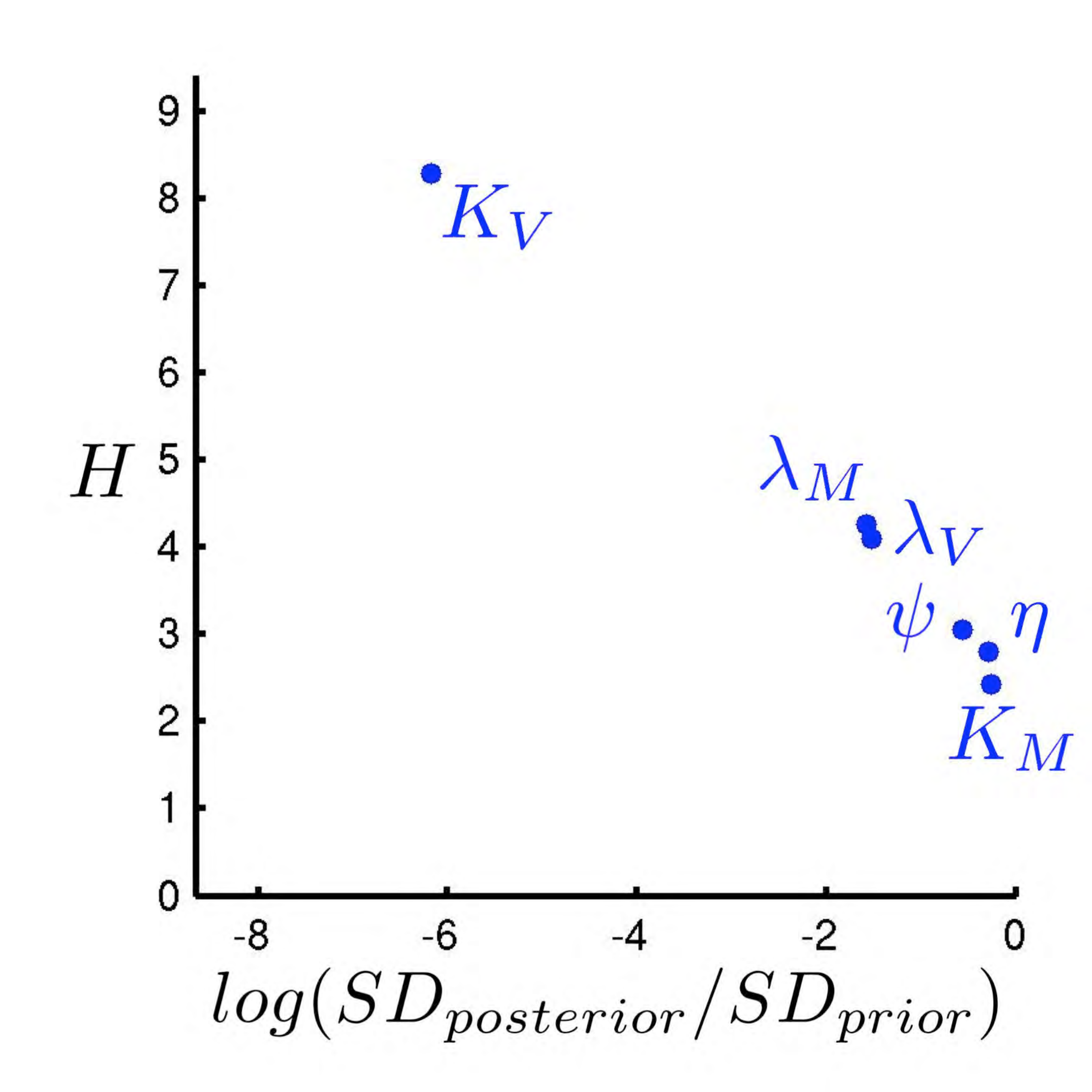} \\
\caption{Macaque 404: Learnability summaries.
Posterior:prior relative entropy $H$ against log of posterior to prior standard deviation, in some
cases implicitly based on
parameters following inverse c.d.f. transform to uniform priors as relevant.}
\end{figure}

\begin{figure}[ht!]
  \centering
 \includegraphics[width=2.0in]{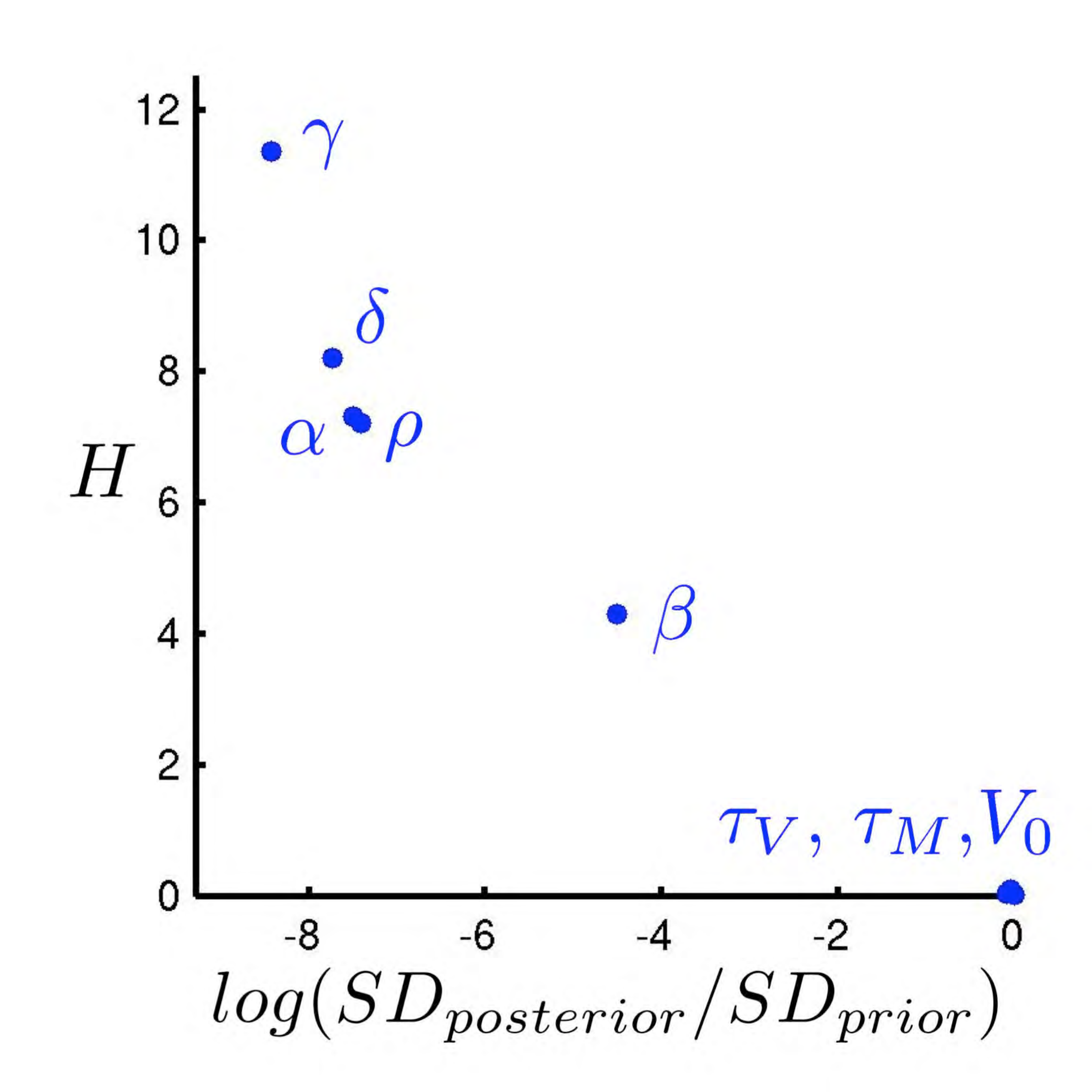}
  \includegraphics[width=2.0in]{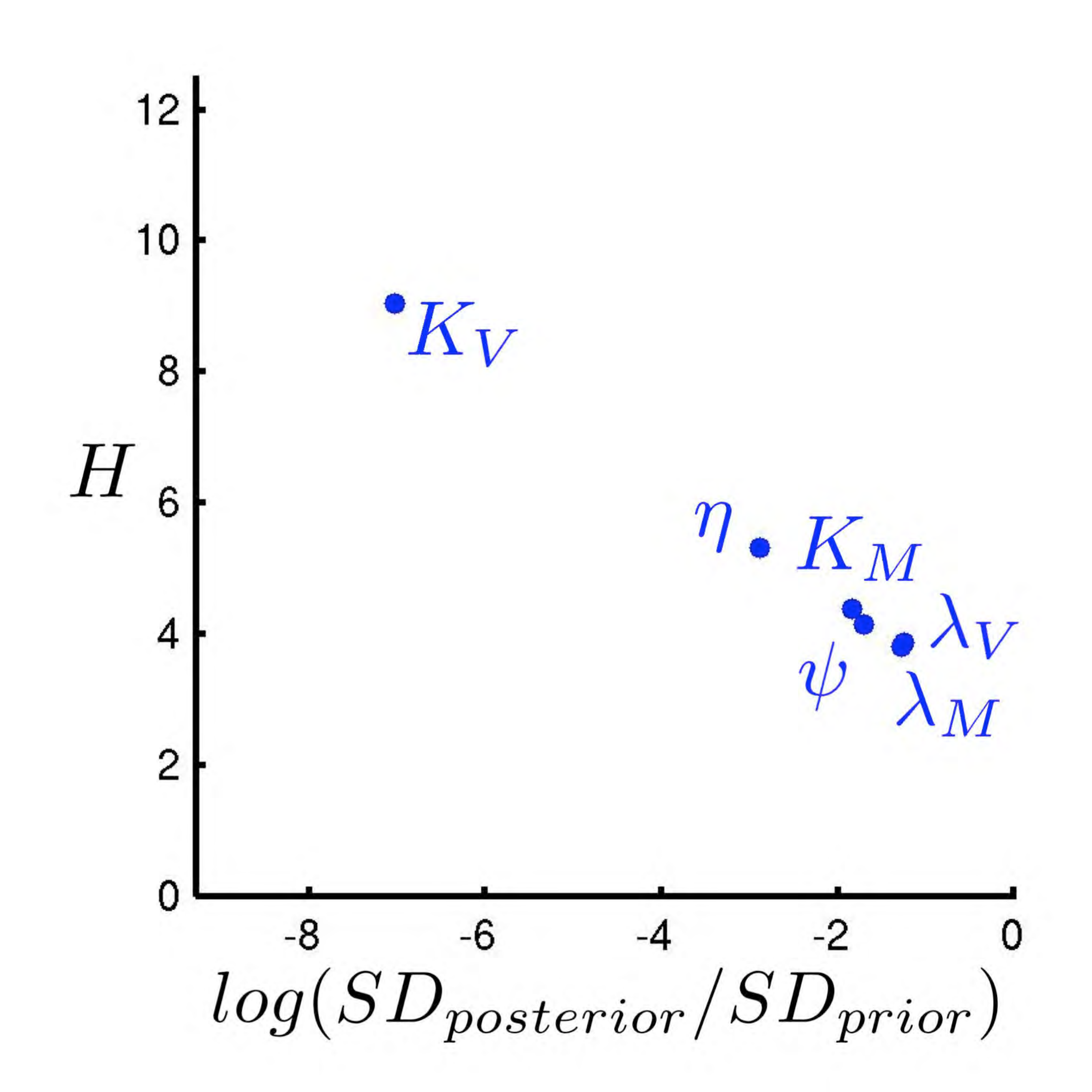} \\
\caption{Macaque 405: Learnability summaries.
Posterior:prior relative entropy $H$ against log of posterior to prior standard deviation, in some
cases implicitly based on
parameters following inverse c.d.f. transform to uniform priors as relevant.}
\end{figure}

\newpage
\begin{figure}[ht!]
  \centering
 \includegraphics[width=2.0in]{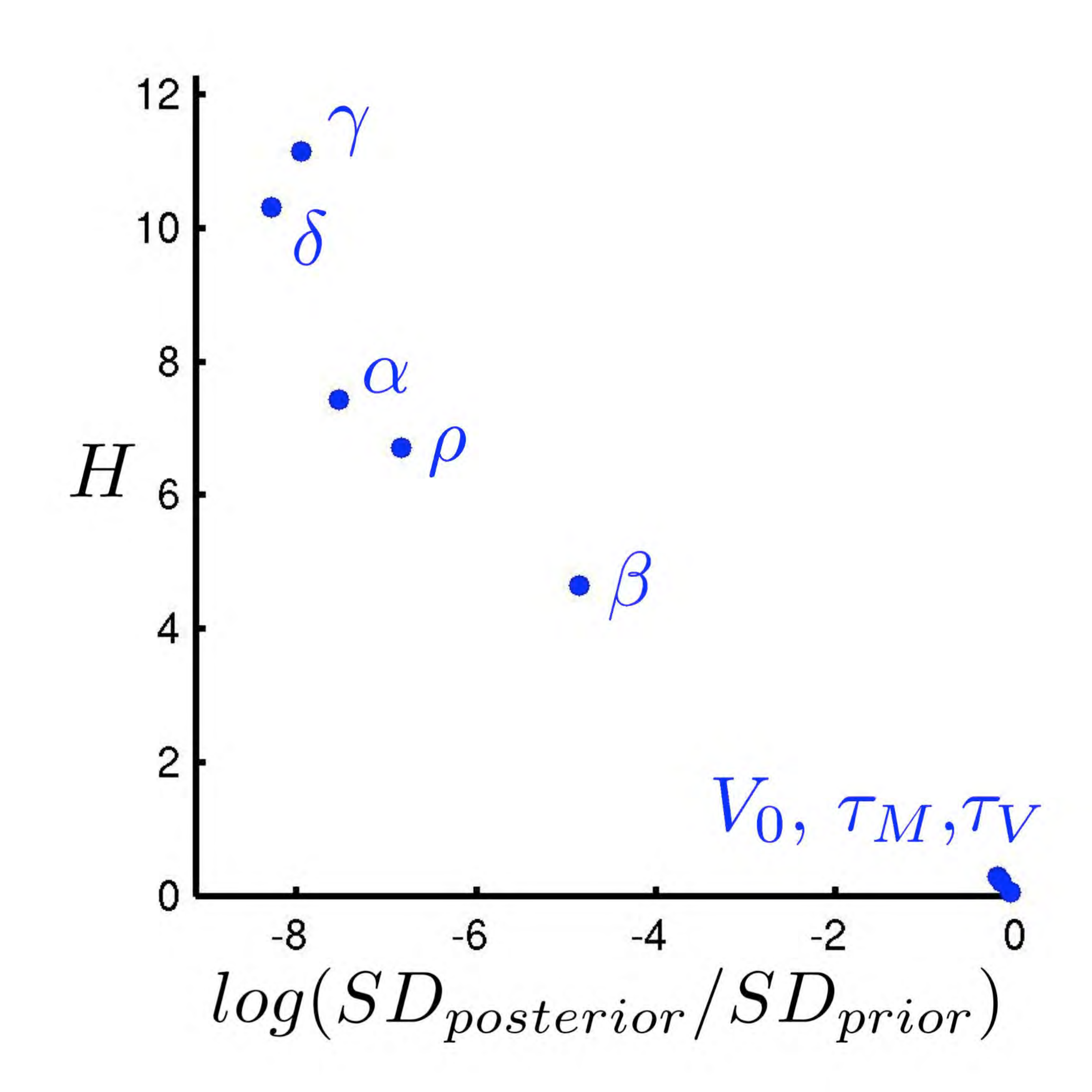}
  \includegraphics[width=2.0in]{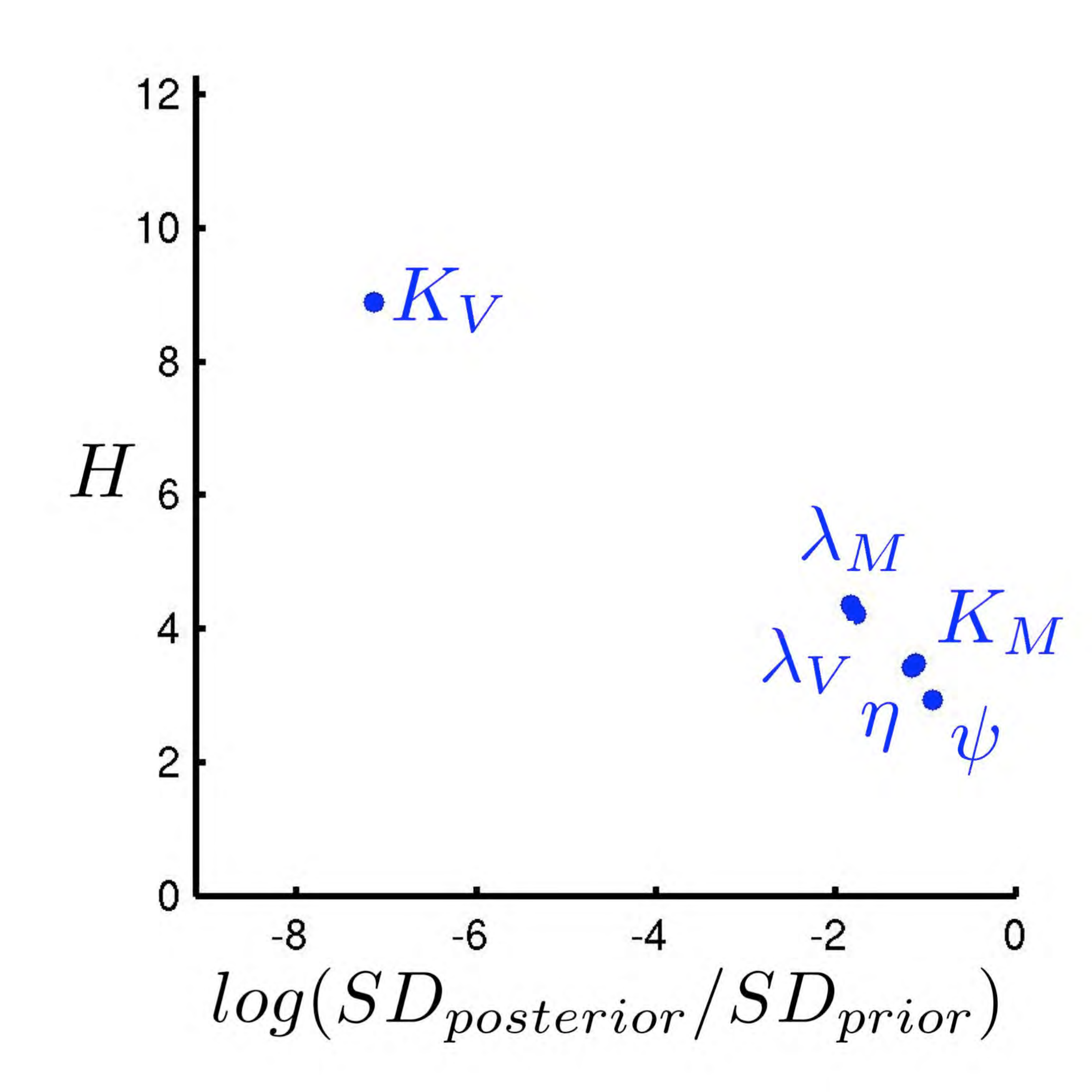} \\
\caption{Macaque 407: Learnability summaries.
Posterior:prior relative entropy $H$ against log of posterior to prior standard deviation, in some
cases implicitly based on
parameters following inverse c.d.f. transform to uniform priors as relevant.}
\end{figure}

\begin{figure}[ht!]
  \centering
 \includegraphics[width=2.0in]{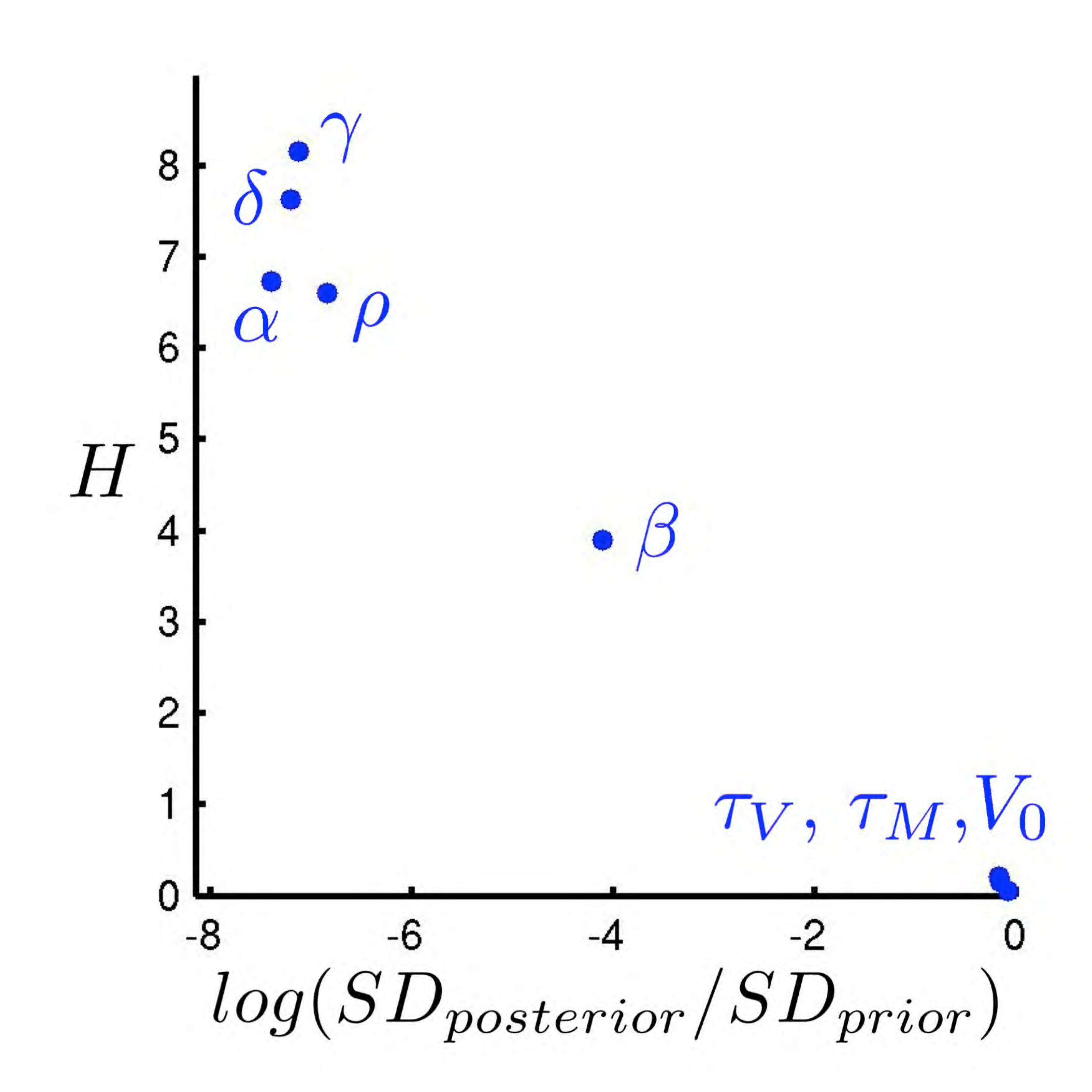}
  \includegraphics[width=2.0in]{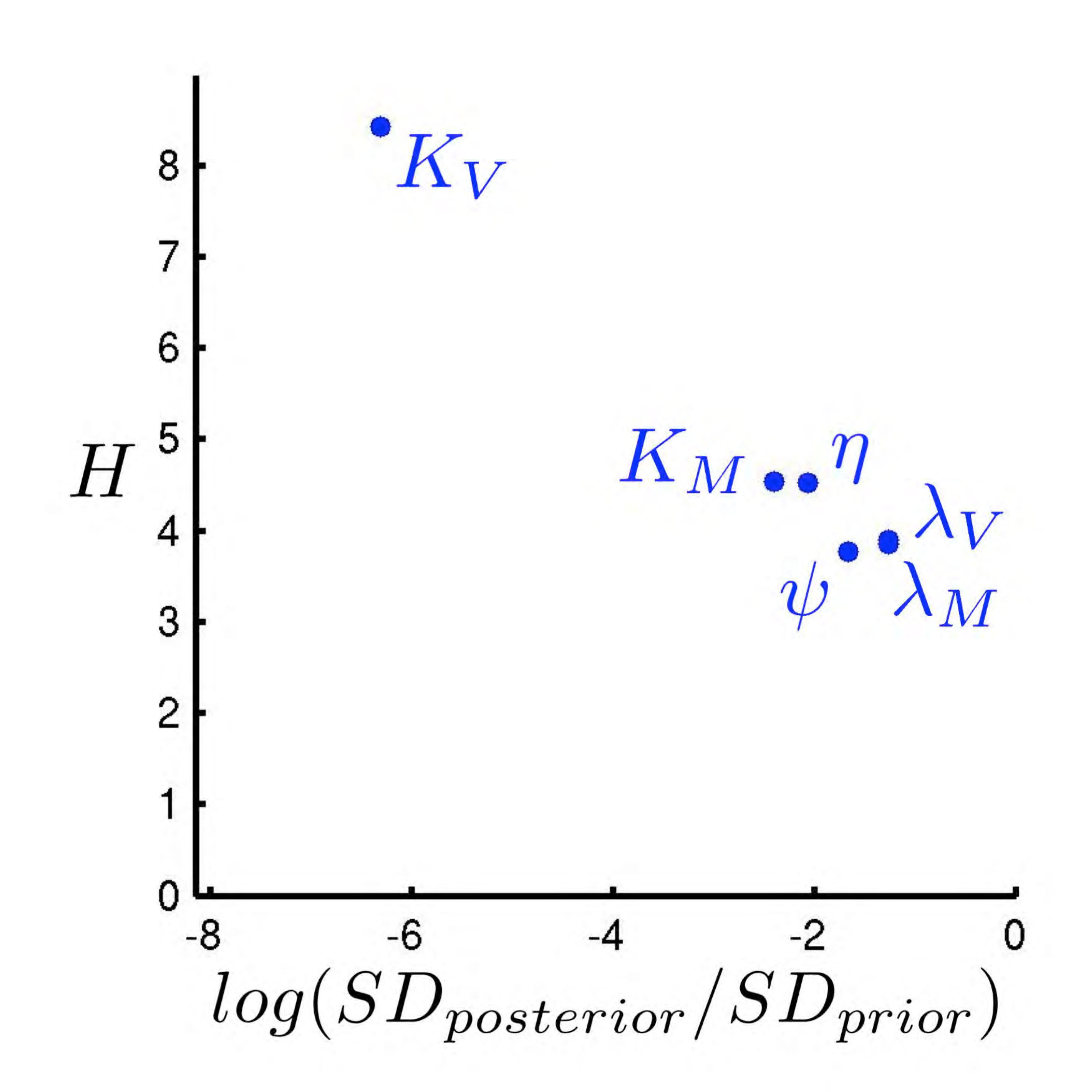} \\
\caption{Macaque 408: Learnability summaries.
Posterior:prior relative entropy $H$ against log of posterior to prior standard deviation, in some
cases implicitly based on
parameters following inverse c.d.f. transform to uniform priors as relevant.}
\end{figure}

\newpage
\begin{figure}[ht!]
  \centering
 \includegraphics[width=2.0in]{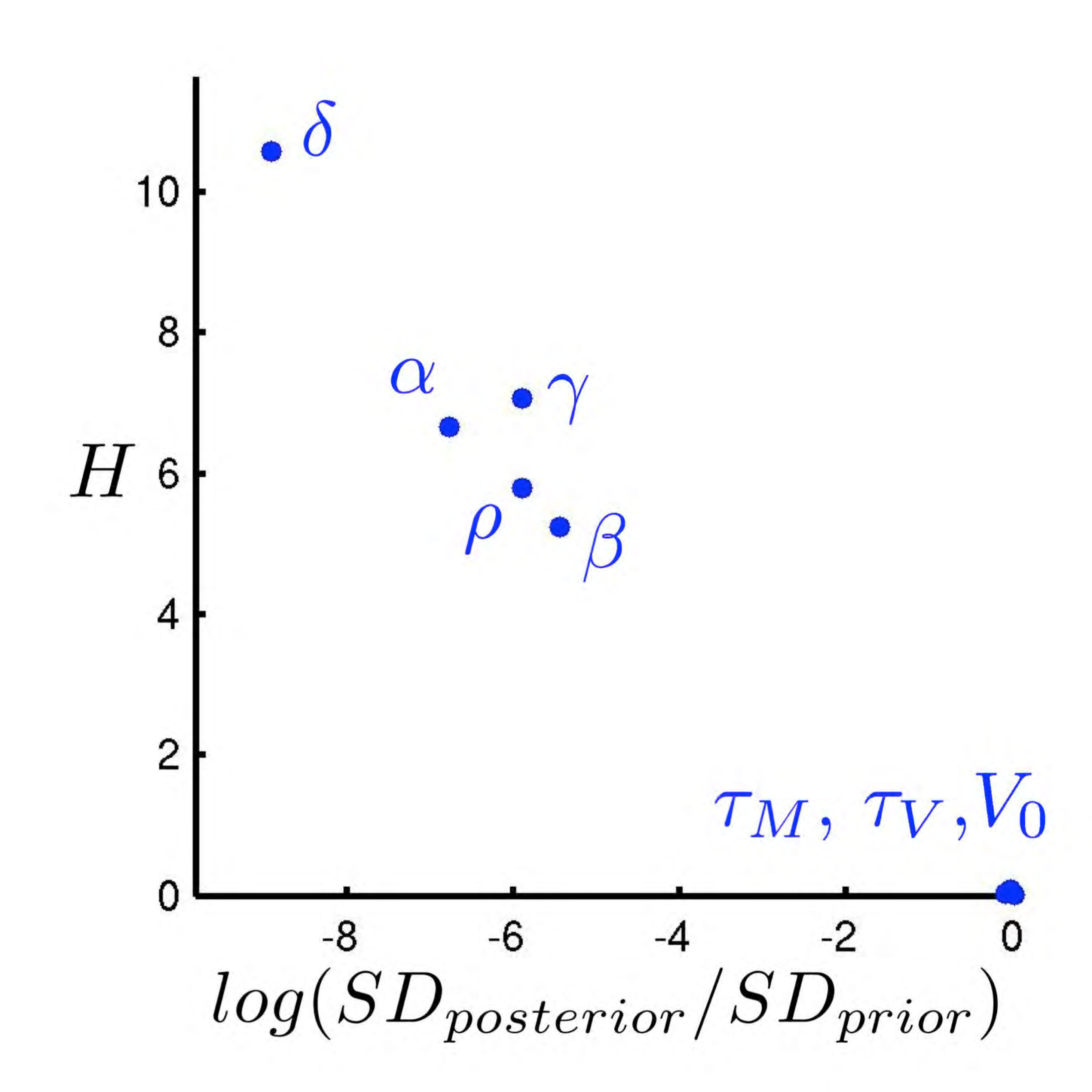}
  \includegraphics[width=2.0in]{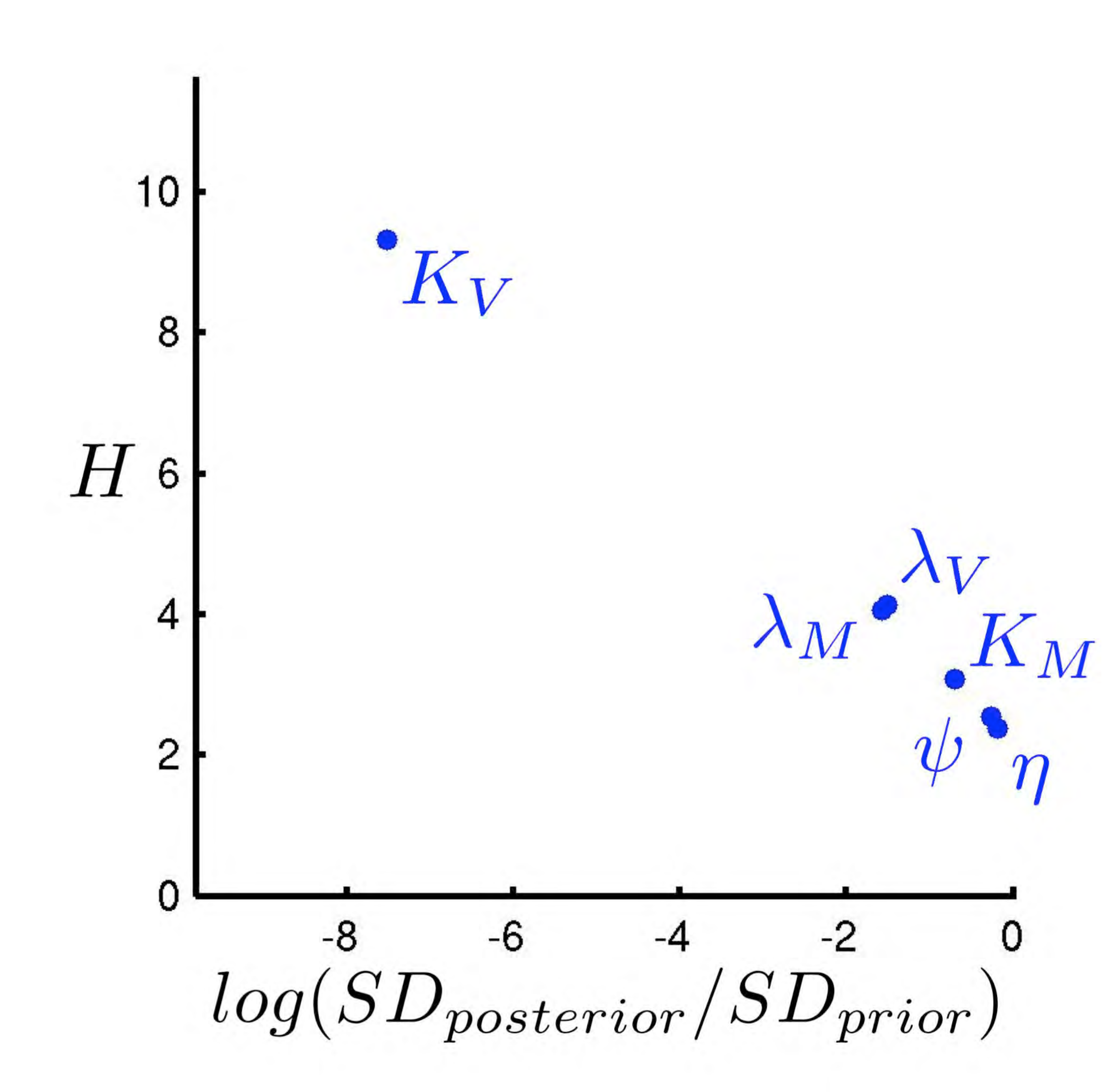} \\
\caption{Macaque 415: Learnability summaries.
Posterior:prior relative entropy $H$ against log of posterior to prior standard deviation, in some
cases implicitly based on
parameters following inverse c.d.f. transform to uniform priors as relevant.}
\end{figure}

\begin{figure}[ht!]
  \centering
 \includegraphics[width=2.0in]{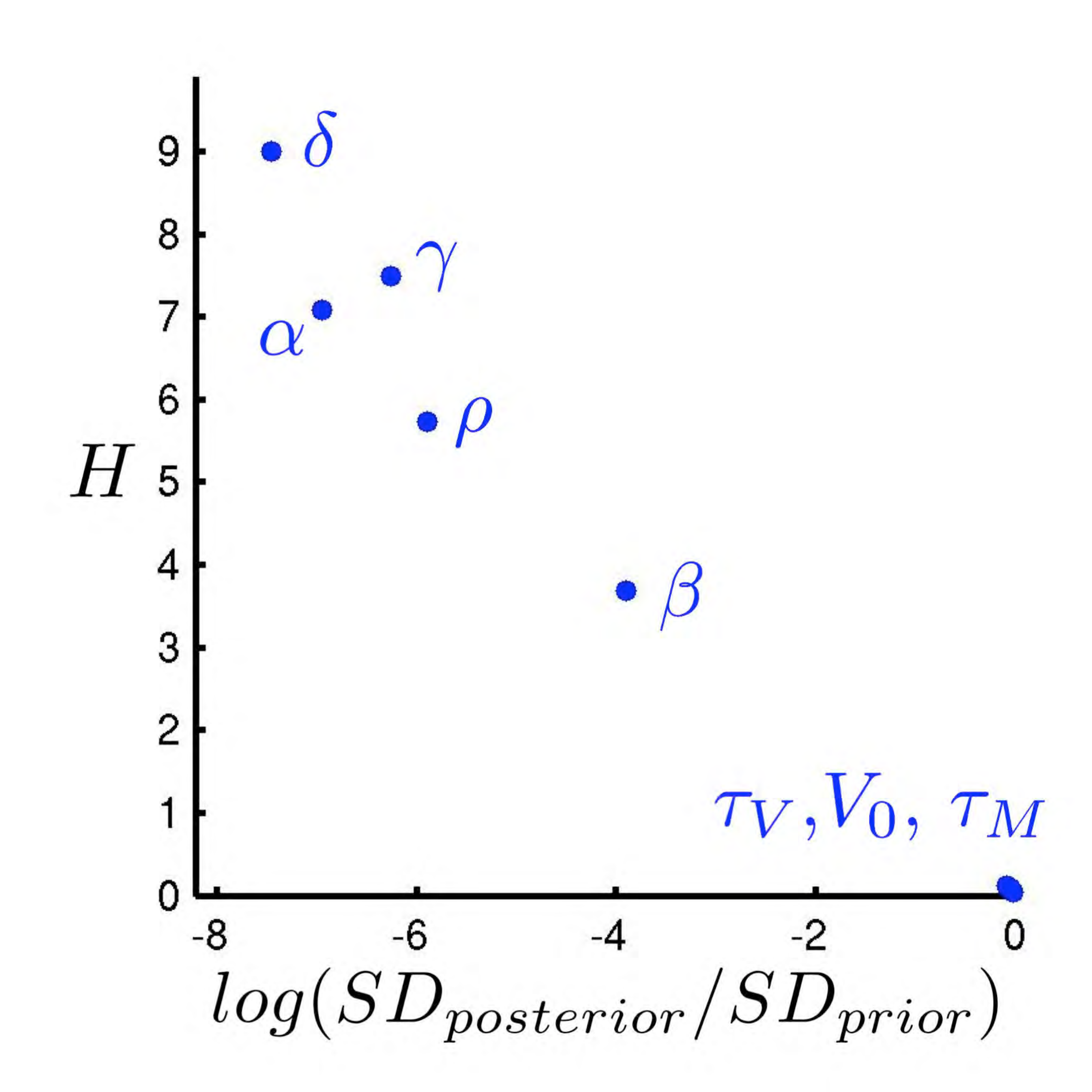}
  \includegraphics[width=2.0in]{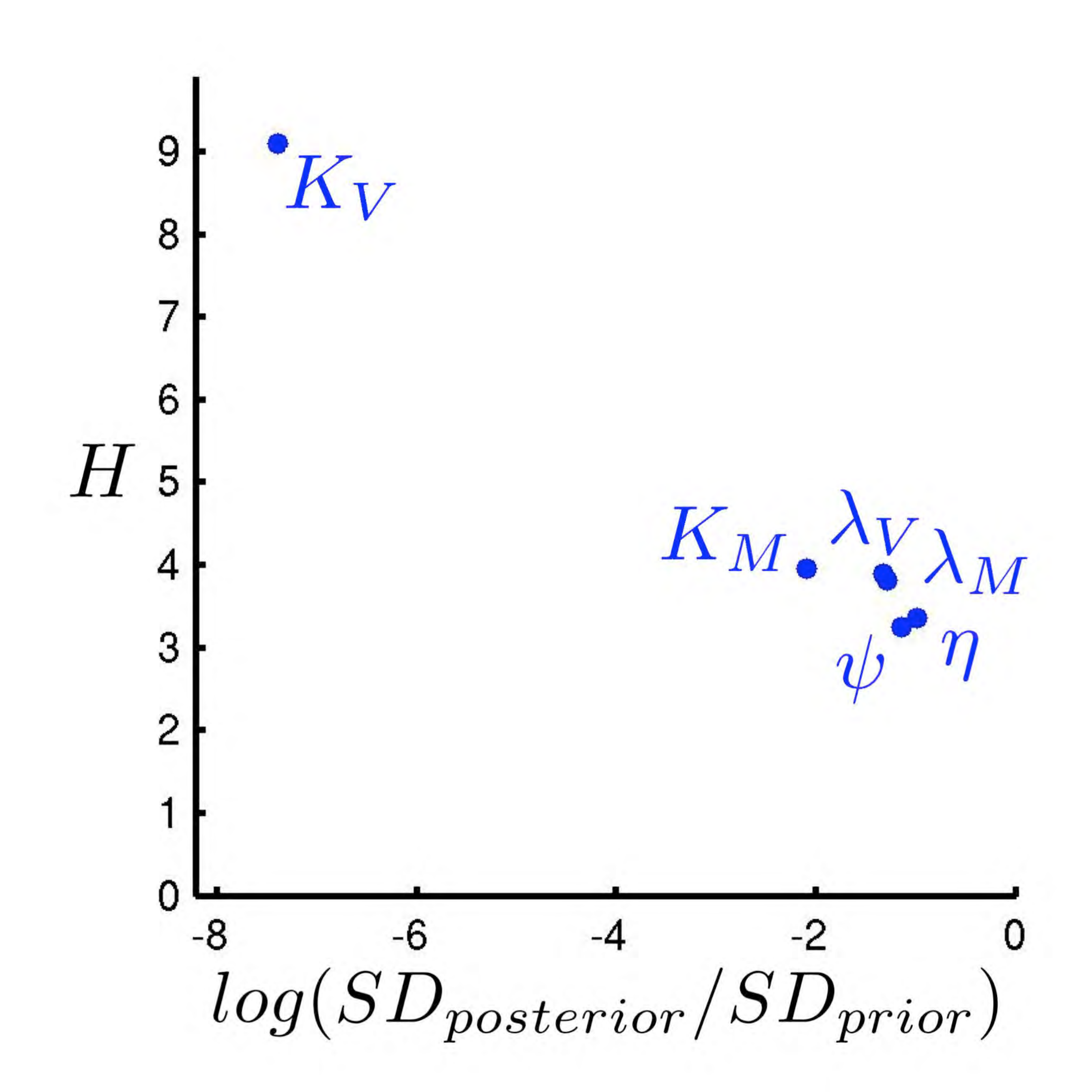} \\
\caption{Macaque 431: Learnability summaries.
Posterior:prior relative entropy $H$ against log of posterior to prior standard deviation, in some
cases implicitly based on
parameters following inverse c.d.f. transform to uniform priors as relevant.}
\end{figure}

\end{document}